\documentclass[paper,11 pt]{article}

\usepackage{etoolbox}
\makeatletter

\apptocmd\@specialoutput{\global\holdinginserts\z@}

\makeatother

\usepackage{cite}
\usepackage{epsfig,amsfonts}
\usepackage{graphicx}
\usepackage{caption}
\usepackage{subcaption}

\usepackage{amsthm,amssymb}

\usepackage{hhline}
\usepackage{braket}
\usepackage{upgreek}
\usepackage[ruled,vlined]{algorithm2e}
\usepackage[linktocpage=true]{hyperref}
\usepackage[dvipsnames]{xcolor}
\usepackage{musicography}
\usepackage[normalem]{ulem}
\definecolor{purple_nice}{rgb}{0.4,0.2,0.7}
\definecolor{fuel_blue}{RGB}{42,162,185}
\definecolor{YInMn_blue}{RGB}{46, 80, 144}
\definecolor{ultramarine}{RGB}{63, 0, 255}
\definecolor{KLEIN_blue}{rgb}{0, 0.18, 0.65}
\hypersetup{
    colorlinks=true,
    linkcolor=ultramarine,
    citecolor=ultramarine,
    filecolor=fuel_blue,
    urlcolor=ultramarine,
}

\usepackage[T1]{fontenc}
\usepackage{amsmath}
\usepackage{mdframed}
\usepackage{graphics}
\usepackage{amsfonts}
\usepackage{amssymb}
\usepackage[multiple]{footmisc}
\usepackage{bm}
\usepackage{amsthm}
\usepackage{slashed}
\usepackage{enumerate,enumitem}
\usepackage{amsbsy}
\usepackage{bbm}
\usepackage{mathrsfs}
\usepackage{url}
\usepackage{bbm} 
\usepackage{color}
\usepackage{framed}
\usepackage{caption}
\usepackage{float}
\usepackage{cancel}
\usepackage{hyperref}
\usepackage{fnpct}
\usepackage{comment}

\newcommand{\tr}{\mathrm{tr}}

\newcommand{\BT}{\mathbf{T}}

\newcommand{\cosm}{\Lambda}

\newcommand{\BA}{\mathbf{A}}

\newcommand{\rad}{z}

\newcommand{\BO}{\mathbf{\Omega}}
\newcommand{\BP}{\bm{\Pi}}

\newcommand{\eigen}{\lambda}
\newcommand{\coupling}{\tau}
\newcommand{\auxcoupling}{\bar{\tau}}
\newcommand{\reigen}{\ell}
\newcommand{\curve}{\Gamma}
\newcommand{\ecoupling}{\varepsilon}
\newcommand{\eauxcoupling}{\bar{\varepsilon}}

\newcommand{\TTb}{\mathrm{T}\overline{\mathrm{T}}}

\usepackage{xcolor}
\usepackage{ulem}
\numberwithin{equation}{section}

\begin{document}
\begin{titlepage}

\title{{\huge \bf  Stress-energy tensor deformations, \\ 
Ricci flows and black holes}}
\author{Nicolò Brizio$^{1}$, Tommaso Morone$^{2}$, and Roberto Tateo$^{3}$\\[0.3cm]}
\date{\small{\textit{Dipartimento di Fisica, Università di Torino and \\ INFN, Sezione di Torino, Via P. Giuria 1, 10125, Torino, Italy}  \\[0.3cm]
$^{1}$\texttt{\href{mailto:nicolo.brizio@unito.it}{nicolo.brizio@unito.it},} $^{2}$\texttt{\href{mailto:tommaso.morone@unito.it}{tommaso.morone@unito.it},}\\ $^{3}$\texttt{\href{mailto:roberto.tateo@unito.it}{roberto.tateo@unito.it}}\\}}
\maketitle
\thispagestyle{empty}

\vspace{3mm}

\begin{abstract}
\vspace{3mm}
\noindent 
This paper reviews and extends the recently discovered connections between marginal and irrelevant stress-energy tensor deformations and gravity theories in arbitrary space-time dimensions.  We start by discussing how $\TTb$ and $\sqrt{\TTb}$ deformations of two-dimensional field theories can be equivalently interpreted as the coupling of the undeformed matter sector to a gravity theory. We then extend this duality to higher-dimensional scenarios by using an approach that relies on the non-trivial eigenvalue degeneracy characterising the energy-momentum tensor of specific physical theories. We also explore incorporating dynamical degrees of freedom in the gravity sector, and show that the deformed space-time geometry induced by the $\TTb$-like deformations defines a Ricci-Bourguignon flow of the metric tensor, which reduces to a Ricci flow in four dimensions. Finally, exploiting a dressing-type mechanism for the action functional characterizing a broad class of $\TTb$-like  deformations we study explicit examples, such as Einstein-Ricci solitons, $(d-1)$-form field theories, and spherically symmetric electrovacuum solutions.

\end{abstract}

\newpage

\end{titlepage}

\newpage
\tableofcontents

\vspace{7mm}

\hrule

\section{Introduction}
Based on earlier results and observations \cite{Zamolodchikov:1991vx, Mussardo:1999aj, Zamolodchikov:2004ce, Dubovsky:2012sh, Dubovsky:2012wk, Caselle:2013dra}, the $\TTb$ flow in the space of two-dimensional quantum field theories was defined in \cite{Cavaglia:2016oda, Smirnov:2016lqw} \footnote{See also the earlier comment at the end of Section  2 in \cite{Caselle:2013dra}.} as being self-consistently driven by a composite irrelevant operator \cite{Zamolodchikov:2004ce} \footnote{As has become customary in the literature, the notation employed for the deforming operators originates from their conformal limit. Consequently, a precise definition of such objects is necessary in a general setup.} 
\begin{equation}\label{ttbar_intro}
  \TTb = \frac{1}{2}\left[(T^\mu{}_\mu)^2-T^{\mu\nu}T_{\mu\nu}\right] = \det T_{\mu\nu}\,,
\end{equation}
where $T_{\mu\nu}$ denotes the stress-energy tensor of the theory. Although the irrelevant nature of the deformation introduces interactions that grow stronger at high energies, possibly leading to a loss of control over the theory's behavior in the ultraviolet regime, many interesting quantities can be computed exactly and explicitly in terms of the data of the undeformed theory, including the $S$-matrix \cite{Mussardo:1999aj, Dubovsky:2012wk, Caselle:2013dra, Cavaglia:2016oda, Smirnov:2016lqw}, the deformed classical action \cite{Cavaglia:2016oda, Bonelli:2018kik, Conti_Iannella}, the finite volume spectrum on an infinite cylinder\cite{Dubovsky:2012wk, Caselle:2013dra, Cavaglia:2016oda, Smirnov:2016lqw} and the torus partition function \cite{Cardy:2018sdv, Datta:2018thy}. Over the past few years, $\TTb$ deformations were shown to be linked to numerous topics in theoretical physics, such as string theory \cite{Baggio:2018gct, Dei:2018jyj, Chakraborty:2019mdf, Callebaut:2019omt}, holography \cite{McGough:2016lol, Giveon:2017nie, Gorbenko:2018oov, Kraus:2018xrn, Hartman:2018tkw, Guica:2019nzm, Jiang:2019tcq,Jafari:2019qns,Griguolo:2021wgy, Fichet:2023xbu, He:2024xbi}, and generalised hydrodynamics \cite{Doyon:2021tzy, Cardy:2020olv, Doyon:2023bvo}. We refer the reader to \cite{Jiang:2019epa} for a pedagogical introduction to the subject. The flow generated by \eqref{ttbar_intro} preserves most of the symmetries of the seed (i.e., undeformed) theory, including integrability.\footnote{This last property is a characteristic of a much larger class of deformations, called double current deformations \cite{Dubovsky:2023lza}, and encompassing all the recent two-dimensional generalisations of the $\TTb$ deformation.} This feature ultimately originates from the geometric nature of the $\TTb$ deformation \cite{Conti:2018tca}, which can be understood in terms of a global, field-dependent change of coordinates: from this viewpoint, the conserved charges of the original theory are mapped into a new set of integrals of motion by the deformation. Alternatively, the geometrical perspective on $\TTb$ flows can be interpreted as a coupling of the undeformed theory to a gravity theory which describes the evolution of an auxiliary space-time metric under the flow triggered by \eqref{ttbar_intro} \cite{Dubovsky_2017,Cardy:2018sdv, Tolley:2019nmm, Caputa:2020lpa}. Let us also mention that a powerful method to compute $\TTb$ deformed actions is given by the light-cone gauge approach developed in \cite{Frolov:2019nrr, Frolov:2019xzi}.

In recent years, new variants of stress tensor flows inspired by the two-dimensional $\TTb$ flow have been studied, mostly at the classical level. Among these several proposals, the two-dimensional $\sqrt{\TTb}$ flow \cite{Conti_2022, Ferko:2022cix, Ferko:2023ruw, Babaei-Aghbolagh:2022uij, Babaei-Aghbolagh:2022leo}, generated by the marginal operator,
\begin{equation}
    \sqrt{\TTb} = \sqrt{\frac{1}{2}(T^\mu{}_\mu)^2- \frac{1}{4}T^{\mu\nu}T_{\mu\nu}}\,,
\end{equation}
have attracted significant attention. While its quantum-mechanical definition remains uncertain, \footnote{There have been, however, attempts towards a quantum definition of $\sqrt{\TTb}$ deformations within the context of two-dimensional CFTs (see, for example, \cite{Hadasz:2024pew}).} the $\sqrt{\TTb}$ deformation displays some rather surprising properties even at the classical level. Particularly, it commutes with the $\TTb$ flow, and it was shown to preserve classical integrability for a class of integrable field theories \cite{Borsato:2022tmu}.

Generalisations of stress-energy tensor flows, akin to those defined through the $\TTb$ operator, have been proposed in various works \cite{Taylor:2018xcy, Conti_2022, Bonelli:2018kik, Conti_Iannella, Cardy:2018sdv, Babaei-Aghbolagh:2020kjg, Hou:2022csf, Ferko:2023sps}, with the aim of extending the $\TTb$ setup — at least at the classical level — to higher space-time dimensions. These investigations, alongside the introduction of the so-called  Modified Maxwell (ModMax) theory \cite{Bandos:2020jsw}, and the discovery that both Born-Infeld and ModMax arise from Maxwell theory through a Lagrangian flow involving $\TTb$ and $\sqrt{\TTb}$-like composite fields respectively \cite{Conti_Iannella, Babaei-Aghbolagh:2020kjg, Ferko:2022iru}, have sparked a revival of interest in non-linear electrodynamics \cite{Bandos:2020jsw, Sorokin:2021tge, Lechner:2022qhb, Ferko:2019oyv}.

Furthermore, several connections between gravity theories and stress tensor deformations of field theories in arbitrary dimensions have recently emerged \cite{Morone:2024ffm, Babaei-Aghbolagh:2024hti, Tsolakidis:2024wut, Floss:2023nod}, revealing numerous links between non-linear theories of matter fields and gravity models beyond General Relativity. The emerging dynamics related to the gravitational degrees of freedom can admit a broad spectrum of equivalent descriptions, both in terms of bimetric gravity theories and via the Palatini framework of Ricci-based gravity theories \cite{Vollick:2003qp, Banados:2008rm, Banados:2008fj,Banados_2010, jimenez_2018, Olmo:2020fnk, Guerrero:2020azx, Nascimento:2019qor, Afonso:2019fzv, Pani:2012qb, Banerjee:2021auy, Olmo:2013gqa, Pereira:2023bxt}. Specifically, a class of four-dimensional $\TTb$-like deformations, frequently associated with the emergence of Born-Infeld functionals, was shown to admit an analogous description in terms of coupling of the seed theory to Eddington-Inspired Born-Infeld gravity \cite{Morone:2024ffm}, described by the Palatini action
\begin{equation}
 S_{EiBI}[g,\Gamma] =  \frac{1}{\coupling}\int \mathrm{d}^4 x \left[\sqrt{-\det\left(g_{\mu\nu}+\coupling R_{\mu\nu}[\Gamma]\right)}-\sqrt{-g}\right]\,,  
\end{equation}
where $\coupling$ denotes the $\TTb$-like coupling parameter.
Further extensions of the correspondence allowed to achieve a gravitational description of marginal $\sqrt{\TTb}$-like flows in arbitrary dimensions \cite{Babaei-Aghbolagh:2024hti, Tsolakidis:2024wut} in terms of bimetric gravity theories. 

In this paper, combining the eigenvalue-based approach to stress tensor deformations of matter action functionals introduced in \cite{Babaei-Aghbolagh:2024hti} with the dressing mechanism associated with $\TTb$-like flows \cite{Conti_2022, Morone:2024ffm}, we propose an intuitive and accessible procedure to recover the gravity functionals associated to a class of deformations. An equivalent Palatini formulation is recovered for the gravity sector corresponding to $\sqrt{\TTb}$ deformations in $d=4$, characterised by the action
\begin{equation}
  S[g,\Gamma] =\frac{1}{2}\int \mathrm{d}^4 x \sqrt{-g} \left[g^{\mu\nu}R_{\mu\nu}[\Gamma]\cosh \frac{\gamma}{2} \pm \sinh \frac{\gamma}{2}\sqrt{4R^{\mu \nu} [\Gamma] R_{\mu \nu}[\Gamma]-\left(g^{\mu \nu} R_{\mu \nu}[\Gamma]\right)^2}\right]\,,   
\end{equation}
where $\gamma$ is the dimensionless $\sqrt{\TTb}$ coupling. The metric flows associated with a broad family of stress tensor deformations are then shown to be linked to Ricci-type flows \cite{Hamilton1982, rbf::15} of the space-time manifold. Particularly, when the deformed matter sector is coupled to gravity in four space-time dimensions, the deformed metric satisfies
\begin{equation}\label{introductionrf}
\frac{\mathrm{d}h^{(\coupling)}_{\mu\nu}}{\mathrm{d}\coupling} = R_{\mu\nu}[h^{(\coupling)}]\,, 
\end{equation}
where $h^{(\coupling)}_{\mu\nu}$ denotes the $\TTb$-like deformed metric, $R_{\mu\nu}[h^{(\coupling)}]$ its Ricci curvature, and $\coupling$ is the flow parameter.

We study solitonic configurations of the Ricci flow \eqref{introductionrf}, which, in their simplest form, are realised by de Sitter and Anti de Sitter spacetimes. Additionally, $(d-1)$-form field theories in arbitrary space-time dimensions are analysed, and we show that when the seed theory has the form $\mathcal{L}^{(0)}_m=K$, where $K \propto F^{\mu_1\dots\mu_d}F_{\mu_1\dots\mu_d}$, the corresponding deformed Lagrangian is given by
\begin{equation}
\mathcal{L}^{(\coupling)}_m=\frac{3}{4\coupling} \left[ {}_3F_2\left(-\frac{1}{2},-\frac{1}{4},\frac{1}{4};\frac{1}{3},\frac{2}{3};\frac{256 }{27}\coupling K\right)-1\right]\,.    
\end{equation}

The $\TTb$/gravity duality is further applied in the study of deformed black hole solutions, whose associated space-time geometries satisfy \eqref{introductionrf}. Within this framework, $\TTb$-like deformations in $d=4$ can be alternatively viewed as a mathematical tool to generate and analyse solutions to Lorentzian Ricci flows.\\

The remainder of this paper is organised as follows. In Section \ref{sec::2d}, we focus on the two-dimensional case, and reproduce the ghost-free massive gravity action associated with the $\TTb$ deformation as introduced by Tolley in 2019 \cite{Tolley:2019nmm}. In the two-dimensional framework, where the conservation laws for the stress-energy tensor allow promoting the local metric deformation to a global change of coordinates, the geometric viewpoint is extended to the quantum level \cite{Aramini:2022wbn}. $\sqrt{\TTb}$ deformations of two-dimensional field theories are discussed from a purely classical perspective, and the metric deformation associated with the marginal flow \cite{Ebert:2022ehb} is employed to recover the gravity functional associated with the deformation \cite{Babaei-Aghbolagh:2024hti}. 
In Section \ref{sec::4d}, resting on the peculiar eigenvalue structure of the energy-momentum tensor which characterises a specific family of physical theories, we lay out a generalisation of the $\TTb$ operator \eqref{ttbar_intro} to higher-dimensional space-times, which reproduces the proposals introduced in \cite{Conti_Iannella, Ferko:2023sps}. By combining the eigenvalue-based approach with a higher-dimensional version of the dressing mechanism, we derive the gravity action associated with a class of $\TTb$-like deformations, which, in four dimensions, is shown to be dynamically equivalent to Eddington-inspired Born-Infeld gravity. We then employ similar strategies to study $\sqrt{\TTb}$-like deformations of higher-dimensional field theories.
In Section \ref{sec:ricciflow}, we show that a broad family of $\TTb$-like deformations can be interpreted in terms of Ricci-Bourguignon flows \cite{rbf::15} of the space-time geometry. In the four-dimensional setting, Ricci flows emerge naturally. We study the simplest instance of Einstein-Ricci solitons, which correspond to a matter sector describing homogeneous vacuum energy contributions, and we discuss singular limits arising when a $\TTb$-like deformation is performed on the model. We also examine the case of $(d-1)$-form field theories in arbitrary space-time dimensions, reproducing and extending the results of \cite{Conti:2018jho}.
In Section \ref{sec:black_holes}, resting on the metric deformations which characterise stress tensor deformations in four dimensions, we study static, spherically symmetric electrovacuum solutions which emerge when the deformed matter sector is coupled to General Relativity. We analyse extremal black hole solutions, showing that their associated Bekenstein-Hawking entropy is linked to the Ricci flow of a two-sphere induced by the $\TTb$-like flow parameter. 
\section{Two-dimensional $\TTb$ deformations}\label{sec::2d}
In this section, we discuss how the established equivalence between $\TTb$ deformations and topological gravity in curved space-times \cite{Tolley:2019nmm} can be linked with the recent interpretation of classical $\TTb$ deformations as adiabatic flows in the configuration space of metrics \cite{Conti_2022, Morone:2024ffm}. The $\TTb$/gravity duality is then uplifted to the quantum level through the existence of a global change of coordinates related to the metric deformation. Two-dimensional $\sqrt{\TTb}$ deformations and their gravity duals are discussed by using similar strategies.
\subsection{The $\TTb$ dressing mechanisms and the gravity picture}
Consider a family of action functionals denoted by $S_m^{(\coupling)}$, which depend on a dimensional parameter $\coupling$, and satisfy the $\TTb$ flow equation 
\cite{Cavaglia:2016oda, Smirnov:2016lqw}
\begin{equation}\label{ttb2}
    \frac{\partial S_m^{(\coupling)}}{\partial\coupling} = \frac{1}{2} \int \mathrm{d}^2x \sqrt{-g} \,\left[\left(T^{(\coupling)\mu}{}_{\mu}\right)^2-T^{(\coupling)\mu\nu}T^{(\coupling)}_{\mu\nu}\right],\quad g := \det g_{\mu\nu}\,,
\end{equation}
where $T^{(\coupling)}_{\mu\nu}$ denotes the Hilbert stress-energy tensor of the theory,
\begin{equation}\label{define_T}
    T_{\mu\nu}^{(\coupling)} = -\frac{2}{\sqrt{-g}}\frac{\delta S_m^{(\coupling)}}{\delta g^{\mu\nu}}\,.
\end{equation}
The theory $S^{(0)}_m$ at $\coupling=0$ is typically referred to as the seed theory of the $\TTb$ flow.  Notably, it has been demonstrated \cite{Conti_2022,Morone:2024ffm} that the data from the seed theory can be used to reconstruct the full set of solutions to equation \eqref{ttb2} via a field-dependent deformation of the metric, along with the addition of a dressing term:
\begin{equation}\label{dressing}
\begin{split}
  S_m^{(\coupling)} [\varphi,h^{(\coupling)}] &= \left.\left\{S_m^{(0)}[\varphi,g] - \frac{\coupling}{2} \int \mathrm{d}^2x \sqrt{-g} \,\left[\left(T^{(0) \mu}{ }_\mu\right)^2-T^{(0)\mu\nu}T^{(0)}_{\mu\nu}\right]\right\}\right|_{g=g(h)}\\
  &=\left.\left\{S_m^{(0)}[\varphi,g] - \frac{\coupling}{2} \int \mathrm{d}^2x \sqrt{-g} \,T^{(0)\mu\nu}\widehat{T}^{(0)}_{\mu\nu}\right\}\right|_{g=g(h)} \,,
\end{split}
\end{equation}
where we have introduced the auxiliary tensor
\begin{equation}
\widehat{T}^{(0)}_{\mu\nu} = g^{\alpha\beta}T^{(0)}_{\alpha\beta}g_{\mu\nu} - T^{(0)}_{\mu\nu}\,.   
\end{equation}
To determine the constraints on the deformed metric $h^{(\coupling)}_{\mu\nu}$ induced by the dynamical identity \eqref{dressing}, we move away from the seed theory, and consider the auxiliary parameter $s := \coupling + c$, where $c$ is a constant. Then, under an infinitesimal deformation $\delta s$ of the parameter $s$, the $\TTb$ flow equation \eqref{ttb2} reads
\begin{equation}
    S^{(s+\delta s)}_m[\varphi, h^{(s)}] = S^{(s)}_m[\varphi, h^{(s)}] + \frac{\delta s}{2}\int \mathrm{d}^2  x \sqrt{-h^{(s)}} \,T^{(s)\mu\nu}\widehat{T}^{(s)}_{\mu\nu}\,,
\end{equation}
where
\begin{equation}
    T_{\mu\nu}^{(s)}:= -\frac{2}{\sqrt{-h^{(s)}}} \frac{\delta S^{(s+\delta s)}_m}{\delta (h^{-1})^{(s)\mu\nu}}\,,
\end{equation}
and $h^{(s)}:= \det h^{(s)}_{\mu\nu}$. Now let $h^{(s+\delta s)}_{\mu\nu} = h^{(s)}_{\mu\nu}+\delta s\, q_{\mu\nu}^{(s)}$, where $q_{\mu\nu}^{(s)}$ is a dynamical deformation of the metric $h^{(s)}_{\mu\nu}$, and consider
\begin{equation}
\begin{split}
 S^{(s+\delta s)}_m[\varphi, h^{(s+\delta s)}] =&\,  S^{(s+\delta s)}_m[\varphi, h^{(s)}] + \frac{\delta s}{2} \int \mathrm{d}^2 x\sqrt{-h^{(s)}}\, q_{\mu\nu}^{(s)} T^{(s)\mu\nu}  \\
 =&\, S^{(s)}_m[\varphi, g] +\frac{\delta s}{2}\int \mathrm{d}^2  x \sqrt{-h^{(s)}} \,T^{(s)\mu\nu}\widehat{T}^{(s)}_{\mu\nu}\\ &+ \frac{\delta s}{2} \int \mathrm{d}^2 x\sqrt{-h^{(s)}}\, T^{(s)\mu\nu} q^{(s)}_{\mu\nu}\,.
\end{split}
\end{equation}
Note that, setting $c =-\coupling$ for some fixed value of the flow parameter, the equivalence \eqref{dressing} is satisfied when $q^{(s)}_{\mu\nu} = -2 \widehat{T}^{(s)}_{\mu\nu}$. Finally, using the identity
\begin{equation}
    S^{(s)}_m[\varphi, h^{(s)} +\delta s\, q_{\mu\nu}^{(s)} ] =  S^{(s)}_m\left[\varphi, h^{(s+\delta s)} -\delta s\left(\frac{\mathrm{d} h^{(s)}}{\mathrm{d}s} -q_{\mu\nu}^{(s)}\right) \right]\,,
\end{equation}
we obtain the condition
\begin{equation}
    \frac{\mathrm{d}h^{(s)}}{\mathrm{d}s} = - 2 \widehat{T}^{(s)}_{\mu\nu}\,.
\end{equation}
In terms of the original coupling $\coupling$, the above equation reduces to
\begin{equation}\label{metric_flow}
    \frac{\mathrm{d}h_{\mu\nu}^{(\coupling)}}{\mathrm{d}\coupling} = -2\left[(h^{-1})^{(\coupling)\alpha\beta}T^{(\coupling)}_{\alpha\beta}h^{(\coupling)}_{\mu\nu}-T^{(\coupling)}_{\mu\nu}\right]=-2\,\widehat{T}^{(\coupling)}_{\mu\nu}\,,\quad h_{\mu\nu}^{(0)} = g_{\mu\nu}\,.
\end{equation}
In two space-time dimensions, a general solution to \eqref{metric_flow} is provided by \cite{Conti:2018tca}
\begin{equation}\label{metric_flow_sol}
    h_{\mu\nu}^{(\coupling)} = g_{\mu\nu}-2\coupling \widehat{T}_{\mu\nu}^{(0)}+\coupling^2 \widehat{T}^{(0)\alpha}_\mu \widehat{T}_{\alpha\nu}^{(0)}\,.
\end{equation}
In what follows, we denote by $\BT^{(0)} = \{g^{\mu\alpha}T^{(0)}_{\alpha\nu}\}_{\mu,\nu = 0,1}$ the matrix associated to the stress tensor of the seed theory, and we assume that $\BT^{(0)}$ can be brought into a diagonal form by a similarity transformation $\mathbf{U}$, such that
\begin{equation}
    \mathbf{T}^{(0)} = \mathbf{U}\begin{pmatrix}
\eigen_1 & 0 \\
0 & \eigen_2
\end{pmatrix}\mathbf{U}^{-1}\,, 
\end{equation}
and the auxiliary matrix $ \mathbf{\widehat{T}}^{(0)}$ can be written in the form:
\begin{equation}\label{diagonal_that}
    \mathbf{\widehat{T}}^{(0)} = \mathbf{U} \left[ \begin{pmatrix}
\eigen_1+\eigen_2 & 0 \\
0 & \eigen_1+\eigen_2
\end{pmatrix}  - \begin{pmatrix}
\eigen_1 & 0 \\
0 & \eigen_2
\end{pmatrix} \right]  \mathbf{U}^{-1} = \mathbf{U}\begin{pmatrix}
\eigen_2 & 0 \\
0 & \eigen_1
\end{pmatrix}\mathbf{U}^{-1}\,.
\end{equation}
Introducing the deformation matrix $\BO = \{g^{\mu\alpha}h_{\alpha\nu}^{(\coupling)}\}_{\mu,\nu = 0,1}$, equation \eqref{metric_flow_sol} can be equivalently written as
\begin{equation}\label{omega_mat_in_2}
    \bm{\mathrm{\Omega}} = \left(\mathbf{1}-\coupling \widehat{\mathbf{T}}_0\right)^2\,.
\end{equation}
Since $\BO$ is a polynomial function of ${\BT}^{(0)}$, the two matrices commute, and they can be simultaneously diagonalised: using \eqref{diagonal_that}, and letting $\omega_1$ and $\omega_2$ denote the eigenvalues of the deformation matrix, we have
\begin{equation}\label{omega_tau}
    \omega_1 =\left(1-\coupling\eigen_2 \right)^2\,,\quad  \omega_2 =\left(1-\coupling\eigen_1 \right)^2\,.
\end{equation}
In addition, since $\tr[\mathbf{T}^{(0)}\widehat{\mathbf{T}}^{(0)}] = d \eigen_1\eigen_2$, the dressing formula \eqref{dressing} can be expressed in terms of the eigenvalues of $\mathbf{T}^{(0)}$ as
\begin{equation}\label{eigen_dress}
 S^{(\coupling)}_m[\varphi,h^{(\coupling)}] \,\,\dot{=}\,\, S^{(0)}_m[\varphi,g] - \coupling\int \mathrm{d}^2x \sqrt{-g} \,\eigen_1\eigen_2\,,
\end{equation}
where we used the $\,\dot{=}\,$ symbol to emphasise that the above equivalence holds provided that \eqref{metric_flow} is satisfied.  Using equation \eqref{omega_tau}, one can explicitly introduce the metric $h_{\mu\nu}^{(\coupling)}$ in \eqref{eigen_dress}, by expressing each $\eigen_i$ in terms of the eigenvalues of $\BO$: 
\begin{equation}\label{2dgrav_omega}
\begin{split}
    S_m^{(\coupling)} [\varphi,h^{(\coupling)},g] &= S_m^{(0)}[\varphi,g] - \frac{1}{\coupling} \int \mathrm{d}^2 x \sqrt{-g}\left(\sqrt{\omega_1}-1\right)\left(\sqrt{\omega_2}-1\right)\\
    &= S_m^{(0)}[\varphi,g] - \frac{1}{\coupling} \int \mathrm{d}^2 x\sqrt{-g} \det\left({\sqrt{\BO}}-\mathbf{1}\right)\,.
\end{split}
  \end{equation}
Equation \eqref{2dgrav_omega} can be more compactly rewritten in terms of the zweibeins $e^a{}_\mu$ and $f^a{}_\mu$ corresponding to the metrics $g_{\mu\nu}$ and $h_{\mu\nu}^{(\coupling)}$, respectively. The result is 
\begin{equation}\label{tolley}
  S_m^{(\coupling)} [\varphi,f,e]= S_m^{(0)}[\varphi,e] - \frac{1}{\coupling} \int \mathrm{d}^2 x \,\det (f-e)\,.   
\end{equation}
The bimetric gravity action \eqref{tolley} reproduces Tolley's ghost-free massive gravity proposal \cite{Tolley:2019nmm}. This demonstrates that a $\TTb$ deformed classical field theory in $d=2$ space-time is equivalent to the original unperturbed model coupled with a topological gravity theory.
\begin{figure}[h]
    \centering
    \includegraphics[scale=0.37]{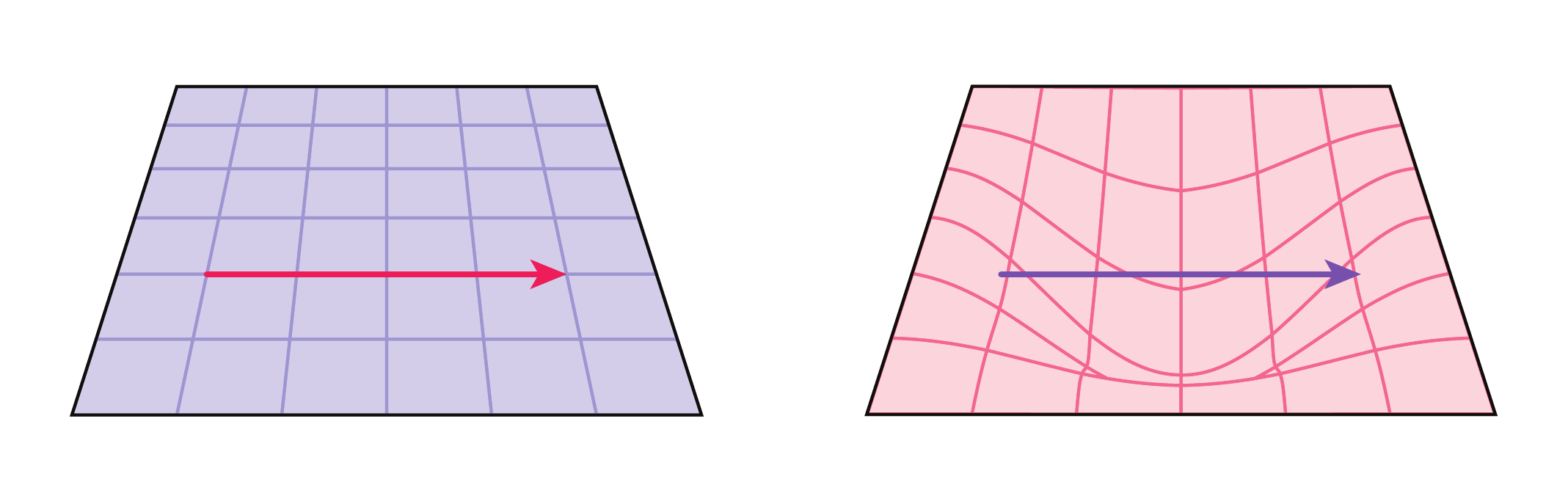}
    \caption{On the left, the trajectory followed by a particle described by the undeformed theory $S^{(0)}_m[\varphi,g]$ over the undeformed space-time geometry defined by $g_{\mu\nu}$. On the right, the trajectory followed by a particle described by the deformed theory $S^{(\coupling)}_m[\varphi,h^{(\coupling)}]$ over the deformed space-time geometry defined by $h_{\mu\nu}^{(\coupling)}$.}
    \label{fig:mesh5}
\end{figure}
\subsection{Global coordinate transformations and the Burgers' equation}
In this section, we briefly discuss an important property that $\TTb$ deformation of two-dimensional field theories exhibits on flat geometries, i.e., when the undeformed metric coincides with the Euclidean metric $g_{\mu\nu}=\delta_{\mu\nu}$. 

In a flat background, the auxiliary metric $h_{\mu\nu}^{(\coupling)}$ is equivalent to a global change of coordinates and,  in this scenario, one can derive the famous Burgers' equation describing the finite-size spectral flow, whose validity extends to the quantum level. The Burgers' equation was initially deduced directly at the quantum level using Zamolodchikov's factorisation theorem \cite{Zamolodchikov:2004ce, Smirnov:2016lqw} or proved using the exact S-matrix combined with thermodynamic Bethe ansatz-type nonlinear integral equation techniques \cite{Cavaglia:2016oda}.

Building on the geometric interpretation of $\TTb$ discussed in previous sections, we take an alternative approach as described in \cite{Aramini:2022wbn}. As shown in \cite{Conti:2018tca}, any solution to the classical equations of motion associated with $\TTb$-deformed Lagrangians can be mapped into the corresponding undeformed solution defined on an auxiliary space-time manifold whose metric is determined by \eqref{metric_flow}. Specifically, from \eqref{metric_flow_sol}, we see that the dynamical change of coordinates $x\mapsto y(x)$ must be induced by the Jacobian\footnote{Since the undeformed metric is the flat space-time, the Jacobian \eqref{coc} and the deformed zweibein $f$ introduced in \eqref{tolley} are the same object.} 
\begin{align}\label{coc}
\frac{\partial y^\mu}{\partial x^\nu}= \delta^\mu{}_\nu- \coupling \widehat{T}^{(0)\mu}{}_\nu(x)\,. 
\end{align}
Moreover, the conservation law for the energy-momentum $\partial_\mu T^{(0)\mu\nu}=0$ implies that the Hessian matrix associated to the local coordinate transformation is symmetric, namely, that
\begin{equation}
 \frac{\partial^2 y^\mu}{\partial x^0\partial x^1}   =  \frac{\partial^2 y^\mu}{\partial x^1\partial x^0}\,,\quad \mu = 0,1\,.
\end{equation}
The above condition ensures that, on-shell, the local change of coordinates can be promoted to a global one \cite{Conti:2018tca}. To simplify our calculations, for the rest of this section, we shall set the mass scale of the theory equal to 1, and take the flow parameter $\coupling$ to be dimensionless. Following the usual conventions, we switch from Cartesian to complex coordinates according to
\begin{align}
&\left\{\begin{array}{l}
w=x^0+i x^1 \\
\bar{w}=x^0-i x^1
\end{array}\,,\right.&\quad&\left\{\begin{array}{l}
z=y^0+i y^1 \\
\bar{z}=y^0-iy^1
\end{array}\right.\,.&
\end{align}
Cartesian and complex components of the stress-energy tensor are related through \cite{DiFrancesco:1997nk} 
\begin{align}
&T_{00}=-\bar{T}-T+2\Theta\,,&\quad &T_{01}=T_{10}=i\bar{T}-i T\,,&\quad
&T_{11}=\bar{T}+T+2\Theta\,.&
\end{align}
Rewriting \eqref{coc} in terms of the holomorphic and anti-holomorphic complex coordinates, one can locally express the change of coordinates $w\mapsto z(w)$ as
\begin{equation} 
\begin{aligned}
&\mathrm{d}z=\mathrm{d}w-2\coupling\left[\Theta^{(0)}(w) \mathrm{d}w+
\bar{T}^{(0)}(w)\mathrm{d}\bar{w}\right],\\
 &\mathrm{d}\bar{z}=\mathrm{d}\bar{w}-2\coupling \left[\Theta^{(0)}(w)\mathrm{d}\bar{w}+ T^{(0)}(w)\mathrm{d}w\right].
 \end{aligned}
\end{equation}
We focus on {finite-size configurations} with a characteristic length $ L$. Using the conservation laws,
\begin{equation}\label{cons:law}
{\partial}_{\bar{\omega}} T^{(0)}=\partial_\omega \Theta^{(0)}\,,\quad\partial_\omega\bar{T}^{(0)}={\partial}_{\bar{\omega}}\Theta^{(0)}\,,
\end{equation}
we can define the ``light-cone components'' of the bi-momentum in the undeformed theory as
\begin{equation}
\begin{aligned}
 &{P}^{(0)}( L)=\frac{1}{2} \int_{\curve}\left[\mathrm{d}w \, T^{(0)}(w) +\mathrm{d}\bar{w} \,  \Theta^{(0)}(w)\right],\\
 &\bar{ {P}}^{(0)}( L)=\frac{1}{2} \int_{\bar{\curve}}\left[\mathrm{d}w \, \Theta^{(0)}(w) +\mathrm{d}\bar{w} \, \bar{T}^{(0)}(w) \right],    
\end{aligned}
\end{equation}
where the path $\curve \subset \mathbb{C}$ is chosen such that the integrals are taken over the whole volume at constant (Euclidean) time. In particular, \begin{equation}
\int_{\bar{\curve}} \mathrm{d}w=\int_{\curve} \mathrm{d}\bar{w}:= 2L\,,\quad\bar{\curve}=\curve^*\,.
\end{equation}
Let us denote the total energy and momentum of the system as ${\bf E}= {P}+\bar{ {P}}$ and ${\bf P}= {P}-\bar{ {P}}$, respectively.  
Our main goal is to investigate how the characteristic length of the system is affected by the deformation. To this end, we will adopt a geometric perspective on the $\TTb$ flow. Since we are ultimately dealing with a global change of coordinates, two equivalent interpretations are possible: one can either make explicit use of the coordinate transformation or implicitly hide the deformation within new integration contours, $\curve^{(\coupling)}$ and $\bar{\curve}^{(\coupling)}$.\footnote{In complete generality, we assume $\bar{\curve}^{(\coupling)}\ne\left(\curve^{(\coupling)}\right)^*$ for excited state configurations.} The deformed integrals of motion are
\begin{align}
 {P}^{(\coupling)}( L)=\frac{1}{2} \int_{\curve}\left[\mathrm{d}z \, T^{(\coupling)}(z) +\mathrm{d}\bar{z} \,  \Theta^{(\coupling)}(z)\right]= \frac{1}{2} \int_{\curve^{(\coupling)}}\left[\mathrm{d}w \, T^{(0)}(w) +\mathrm{d}\bar{w} \,  \Theta^{(0)}(w)\right],
\end{align}
with a similar formula holding for $\bar{{P}}^{(\coupling)}$.

As shown in \cite{Aramini:2022wbn}, one can express the net effect of the $\TTb$ deformation over the characteristic scale of the system in terms of a rescaling plus a rotation. More precisely:
\begin{equation}
\left( L, L\right)\mapsto\left(\tilde{L}e^{-\vartheta},\tilde{L}e^{\vartheta}\right)\,,
\end{equation}
where
\begin{equation}
\begin{aligned}
&2\tilde{L}e^{-\vartheta}=\int_{\bar{\curve}^{(\coupling)}}\mathrm{d}w=\int_{\bar{\curve}} \left[\mathrm{d}z + 2\coupling \left( \Theta^{(\coupling)}(z) \mathrm{d}z + \bar{T}^{(\coupling)}(z)\mathrm{d}\bar{z}\right)\right]\,,\\
&2\tilde{L}e^{\vartheta}=\int_{\curve^{(\coupling)}}\mathrm{d}\bar{w}=\int_\curve \left[\mathrm{d}\bar{z}+2\coupling  \left(\Theta^{(\coupling)}(z)\mathrm{d}\bar{z} + T^{(\coupling)}(z)\mathrm{d}z \right)\right]\,.
\end{aligned}
\end{equation}
It is now sufficient to use the inverse map, $(\tilde{L},\tilde{L})\mapsto\left( Le^{\vartheta}, Le^{-\vartheta}\right)$, and the related inverse coordinate transformation, $z = z(w)$, to obtain the relationship between the dynamics in the undeformed and perturbed theory. Explicitly, one has
\begin{align}\label{realtions_in_2d}
&\begin{cases}
\tilde{L}e^{-\vartheta}= L+\coupling {\bf E}^{(\coupling)}(L)-\coupling{\bf P}^{(\coupling)}(L)\\
\tilde{L}e^{\vartheta}= L+\coupling {\bf E}^{(\coupling)}( L)+\coupling{\bf P}^{(\coupling)}( L)
\end{cases}&\quad
&\begin{cases}
Le^{\vartheta}=\tilde{L}-\coupling {\bf E}^{(0)}(\tilde{L})+\coupling{\bf P}^{(0)}(\tilde{L})\\
Le^{-\vartheta}=\tilde{L}-\coupling{\bf E}^{(0)}(\tilde{L})-\coupling{\bf P}^{(0)}(\tilde{L})
\end{cases}\,.&
\end{align}
Simple manipulations of the relations \eqref{realtions_in_2d} yield the equation \cite{Aramini:2022wbn}
\begin{equation}\label{burgers}
\frac{\partial {\bf E}^{(\coupling)}( L)}{\partial \coupling} = {\bf E}^{(\coupling)}( L)\frac{\partial {\bf E}^{(\coupling)}( L)}{\partial  L}+\frac{{\bf P}^{(0)}( L)^2}{ L}\,,
\end{equation}
where the total momentum ${\bf P}$ is explicitly independent of the deformation parameter. This independence arises directly from the fact that the $\TTb$ deformation is a zero-spin perturbation. The hydrodynamic-type equation \eqref{burgers} is the celebrated (inviscid) Burgers' equation \cite{Griffiths}. 

\begin{figure}[h]
    \centering
    \includegraphics[scale=0.36]{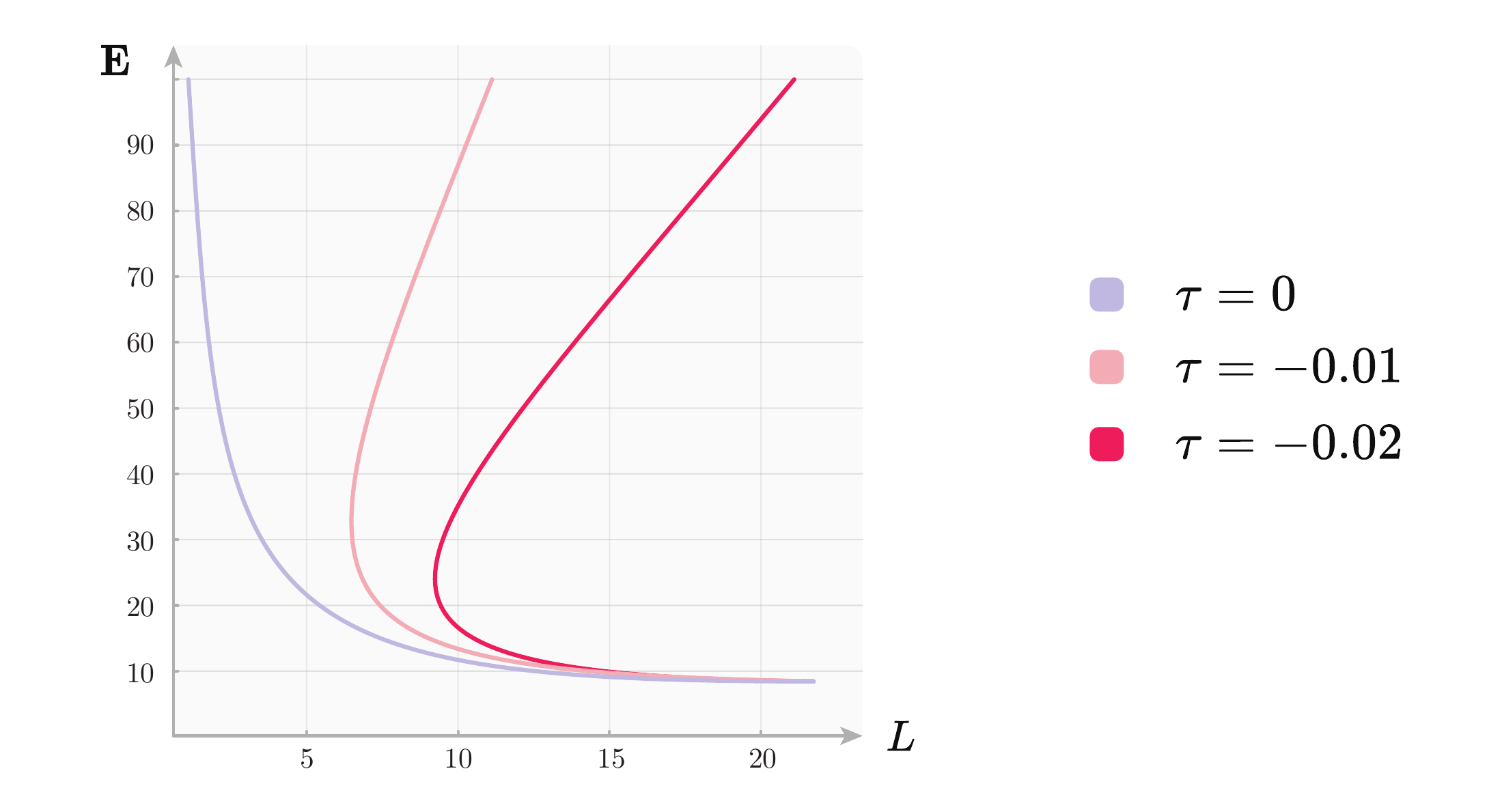}
    \caption{The elliptic solution of the sine-Gordon model has been investigated in detail \cite{Hirota:1971zz, Faddeev:1974em}. The $\TTb$-deformed spectrum, as a function of the characteristic length \( L \), is presented for various values of the deformation parameter \cite{Conti:2018jho, Conti:2019dxg}. The study highlights the emergence of critical phenomena within the classical solutions, such as shock-wave singularities and square root-type transitions.}
    \label{fig:mesh62}
\end{figure}

The validity of the Burgers' equation at both classical and quantum levels is a key insight, suggesting that certain $\TTb$ phenomenology, such as the presence of branch singularities in the spectrum, is not confined to the quantum regime. These features are, in fact, already present at the classical level, at least for the excited states. This was explicitly observed in specific solutions of the deformed sine-Gordon model \cite{Conti:2018jho, Conti:2019dxg}. Figure \ref{fig:mesh62} displays the typical behaviour observed for the so-called finite-gap solutions of the deformed sine-Gordon model.

In addition, the proposed derivation is quite general and finds an intriguing realization within the framework of the so-called ODE/IM correspondence (see \cite{Dorey:1998pt, Bazhanov:1998wj, Dorey:2007zx, Dorey:2019ngq}). In its off-critical variant, this correspondence establishes a precise dictionary between classical and quantum quantities in two-dimensional integrable field theories \cite{Lukyanov:2010rn}.
As demonstrated in \cite{Aramini:2022wbn}, and in agreement with the previous analysis, this quantum/classical aspect of the ODE/IM correspondence is preserved under deformation.
\subsection{Combining $\TTb$ and $\sqrt{\TTb}$ flows}
A different deformation of two-dimensional field theories, driven by the so-called $\sqrt{\TTb}$ (root--$\TTb$) marginal operator, has recently attracted increasing interest. The $\sqrt{\TTb}$ deformation is characterised by the flow equation,
\begin{equation}\label{rootttb2}
    \frac{\partial S_m^{(\gamma)}}{\partial\gamma} =  \int \mathrm{d}^2x \sqrt{-g} \,\sqrt{\frac{1}{2} T^{(\gamma) \mu \nu} T_{\mu \nu}^{(\gamma)}-\frac{1}{4}\left(T^{(\gamma) \mu}{ }_\mu\right)^2}\,,
\end{equation}
where $\gamma$ is the dimensionless coupling. Similar to the $\TTb$ case, the solutions to the flow equation \eqref{rootttb2} at finite coupling can be interpreted as a field-dependent deformation of the original background metric \cite{Ebert:2023tih}. Introducing the auxiliary tensor
\begin{equation}
 \widetilde{T}^{(0)}_{\mu\nu} = \frac{T^{(0)}_{\mu\nu}-\frac{1}{2}T^{(0)\alpha}{}_\alpha g_{\mu\nu}}{\sqrt{\frac{1}{2} T^{(0) \beta \delta} T_{\beta \delta}^{(0)}-\frac{1}{4}\left(T^{(0) \beta}{ }_\beta\right)^2}} \,,
\end{equation}
it is possible to verify that 
\begin{equation}\label{root_dress}
     S_m^{(\gamma)} [\varphi,k^{(\gamma)}] = \left.\left\{S_m^{(0)}[\varphi,g] \right\}\right|_{g=g(k)}\,,
\end{equation}
provided that the deformed metric $k_{\mu\nu}^{(\gamma)}$ satisfies \cite{Ebert:2023tih} \footnote{The notation used in this paper differs from the one used in \cite{Ebert:2023tih}, where $\widetilde{T}^{(0)}_{\mu\nu}$ denotes the traceless part of the stress-energy tensor of the seed theory.} 
\begin{equation}\label{def_root_chris}
    k_{\mu\nu}^{(\gamma)} = \cosh \gamma\, g_{\mu\nu}+{\sinh\gamma}\,\widetilde{T}^{(0)}_{\mu\nu} \,.
\end{equation}
Note that, differently from the standard $\TTb$ case (cf. equation \eqref{dressing}), the $\sqrt{\TTb}$ deformation does not involve any additional dressing operator in \eqref{root_dress}.

Let $\widetilde{\BT}^{(0)} = \{g^{\mu\alpha}\widetilde{T}^{(0)}_{\alpha\nu}\}_{\mu,\nu = 0,1}$ denote the matrix canonically associated to the auxiliary tensor $\widetilde{T}_{\mu\nu}^{(0)}$, and let $\BP = \{g^{\mu\alpha}k_{\alpha\nu}^{(\gamma)}\}_{\mu,\nu = 0,1}$ denote the deformation matrix associated to the $\sqrt{\TTb}$-deformed metric $k_{\mu\nu}^{(\gamma)}$. In virtue of \eqref{def_root_chris}, the functional relationship between the above two matrices can be written as
\begin{equation}\label{pmat}
   \BP = \cosh \gamma\, \mathbf{1} + \sinh \gamma\, \widetilde{\BT}^{(0)} = \exp \left(\gamma  \widetilde{\BT}^{(0)}\right)\,.    
\end{equation}
As before, we assume that ${\BT}^{(0)} = \{g^{\mu\alpha}{T}^{(0)}_{\alpha\nu}\}_{\mu,\nu = 0,1}$ can be diagonalised through a similarity transformation, such that 
\begin{equation}
\mathbf{T}^{(0)} = \mathbf{U}\begin{pmatrix}
\eigen_1 & 0 \\
0 & \eigen_2
\end{pmatrix}\mathbf{U}^{-1}\,.  
\end{equation}  
Correspondingly, we also have:
\begin{equation}\label{diag_sgn}
\widetilde{\mathbf{T}}^{(0)} = \mathbf{U}\begin{pmatrix}
\reigen_+ & 0 \\
0 & \reigen_-
\end{pmatrix}\mathbf{U}^{-1}\,, \quad \reigen_{\pm} = \pm \operatorname{sgn} (\eigen_1-\eigen_2)\,,  
\end{equation}
where $\operatorname{sgn}(x) := x/|x|$. Assuming $\eigen_1 \neq \eigen_2$,\footnote{Note that, when $\eigen_1 = \eigen_2$, the deforming operator vanishes at order $\gamma$, and the flow is trivial.} and, without loss of generality, choosing $\mathbf{U}$ such that $\eigen_1 > \eigen_2$, equation
\eqref{diag_sgn} reduces to 
\begin{equation}
 \widetilde{\mathbf{T}}^{(0)} = \mathbf{U}\,\bm{\sigma}_3\mathbf{U}^{-1}\,,   
\end{equation}
where $\bm{\sigma}_3 = \operatorname{diag}(1,-1)$ is the third Pauli matrix. Plugging the above equation into \eqref{pmat}, the eigenvalues of the deformation matrix $\BP$ are obtained as $ \pi_{\pm} = e^{\pm \gamma}$. Note that, in the case $\eigen_1=\eigen_2$, the $\sqrt{\TTb}$ operator identically vanishes, and the seed theory is unaffected by the deformation.

As far as the gravity interpretation of the pure $\sqrt{\TTb}$ deformation is concerned, the absence of a dressing term in \eqref{root_dress} precludes us from directly paralleling the steps adopted in the $\TTb$ case, and the perturbation only amounts to a dynamical modification of the metric.

However, combining the marginal flow \eqref{rootttb2} with the $\TTb$ flow \eqref{ttb2} can provide a more interesting perspective. Notably, the two deformations commute and the whole dressing mechanism is implemented by considering
\begin{equation}\label{combined_dressing}
    \begin{split}
  S_m^{(\coupling,\gamma)} [\varphi,h^{(\coupling,\gamma)}] &= \left.\left\{S_m^{(0,0)}[\varphi,g] - \frac{\coupling}{2} \int \mathrm{d}^2x \sqrt{-g} \,\left[\left(T^{(0,0) \mu}{ }_\mu\right)^2-T^{(0,0)\mu\nu}T^{(0,0)}_{\mu\nu}\right]\right\}\right|_{g=g(h)} ,
\end{split}
\end{equation}
with the deformed metric $h_{\mu\nu}^{(\coupling,\gamma)}$ obtained from $g_{\mu\nu}$ via the action of the composite deformation matrix $\BA = \{g^{\mu\alpha}h_{\alpha\nu}^{(\coupling,\gamma)}\}_{\mu,\nu = 0,1}$ defined by
\begin{equation}
\BA := \BP\circ \BO = \exp \left(\gamma  \widetilde{\BT}^{(0,0)}\right)\left(\mathbf{1}-\coupling \widehat{\BT}^{(0,0)}\right)^2 \,. 
\end{equation}
Denoting by $\{\eigen_1,\eigen_2\}$ the eigenvalues of $\BT^{(0,0)}$, the dressing formula \eqref{combined_dressing} reduces to
\begin{equation}\label{dyn_root}
  S_m^{(\coupling,\gamma)} [\varphi,h^{(\coupling,\gamma)}]  \,\, \dot{=}\,\,  S^{(0,0)}_m[\varphi,g] - \coupling\int \mathrm{d}^2x \sqrt{-g} \,\eigen_1\eigen_2\,,
\end{equation}
where, this time, the dynamical equivalence is verified, provided that
\begin{equation}\label{solve_root_eig}
    \alpha_1 = e^\gamma \left(1-\coupling \eigen_2\right)^2\,,\quad
      \alpha_2 = e^{-\gamma} \left(1-\coupling \eigen_1\right)^2\,,
\end{equation}
with $\alpha_1$ and $\alpha_2$ denoting the eigenvalues of the composite deformation matrix $\BA$. Solving equation \eqref{solve_root_eig} in terms of $\eigen_1$ and $\eigen_2$, and plugging the result back into \eqref{dyn_root}, one finally obtains 
\begin{equation}\label{rootgrav}
  S_m^{(\coupling,\gamma)} [\varphi,h^{(\coupling,\gamma)},g]  =  S^{(0,0)}_m[\varphi,g]   -\frac{1}{\coupling} \int \mathrm{d}^2 x \sqrt{-g}\left(\sqrt{\alpha_1}-e^{\frac{\gamma}{2}}\right)\left(\sqrt{\alpha_2}-e^{-\frac{\gamma}{2}}\right)\,.
\end{equation}
The gravity sector of \eqref{rootgrav} can be written in terms of the zweibeins $e^a_\mu$ and $f_\mu^a$ associated to the metrics $g_{\mu\nu}$ and $h_{\mu\nu}^{(\coupling,\gamma)}$ respectively: with the notation $y_1 := \tr[e^{-1}f]$, $y_2 := \tr[(e^{-1}f)^2]$, one has
\begin{equation}
 \left(\sqrt{\alpha_1}-e^{\frac{\gamma}{2}}\right)\left(\sqrt{\alpha_2}-e^{-\frac{\gamma}{2}}\right)=1 \pm\sqrt{2 y_2-y_1^2} \sinh \frac{\gamma}{2} - y_1 \cosh\frac{\gamma}{2}+\frac{1}{2}(y_1^2-y_2)\,.
\end{equation}
The above gravity functional reproduces the proposals from \cite{Babaei-Aghbolagh:2024hti, Tsolakidis:2024wut}.
As expected, in the limit $\gamma \to 0$, the gravity action \eqref{tolley} associated with the pure $\TTb$ deformation is recovered. 
\section{An eigenvalue-based approach to higher-dimensional stress tensor flows}\label{sec::4d}
Since the introduction of $\TTb$ deformations in two-dimensional field theories, various generalizations to higher dimensions have been proposed. These generalizations are motivated by different physical setups and have been studied using diverse mathematical methods and techniques \cite{Bonelli:2018kik, Taylor:2018xcy, Conti_Iannella, Conti_2022}. In this section, we will use the eigenvalue-based method outlined in Section \ref{sec::2d} to study the higher-dimensional extensions put forward in   \cite{Conti_Iannella, Ferko:2023sps}. We then proceed by recovering the gravity theories corresponding to a specific set of stress tensor deformations using the dressing mechanism.
\subsection{Defining $\TTb$-like deformations in higher dimensions}
Suppose that the stress-energy tensor of the seed theory admits $2$ degenerate eigenvalues $\eigen_1$ and $\eigen_2$, of multiplicity $n$ and $m = d-n$ respectively, such that
\begin{equation}\label{eig_degeneracy_generic}
    \BT^{(0)} = \mathbf{U} \begin{pmatrix}
\eigen_1 \mathbf{1}_{n\times n} & 0 \\
0 & \eigen_2 \mathbf{1}_{m\times m}
\end{pmatrix}\mathbf{U}^{-1} \,,
\end{equation}
where $\mathbf{1}_{j\times j}$ denotes the $j\times j$ identity matrix. The framework described in the previous sections can then be partially generalised to higher space-time dimensions through the introduction of an auxiliary tensor
\begin{equation}\label{hat_T_in_generic_dim}
  \widehat{T}_{\mu\nu}^{(0)} := \frac{2}{d}T^{(0)\alpha}{}_\alpha g_{\mu\nu} - T^{(0)}_{\mu\nu} + \frac{(m-n)}{\sqrt{m n d}} \sqrt{T^{(0)\alpha\beta}T_{\alpha\beta}^{(0)}-\frac{1}{d}\left(T^{(0)\alpha}{}_\alpha\right)^2} g_{\mu\nu}\,.      
\end{equation}
Assuming, without loss of generality, that $\eigen_1 > \eigen_2$, we notice that the contribution from the square root term appearing in \eqref{hat_T_in_generic_dim} becomes
\begin{equation}
\frac{(m-n)}{\sqrt{m n d}} \sqrt{T^{(0)\alpha\beta}T_{\alpha\beta}^{(0)}-\frac{1}{d}\left(T^{(0)\alpha}{}_\alpha\right)^2} = \frac{m-n}{d} (\eigen_1 - \eigen_2)\,.    
\end{equation}
Furthermore, since $T^{(0)\alpha}{}_\alpha = n\eigen_1 + m \eigen_2$, we have
\begin{equation}\label{root_as_lambda}
   \frac{2}{d}T^{(0)\alpha}{}_\alpha +\frac{(m-n)}{\sqrt{m n d}} \sqrt{T^{(0)\alpha\beta}T_{\alpha\beta}^{(0)}-\frac{1}{d}\left(T^{(0)\alpha}{}_\alpha\right)^2}  = \eigen_1 + \eigen_2\,.
\end{equation}
Using \eqref{root_as_lambda}, one can verify that, in matrix notation, the right-hand side of equation \eqref{hat_T_in_generic_dim} reduces to
\begin{equation}\label{hat_T_in_generic_dim_diagonal}
 \widehat{\BT}^{(0)} = \mathbf{U} \begin{pmatrix}
\eigen_2 \mathbf{1}_{n\times n} & 0 \\
0 & \eigen_1 \mathbf{1}_{m\times m}
\end{pmatrix}\mathbf{U}^{-1} \,.   
\end{equation}
Building on the analogy with the two-dimensional case, it is natural to consider stress tensor flows of the form
\begin{equation}\label{ttbd}
\begin{split}
    \frac{\partial S_m^{(\coupling)}}{\partial\coupling} = \frac{1}{d} \int \mathrm{d}^dx \sqrt{-g}\,T^{(\coupling)\mu\nu}\widehat{T}^{(\coupling)}_{\mu\nu}\,.    
\end{split}
\end{equation}
Note that, due to the boundary conditions \eqref{eig_degeneracy_generic} and \eqref{hat_T_in_generic_dim_diagonal}, the deforming operator $T^{(\coupling)\mu\nu}\widehat{T}^{(\coupling)}_{\mu\nu}$ is simply proportional to the product of the two eigenvalues of the deformed stress tensor. 
When $n=d-1$, the structure of the stress-energy tensor \eqref{eig_degeneracy_generic} reproduces the characteristic degeneracy of standard bosonic field theories, and the flow equation \eqref{ttbd} coincides with the proposal of \cite{Ferko:2023sps}.

Moreover, it is evident from equation \eqref{hat_T_in_generic_dim} that significant simplifications occur in even-dimensional space-times when $n = m = d/2$. In this case,  the auxiliary tensor \eqref{hat_T_in_generic_dim} reduces to
\begin{equation}
 \widehat{T}^{(0)}_{\mu\nu} = \frac{2}{d}g^{\alpha\beta}T^{(0)}_{\alpha\beta}g_{\mu\nu} - T^{(0)}_{\mu\nu}\,,   
\end{equation}
and the flow \eqref{ttbd} can be shown to admit a dressing-type formulation in terms of a deformed metric $h_{\mu\nu}^{(\coupling)}$ as \cite{Conti_2022, Morone:2024ffm}
\begin{equation}\label{dressing_4}
\begin{split}
  S_m^{(\coupling)} [\varphi,h^{(\coupling)}] &= \left.\left\{S_m^{(0)}[\varphi,g] - \frac{\coupling}{d} \int \mathrm{d}^dx \sqrt{-g} \,\left[\frac{2}{d}\left(T^{(0) \mu}{ }_\mu\right)^2-T^{(0)\mu\nu}T^{(0)}_{\mu\nu}\right]\right\}\right|_{g=g(h)}\\
  &=\left.\left\{S_m^{(0)}[\varphi,g] - \frac{\coupling}{d} \int \mathrm{d}^d x \sqrt{-g} \,T^{(0)\mu\nu}\widehat{T}^{(0)}_{\mu\nu}\right\}\right|_{g=g(h)} \,.
\end{split}
\end{equation}
For the dynamical equivalence \eqref{dressing_4} to hold, the auxiliary metric $h_{\mu\nu}^{(\coupling)}$ must obey the flow equation
\begin{equation}\label{flowh4d}
    \frac{\mathrm{d} h^{(\coupling)}_{\mu\nu}}{\mathrm{d}\coupling} = -\frac{4}{d}\widehat{T}^{(\coupling)}_{\mu\nu}\,,\quad  h_{\mu\nu}^{(0)} = g_{\mu\nu}\,.
\end{equation}
In even space-time dimensions, the set of differential equations \eqref{flowh4d} can be integrated \cite{Conti_2022}. The result can be summarised via the formal expression \footnote{Note that equation \eqref{integrate_d} should be understood as a series expansion in $\coupling$.}
\begin{equation}\label{integrate_d}
    h_{\mu\nu}^{(\coupling)} = \left[\left(g -\coupling \widehat{T}^{(0)}\right)^{\frac{4}{d}}\right]_{\mu\nu}\,.
\end{equation}
Moreover, note that, differentiating \eqref{integrate_d} with respect to $\coupling$, and comparing the result with \eqref{flowh4d}, we obtain
\begin{equation}
    \frac{\mathrm{d} h_{\mu\nu}^{(\coupling)}}{\mathrm{d} \coupling} = -\frac{4}{d}\widehat{T}_\mu^{(0)\alpha}\left[\left(g -\coupling \widehat{T}^{(0)}\right)^{\frac{4}{d}-1}\right]_{\alpha\nu} = -\frac{4}{d}  \widehat{T}^{(\coupling)}_{\mu\nu} \,,
\end{equation}
yielding the consistency condition
\begin{equation}
    \widehat{T}^{(\coupling)}_{\mu\nu} = \widehat{T}_\mu^{(0)\alpha}\left[\left(g -\coupling \widehat{T}^{(0)}\right)^{\frac{4}{d}-1}\right]_{\alpha\nu}\,.
\end{equation}
As for the two-dimensional case, we introduce a deformation matrix $\BO = \{g^{\mu\alpha}h_{\alpha\nu}^{(\coupling)}\}_{\mu,\nu = 0,\dots,d-1}$, and, denoting its eigenvalues by $\omega_i$, we have
\begin{equation}\label{omega4}
    \bm{\mathrm{\Omega}} = \left(\mathbf{1}-\coupling \widehat{\mathbf{T}}^{(0)}\right)^{\frac{4}{d}}\,, \quad   \omega_{1,2} =\left(1-\coupling \eigen_{2,1}\right)^{\frac{4}{d}} \,.
\end{equation}
In terms of the eigenvalues of $\BT^{(0)}$, the dressing formula \eqref{dressing_4} becomes:
\begin{equation}\label{eigen_dress_4}
 S^{(\coupling)}_m[\varphi,h^{(\coupling)}] \,\,\dot{=}\,\, S^{(0)}_m[\varphi,g] - \coupling\int \mathrm{d}^dx \sqrt{-g} \,\eigen_1\eigen_2\,.
\end{equation}
Again, the dressing contributions reproduce those of the two-dimensional case. At this point, one can explicitly introduce the auxiliary metric $h_{\mu\nu}^{(\coupling)}$ into \eqref{eigen_dress_4} through \eqref{omega4}, obtaining 
\begin{equation}\label{4dgrav_omega_gen}
\begin{split}
    S_m^{(\coupling)} [\varphi,h^{(\coupling)},g] &= S_m^{(0)}[\varphi,g] - \frac{1}{\coupling} \int \mathrm{d}^d x \sqrt{-g}\left(\omega_1^{\frac{d}{4}}-1\right)\left(\omega_2^{\frac{d}{4}}-1\right)\,.
\end{split}
\end{equation}
Defining the Lorentz invariants $x_1 := \tr[\BO]$, $x_2 = \tr[\BO^2]$, the gravity functional on the right-hand side of \eqref{4dgrav_omega_gen} can be equivalently written as
\begin{equation}
    - \frac{1}{\coupling} \int \mathrm{d}^d x \sqrt{-g} \left[d^{\frac{d}{2}}  \left(x_1-\sqrt{d x_2-x_1^2}\right)^{\frac{d}{4}}-1\right] \left[d^{\frac{d}{2}}  \left(x_1+\sqrt{d x_2-x_1^2}\right)^{\frac{d}{4}}-1\right]\,.
\end{equation}
In four space-time dimensions, substantial simplifications allow recasting \eqref{4dgrav_omega_gen} into the following compact form:
\begin{equation}
    S_m^{(\coupling)} [\varphi,h^{(\coupling)},g] = S_m^{(0)}[\varphi,g] - \frac{1}{\coupling} \int \mathrm{d}^4 x  \sqrt{\det(h^{(\coupling)}-g)}\,,
\end{equation}
reproducing the gravity sector introduced in \cite{Babaei-Aghbolagh:2024hti}.
\subsection{Dynamical gravity theories and the Palatini frame}
So far, after introducing a higher-dimensional analogue of the $\TTb$ operator, we have been following a similar approach to the two-dimensional setup. However, unlike in two and three-dimensional scenarios, gravity theories in space-times with dimensions greater than or equal to four can have dynamical degrees of freedom.  To incorporate these into the theory, we minimally couple the (deformed) theory in the $h_{\mu\nu}^{(\coupling)}$ frame to General Relativity via the Einstein-Hilbert term
\begin{equation}\label{total_action}
  S[\varphi,h^{(\coupling)}] =S_{EH}[h^{(\coupling)}] + S_m^{(\coupling)}[\varphi,h^{(\coupling)}]\,,\quad    S_{EH}[h^{(\coupling)}] = \frac{1}{2}\int \mathrm{d}^d x \sqrt{-h^{(\coupling)}} R[h^{(\coupling)}]\,,
\end{equation}
where $R[h^{(\coupling)}] = (h^{-1})^{(\coupling)\mu\nu}R_{\mu\nu}[h^{(\coupling)}]$ is the scalar curvature associated to the metric tensor $h_{\mu\nu}^{(\coupling)}$, and $R_{\mu\nu}[h^{(\coupling)}]$ is the Ricci tensor, defined through the Levi-Civita connection $\Gamma[h^{(\coupling)}]$ as
\begin{equation}
R_{\mu\nu}[h^{(\coupling)}] :=  \partial_\alpha \Gamma_{\nu \mu}^\alpha[h^{(\coupling)}]-\partial_\nu \Gamma_{\alpha \mu}^\alpha[h^{(\coupling)}]+\Gamma_{\alpha \beta}^\alpha [h^{(\coupling)}]\Gamma_{\nu \mu}^\beta[h^{(\coupling)}]-\Gamma_{\nu \beta}[h^{(\coupling)}]^\alpha \Gamma_{\alpha \mu}^\beta[h^{(\coupling)}]\,.    
\end{equation}
The Einstein field equations for \eqref{total_action} are given by
\begin{equation}\label{einstein_equations}
    R_{\mu\nu}[h^{(\coupling)}]-\frac{1}{2}R[h^{(\coupling)}]h^{(\coupling)}_{\mu\nu} = T^{(\coupling)}_{\mu\nu}\,.
\end{equation}
For the rest of this section, we will focus on the $n = m = d/2$ case in four space-time dimensions, where further simplifications arise due to the linear structure of \eqref{integrate_d}.
In $d=4$, such eigenvalue structure is typical of Abelian gauge theories \cite{Ferko:2022iru}, three-form field theories \cite{Mitskievich:1998uk}, and even symmetry breaking models with non-canonical kinetic terms \cite{Nascimento:2019qor}, whose stress-energy tensor can be written as
\begin{equation}\label{eig_degeneracy}
    \BT^{(0)} = \mathbf{U} \begin{pmatrix}
\eigen_1 \mathbf{1}_{2\times 2} & 0 \\
0 & \eigen_2 \mathbf{1}_{2\times 2}
\end{pmatrix}\mathbf{U}^{-1} \,,
\end{equation}
while the auxiliary tensor \eqref{hat_T_in_generic_dim} reduces to
\begin{equation}\label{hatT}
    \widehat{T}^{(0)}_{\mu\nu} = \frac{1}{2}g^{\alpha\beta}T^{(0)}_{\alpha\beta}g_{\mu\nu} - T^{(0)}_{\mu\nu}\,.
\end{equation}
The deformed metric simply reads  
\begin{equation}\label{integrate4}
    h^{(\coupling)}_{\mu\nu} = g_{\mu\nu} -\coupling \widehat{T}_{\mu\nu}^{(0)}\,,
\end{equation}
and, after differentiating \eqref{integrate4} with respect to $\coupling$ and comparing the result with \eqref{flowh4d}, one obtains the consistency relation
\begin{equation}\label{tthat}
    \widehat{T}^{(0)}_{\mu\nu} =  \widehat{T}^{(\coupling)}_{\mu\nu}\,.
\end{equation}
Using the above results in \eqref{einstein_equations}, we have
\begin{equation}\label{eoms_einst}
    R_{\mu\nu}[h^{(\coupling)}] = -\widehat{T}_{\mu\nu}^{(\coupling)} = - \widehat{T}_{\mu\nu}^{(0)}\,.
\end{equation}
Introducing the matrix $\mathbf{R} = \{g^{\mu\alpha}R_{\alpha\nu}[h^{(\coupling)}]\}_{\mu,\nu = 0,\dots,3}$, and denoting by $\varrho_i$ its eigenvalues, the above equation takes the diagonal form $  \varrho_{1,2} = -\eigen_{2,1}$. This implies that the eigenvalues of the deformation matrix $\mathbf{O}$ are related to those of the Ricci curvature matrix $\mathbf{R}$:
\begin{equation}\label{omegarho}
    \omega_i =  1+\coupling\varrho_i\,, \quad i = 1,2\,.
\end{equation}
Considering the additional Einstein-Hilbert term, the dynamical equivalence \eqref{4dgrav_omega_gen} in $d=4$ now reads
\begin{equation}\label{eh+mat_dyn}
    S_{EH}[h^{(\coupling)},g]+ S^{(\coupling)}_m[\varphi,h^{(\coupling)},g] =  S_{EH}[h^{(\coupling)}]+S^{(0)}_m[\varphi,g] - \frac{1}{\coupling} \int \mathrm{d}^4 x \sqrt{-g}\left(\omega_1-1\right)\left(\omega_2-1\right)\,,
\end{equation}
where $S_{EH}[h^{(\coupling)},g]$ is simply obtained by rewriting the Einstein-Hilbert action in the $h^{(\coupling)}_{\mu\nu}$ frame in terms of the eigenvalues the deformation matrix $\BO$:
\begin{equation}
    S_{EH}[h^{(\coupling)},g] = \int \mathrm{d}^4 x \sqrt{-g} \left(\varrho_1\omega_2 + \varrho_2\omega_1\right)\,.
\end{equation}
Substituting \eqref{omegarho} back into \eqref{eh+mat_dyn}, one obtains
\begin{equation}\label{final}
\begin{split}
  & \,\,S_{EH}[h^{(\coupling)},g]+ S^{(\coupling)}_m[\varphi,h^{(\coupling)},g] \\ =  &\int \mathrm{d}^4 x \sqrt{-g} \left(\varrho_1+\varrho_2+\coupling\varrho_1\varrho_2\right)+S^{(0)}_m[\varphi,g] \\
 = & \int \mathrm{d}^4 x \sqrt{-g}\left[\frac{1}{2} g^{\mu \nu} R_{\mu \nu}[h^{(\coupling)}]+\frac{\coupling}{4}\left(\frac{1}{2}\left(g^{\mu \nu} R_{\mu \nu}[h^{(\coupling)}]\right)^2-R^{\mu \nu} [h^{(\coupling)}] R_{\mu \nu}[h^{(\coupling)}]\right)\right]+ S^{(0)}_m[\varphi,g]\,.
\end{split}
\end{equation}
The gravity action in \eqref{final} mirrors the standard Einstein-Hilbert term, plus higher order corrections, albeit with a significant difference: the curvature terms do not depend on the metric $g_{\mu\nu}$, yet on some auxiliary metric $h_{\mu\nu}^{(\coupling)}$. Interestingly, since $\mathbf{R}$ only admits two independent eigenvalues $\varrho_1$ and $\varrho_2$, the above action can be equivalently written as
\begin{equation}\label{new_action}
 S_{EH}[h^{(\coupling)},g]+ S^{(\coupling)}_m[\varphi,h^{(\coupling)},g] = \frac{1}{\coupling}\int \mathrm{d}^4 x \left[\sqrt{-\det\left(g_{\mu\nu}+\coupling R_{\mu\nu}[h^{(\coupling)}]\right)}-\sqrt{-g}\right] +  S^{(0)}_m[\varphi,g]\,.  
\end{equation}
Note that the auxiliary metric $h_{\mu\nu}^{(\coupling)}$ only enters the action \eqref{new_action} through its Levi-Civita connection $\Gamma^{\lambda}_{\mu\nu}$. For this reason, the gravity sector on the right-hand side of the above equation can be expressed as the Palatini action \footnote{For a rigorous proof of this argument, see \cite{Morone:2024ffm}.}
\begin{equation}\label{eibi}
    S_{EiBI}[g,\Gamma] =  \frac{1}{\coupling}\int \mathrm{d}^4 x \left[\sqrt{-\det\left(g_{\mu\nu}+\coupling R_{\mu\nu}[\Gamma]\right)}-\sqrt{-g}\right]\,, 
\end{equation}
with the equations of motion for the independent connection implementing its $h_{\mu\nu}^{(\coupling)}$-compatibility:
\begin{equation}\label{compatible}
\Gamma^\lambda_{\mu\nu} =\frac12 (h^{-1})^{(\coupling)\lambda\alpha} \left(\partial_\nu h_{\mu\alpha}^{(\coupling)} + \partial_\mu h_{\alpha\nu}^{(\coupling)} - \partial_\alpha h_{\mu\nu}^{(\coupling)}\right)\,.  \end{equation}
The gravity theory defined by \eqref{eibi} is known as asymptotically flat Eddington-inspired Born-Infeld gravity, and we refer the reader to \cite{BeltranJimenez:2017doy} for a comprehensive review of the subject. Note that, despite the contributions of higher-order curvature invariants, the theory is ghost-free \cite{BeltranJimenez:2019acz}. The above discussion indicates that, in four space-time dimensions, as long as the stress tensor structure of the field theory obeys \eqref{eig_degeneracy}, the equations of motion resulting from $\TTb$-deformed action functionals coupled to standard General Relativity are equivalent to those obtained by coupling the undeformed action to the Palatini gravity theory \eqref{eibi}.

It is interesting to have a closer look at matter sectors of the form 
\begin{equation}\label{S_cosmo_2}
 {\underline{S}}^{(0)}_m[\varphi,g] = S^{(0)}_m [\varphi,g] - \int \mathrm{d}^4x \sqrt{-g}\cosm\,,
\end{equation}
where the additional term involving $\cosm$ accounts for vacuum energy contributions. Since the cosmological term does not spoil the desired eigenvalue degeneracy of the stress-energy tensor, the previously discussed results remain applicable in this scenario, with
\begin{equation}
    h_{\mu\nu}^{(\coupling)} = g_{\mu\nu} -\coupling \underline{T}^{(0)}_{\mu\nu}\,,\quad \underline{T}^{(0)}_{\mu\nu} := -\frac{2}{\sqrt{-g}}\frac{\delta  {\underline{S}}^{(0)}_m}{g^{\mu\nu}}\,.
\end{equation}
Alternatively, since $\underline{T}^{(0)}_{\mu\nu} = {T}^{(0)}_{\mu\nu} + \cosm g_{\mu\nu}$, where ${T}^{(0)}_{\mu\nu}$ denotes the energy-momentum tensor associated to $S^{(0)}_m [\varphi,g]$, one may alternatively move the cosmological term to the gravity sector of the theory, and consider the equivalent metric deformation
\begin{equation}
    h_{\mu\nu}^{(\coupling)} = \left(1+\coupling \cosm\right) g_{\mu\nu} - \coupling \widehat{T}^{(0)}_{\mu\nu} := \eta g_{\mu\nu} - \coupling\widehat{T}^{(0)}_{\mu\nu}\,.
\end{equation}
In this way, it is easy to include the cosmological contributions into the gravity sector \eqref{eibi} associated with the $\TTb$-like deformation, which now reads
\begin{equation}\label{new_action_2}
\begin{split}
 S_{EH}[h^{(\coupling)}]+ \underline{S}^{(\coupling)}_m[\varphi,h^{(\coupling)}] \simeq \frac{1}{\coupling}\int \mathrm{d}^4 x \left[\sqrt{-\det\left(g_{\mu\nu}+\coupling R_{\mu\nu}[\Gamma]\right)}-\eta\sqrt{-g}\right] +  {S}^{(0)}_m[\varphi,g]\,,
\end{split}
\end{equation}
where the notation $\simeq$ is used to emphasise that the equivalence \eqref{new_action_2} holds when the equations of motion for the dynamical connection are satisfied.
The Palatini action appearing in \eqref{new_action_2} reproduces the Eddington-inspired Born-Infeld functional with a cosmological term, as discussed in \cite{BeltranJimenez:2017doy}. Note that, in terms of the auxiliary parameter $\eta$, one has
\begin{equation}
    \cosm = \frac{\eta-1}{\coupling}\,,
\end{equation}
which is reminiscent of its two-dimensional analogue introduced in \cite{Gorbenko:2018oov}.

\subsection{Higher-dimensional $\sqrt{\TTb}$-like flows}
Following the steps of the previous section, and assuming that the stress-energy tensor of the seed theory admits $2$ degenerate eigenvalues $\eigen_1$ and $\eigen_2$, respectively of multiplicity $n$ and $m$, such that it can be diagonalised as 
\begin{equation}
 \BT^{(0)} = \mathbf{U} \begin{pmatrix}
\eigen_1 \mathbf{1}_{n\times n} & 0 \\
0 & \eigen_2 \mathbf{1}_{m\times m}
\end{pmatrix}\mathbf{U}^{-1}\,,   
\end{equation}
we uplift two-dimensional $\sqrt{\TTb}$ flows to higher-dimensional space-times by introducing the auxiliary matrix
\begin{equation}\label{4droot_mat}
    \widetilde{\BT}^{(0)} := \mathbf{U} \begin{pmatrix}
\frac{d}{2n} \mathbf{1}_{n\times n} & 0 \\
0 & -\frac{d}{2m} \mathbf{1}_{m\times m}
\end{pmatrix}\mathbf{U}^{-1}\,.
\end{equation}
Assuming $\eigen_1 > \eigen_2$, the matrix \eqref{4droot_mat} can be obtained according to the following definition:
\begin{equation}\label{root_tens_4:NG}
    \widetilde{T}^{(0)}_{\mu\nu} = \frac{\left(d\,T^{(0)}_{\mu\nu}-T^{(0)\alpha}{}_\alpha g_{\mu\nu}\right)}{2\sqrt{mn}\sqrt{ d\,T^{(0) \beta \delta} -T_{\beta \delta}^{(0)}\left(T^{(0) \beta}{ }_\beta\right)^2}} \,,
\end{equation}
ensuring, as for the $d=2$ case,
\begin{equation}
\frac{1}{d}T^{(0)\mu\nu}\widetilde{T}^{(0)}_{\mu\nu} = \frac{1}{2}(\eigen_1-\eigen_2)\,.    
\end{equation}
When $n=m=d/2$, the auxiliary tensor \eqref{root_tens_4:NG} reduces to
\begin{equation}\label{root_tens_4}
    \widetilde{T}^{(0)}_{\mu\nu} = \frac{\sqrt{d}\left(T^{(0)}_{\mu\nu}-\frac{1}{d}T^{(0)\alpha}{}_\alpha g_{\mu\nu}\right)}{\sqrt{ T^{(0) \beta \delta} T_{\beta \delta}^{(0)}-\frac{1}{d}\left(T^{(0) \beta}{ }_\beta\right)^2}} \,,
\end{equation}
and its associated matrix is simply
\begin{equation}
     \widetilde{\BT}^{(0)} = \mathbf{U}\left(\bm{\sigma}_3 \otimes \mathbf{1}_{\frac{d}{2}\times \frac{d}{2}}\right)\mathbf{U}^{-1}\,.
\end{equation}
For the rest of this section, we will focus on the $n=m=d/2$ case, whose associated metric deformations have been studied in \cite{Ferko:2024yhc} for $d=4$, but a similar approach can be easily applied to encompass $\sqrt{\TTb}$-like deformations in arbitrary space-time dimensions. The marginal flow associated to the auxiliary tensor \eqref{root_tens_4} reads
\begin{equation}\label{rootttb4}
\begin{split}
     \frac{\partial S_m^{(\gamma)}}{\partial\gamma} &=  \int \mathrm{d}^dx \sqrt{-g} \,\sqrt{ \frac{1}{d}T^{(\gamma) \mu \nu} T_{\mu \nu}^{(\gamma)}-\frac{1}{d^2}\left(T^{(\gamma) \mu}{ }_\mu\right)^2}\\
     &=\frac{1}{d} \int \mathrm{d}^dx \sqrt{-g} T^{(\gamma)\mu\nu}\widetilde{T}^{(\gamma)}_{\mu\nu}\,\,.  
\end{split}
\end{equation}
As in the two-dimensional setting, the solutions to the above equation can be obtained from the seed theory data through a field-dependent metric deformation, with
\begin{equation}\label{root_dress4d}
     S_m^{(\gamma)} [\varphi,k^{(\gamma)}] = \left.\left\{S_m^{(0)}[\varphi,g] \right\}\right|_{g=g(k)}\,,
\end{equation}
provided that the deformed metric $k_{\mu\nu}^{(\gamma)}$ is related to the undeformed one by \cite{Ferko:2024yhc}
\begin{equation}
    k_{\mu\nu}^{(\gamma)} = \cosh \frac{2\gamma}{d}\, g_{\mu\nu}+{\sinh\frac{2\gamma}{d}}\,\widetilde{T}^{(0)}_{\mu\nu} \,.
\end{equation}
Similarly to the $d=2$ case, the associated deformation matrix  $\BP = \{g^{\mu\alpha}k^{(\gamma)}_{\alpha\nu}\}_{\mu,\nu = 0,\dots,d-1}$ is given by
\begin{equation}\label{pimatr_4d}
  \BP = \cosh \frac{2\gamma}{d}\, \mathbf{1} + \sinh \frac{2\gamma}{d}\, \widetilde{\BT}^{(0)} = \exp \left(\frac{2\gamma}{d}  \widetilde{\BT}^{(0)}\right)\,.   
\end{equation}
Note that, since $\det \BP = 1$, we have $\sqrt{-g}$ = $\sqrt{-k^{(\gamma)}}$. Again, the lack of a dressing term in \eqref{root_dress4d} requires coupling the $\sqrt{\TTb}$ deformation to an irrelevant $\TTb$-like deformation to compute the associated gravity sector. Combining \eqref{root_dress4d} with \eqref{dressing_4}, the dressing mechanism for the higher-dimensional $\sqrt{\TTb}+\TTb$-like deformation reads \begin{equation}\label{combined_dressing_4d}
    \begin{split}
  S_m^{(\coupling,\gamma)} [\varphi,h^{(\coupling,\gamma)}] &= \left.\left\{S_m^{(0,0)}[\varphi,g] - \frac{\coupling}{d} \int \mathrm{d}^4x \sqrt{-g} \,\left[\frac{2}{d}\left(T^{(0,0) \mu}{ }_\mu\right)^2-T^{(0,0)\mu\nu}T^{(0,0)}_{\mu\nu}\right]\right\}\right|_{g=g(h)} ,
\end{split}
\end{equation}
with the deformed metric $h_{\mu\nu}^{(\coupling,\gamma)}$ obtained from $g_{\mu\nu}$ via the action of the combined deformation matrix $\BA = \{g^{\mu\alpha}h^{(\coupling,\gamma)}_{\alpha\nu}\}_{\mu,\nu = 0,\dots, d-1}$ defined by
\begin{equation}\label{a4d}
\BA := \BP\circ \BO = \exp \left(\frac{2\gamma}{d}  \widetilde{\BT}^{(0,0)}\right)\left(\mathbf{1}-\coupling \widehat{\BT}^{(0,0)}\right)^{\frac{d}{4}} \,. 
\end{equation}
In the diagonal frame, equation \eqref{combined_dressing_4d} can be equivalently written as
\begin{equation}\label{eig_4d_root}
S^{(\coupling,\gamma)}_m[\varphi,h^{(\coupling,\gamma)}] \,\,\dot{=}\,\, S^{(0,0)}_m[\varphi,g] - \coupling \int \mathrm{d}^4x \sqrt{-g} \,\eigen_1\eigen_2\,.   
\end{equation}
Solving \eqref{a4d} in terms of $\eigen_1$ and $\eigen_2$, and plugging the result back into \eqref{eig_4d_root}, we obtain
\begin{equation}\label{root_gravity_in_d}
 S^{(\coupling,\gamma)}_m[\varphi,h^{(\coupling,\gamma)},g] = S^{(0,0)}_m[\varphi,g] - \frac{1}{\coupling}\int \mathrm{d}^dx \sqrt{-g}\left(\alpha_1^{\frac{d}{4}} - e^{\frac{\gamma}{2}}\right)  \left(\alpha_2 - e^{-\frac{\gamma}{2}}\right) \,,
\end{equation}
where $\alpha_1$ and $\alpha_2$ denote the eigenvalues of $\BA$. Introducing the quantities $x_1 := \tr[\BA]$, $x_2 = \tr[\BA^2]$, the bimetric gravity sector in \eqref{root_gravity_in_d} can be equivalently written as
\begin{equation}
    - \frac{1}{\coupling} \int \mathrm{d}^d x \sqrt{-g} \left[d^{\frac{d}{2}}  \left(x_1\mp\sqrt{d x_2-x_1^2}\right)^{\frac{d}{4}}-e^{\frac{\gamma}{2}}\right] \left[d^{\frac{d}{2}}  \left(x_1\pm\sqrt{d x_2-x_1^2}\right)^{\frac{d}{4}}-e^{-\frac{\gamma}{2}}\right]\,.
\end{equation}
As in the previous section, particularly in $d=4$, algebraic simplifications make it possible to include dynamical degrees of freedom for the gravity theory and recast the bimetric functionals appearing in \eqref{root_gravity_in_d} within a Palatini framework. To this purpose, we consider the total action
\begin{equation}\label{total_action_root}
  S[\varphi,h^{(\coupling,\gamma)}] =S_{EH}[h^{(\coupling,\gamma)}] + S_m^{(\coupling,\gamma)}[\varphi,h^{(\coupling,\gamma)}]\,,\quad    S_{EH}[h^{(\coupling,\gamma)}] = \frac{1}{2}\int \mathrm{d}^4 x \sqrt{-h^{(\coupling,\gamma)}} R[h^{(\coupling,\gamma)}]\,,
\end{equation}
with the Einstein field equations reading
\begin{equation}
    R_{\mu\nu}[h^{(\coupling,\gamma)}] = -\widehat{T}^{(\coupling,\gamma)}_{\mu\nu} =-\widehat{T}^{(0,\gamma)}_{\mu\nu} \,.
\end{equation}
Since the two deformations commute, we can first pull $\widehat{T}^{(0,\gamma)}_{\mu\nu} $ back to $\gamma =0$, and subsequently deform the equations of motion. With this idea in mind, the combined deformation matrix $\BA = \BP \circ \BO$ must have the following eigenvalue structure:
\begin{equation}
    \alpha_1 = e^{\frac{\gamma}{2}} \left(1+\coupling \varrho_1\right)\,,\quad \alpha_2 = e^{-\frac{\gamma}{2}} \left(1+\coupling \varrho_2\right)\,.
\end{equation}
Reproducing the steps outlined in the previous section, the full action can be written as
\begin{equation}\label{final_root}
\begin{split}
  & \,\,S_{EH}[h^{(\coupling,\gamma)},g]+ S^{(\coupling,\gamma)}_m[\varphi,h^{(\coupling,\gamma)},g] \\ =  &\int \mathrm{d}^4 x \sqrt{-g} \left[e^{-\frac{\gamma}{2}} \varrho_1+e^{\frac{\gamma}{2}} \varrho_2+\coupling \left(2 \cosh \frac{\gamma}{2} -1\right) \varrho_1\varrho_2\right]+S^{(0,0)}_m[\varphi,g]\,. \\
\end{split}
\end{equation}
More explicitly, the above functional corresponds to the following action:
\begin{equation}\label{root_grav_1}
    \begin{split}
        \frac{1}{2} \int \mathrm{d}^4 x \sqrt{-g} &\left[ g^{\mu\nu}R_{\mu\nu}[h^{(\coupling,\gamma)}]\cosh \frac{\gamma}{2} +\coupling \left(\cosh \frac{\gamma}{2} -\frac{1}{2}\right)\bigg(\frac{1}{2}\left(g^{\mu \nu} R_{\mu \nu}[h^{(\coupling,\gamma)}]\right)^2\right.\\ &-R^{\mu \nu} [h^{(\coupling,\gamma)}] R_{\mu \nu}[h^{(\coupling,\gamma)}]\bigg)  \left.\pm \sinh \frac{\gamma}{2}\sqrt{4R^{\mu \nu} [h^{(\coupling,\gamma)}] R_{\mu \nu}[h^{(\coupling,\gamma)}]-\left(g^{\mu \nu} R_{\mu \nu}[h^{(\coupling,\gamma)}]\right)^2}\right]\\ &+S^{(0,0)}_m[\varphi,g]\,.
    \end{split}
\end{equation}
As in \eqref{new_action}, since the deformed metric $h^{(\coupling,\gamma)}_{\mu\nu}$ contributes to the above functional solely through its Levi-Civita connection, the action \eqref{root_grav_1} can be equivalently expressed in the Palatini formalism as
\begin{equation}\label{root_grav_palatini}
    \begin{split}
         \frac{1}{2}\int \mathrm{d}^4 x \sqrt{-g} &\left[g^{\mu\nu}R_{\mu\nu}[\Gamma]\cosh \frac{\gamma}{2} +{\coupling} \left( \cosh \frac{\gamma}{2} -\frac{1}{2}\right)\left(\frac{1}{2}\left(g^{\mu \nu} R_{\mu \nu}[\Gamma]\right)^2-R^{\mu \nu} [\Gamma] R_{\mu \nu}[\Gamma]\right)  \right.\\
         & \left.\pm \sinh \frac{\gamma}{2}\sqrt{4R^{\mu \nu} [\Gamma] R_{\mu \nu}[\Gamma]-\left(g^{\mu \nu} R_{\mu \nu}[\Gamma]\right)^2}\right]+S^{(0,0)}_m[\varphi,g]\,.
    \end{split}
\end{equation}
Moreover, introducing the auxiliary parameter
\begin{equation}
    \chi := \coupling \left(2\cosh\frac{\gamma}{2}-1\right)\,,
\end{equation}
together with the tensor
\begin{equation}
    P_{\mu\nu}[\Gamma] = \cosh{\frac{\gamma}{2}} R_{\mu\nu}[\Gamma] - 2\sinh \frac{\gamma}{2}\frac{R_{\mu\alpha}[\Gamma]g^{\alpha\beta}R_{\beta\nu}[\Gamma]-\frac{1}{4}g^{\alpha\beta}R_{\alpha\beta}[\Gamma]R_{\mu\nu}[\Gamma]}{\sqrt{R^{\delta\eta} [\Gamma] R_{\delta\eta}[\Gamma]-\frac{1}{4}\left(g^{\delta\eta} R_{\delta\eta}[\Gamma]\right)^2}} \,,
\end{equation}
equation \eqref{root_grav_palatini} can be written in a form which is reminiscent of the Eddington-inspired Born-Infeld functional \eqref{eibi}:
\begin{equation}\label{rooteibi}
 S_{EH}[h^{(\coupling,\gamma)}]+ S^{(\coupling,\gamma)}_m[\varphi,h^{(\coupling,\gamma)}] \simeq \frac{1}{\chi}\int \mathrm{d}^4 x \left[\sqrt{-\det\left(g_{\mu\nu}+\chi P_{\mu\nu}[\Gamma]\right)}-\sqrt{-g}\right] +  S^{(0,0)}_m[\varphi,g]\,.  
\end{equation}
As a final remark, recall that, in \eqref{combined_dressing_4d}, we combined the $\sqrt{\TTb}$ and the $\TTb$ deformation to provide an explicit dressing sector for the marginal flow. However, setting $\coupling=0$ in \eqref{root_grav_palatini}, it is now trivial to single out the pure $\sqrt{\TTb}$-like contributions to the gravity theory, which yield
\begin{equation}\label{root_gravity:palatini}
  \frac{1}{2}\int \mathrm{d}^4 x \sqrt{-g} \left[g^{\mu\nu}R_{\mu\nu}[\Gamma]\cosh \frac{\gamma}{2} \pm \sinh \frac{\gamma}{2}\sqrt{4R^{\mu \nu} [\Gamma] R_{\mu \nu}[\Gamma]-\left(g^{\mu \nu} R_{\mu \nu}[\Gamma]\right)^2}\right]\,.    
\end{equation}
Interestingly, the gravity action \eqref{root_gravity:palatini} reproduces the typical structure of $\sqrt{\TTb}$-deformed matter theories in $d=4$ space-time (see, for example, the ModMax action \eqref{modmax_action}).
\section{Ricci flows in the space of metrics}\label{sec:ricciflow}
Ricci flows, first introduced by Hamilton in 1982 \cite{Hamilton1982}, describe the evolution of space-time manifolds under the geometric flow
\begin{equation}\label{ricci_intro}
    \frac{\mathrm{d}h_{\mu\nu}^{(\sigma)}}{\mathrm{d} \sigma} = - 2 R_{\mu\nu}[h^{(\sigma)}]\,,
\end{equation}
where $h_{\mu\nu}^{(\sigma)}$ denotes the metric of the manifold, which depends on some flow parameter $\sigma$, and $R_{\mu\nu}[h^{(\sigma)}]$ is its Ricci curvature. The Ricci flow has been critically important in advancing our understanding of the geometric structure of manifolds: one of its most celebrated applications is Perelman's resolution of the Poincaré conjecture \cite{2002math11159P}. Further generalisations of the flow \eqref{ricci_intro} have been proposed in the past years, such as the Ricci-Bourguignon flow \cite{rbf::15}
\begin{equation}\label{ric-bou-intro}
\frac{\mathrm{d}h_{\mu\nu}^{(\sigma)}}{\mathrm{d} \sigma} = - 2 R_{\mu\nu}[h^{(\sigma)}] +2 \kappa R[h^{(\sigma)}] h^{(\sigma)}_{\mu\nu}\,, \end{equation}
where $\kappa$ is an arbitrary parameter. \footnote{The choice of $\kappa$ allows interpolating between the Ricci flow \eqref{ricci_intro}, realised for $\kappa=0$, and the Yamabe flow, obtained from \eqref{ric-bou-intro} in the limit of large $\kappa$.}
In this section, we show how Ricci-type flow equations naturally arise from the metric flows induced by $\TTb$-like deformations when the associated matter sector is coupled to General Relativity.
\subsection{Ricci flows from $\TTb$-like flows}
In Section \ref{sec::4d}, we focused on the class of functionals
    \begin{equation}\label{total_action_2}
  S[\varphi,h^{(\coupling)}] =S_{EH}[h^{(\coupling)}] + S_m^{(\coupling)}[\varphi,h^{(\coupling)}]\,,
\end{equation}
where the matter sector $S_m^{(\coupling)}[\varphi,h^{(\coupling)}]$ satisfies the $\TTb$-like flow 
\begin{equation}\label{new_ttb4}
      \frac{\mathrm d S_m^{(\coupling)}}{\mathrm{d}\coupling} = \frac{1}{d} \int \mathrm{d}^d x \sqrt{-h^{(\coupling)}}\,\left[\frac{2}{d}\left(T^{(\coupling)\mu}{}_{\mu}\right)^2-T^{(\coupling)\mu\nu}T^{(\coupling)}_{\mu\nu}\right]
      \end{equation}
in arbitrary dimensions. 
In addition, we also looked into how the deformation of the matter sector can be interpreted as a distortion of the space-time geometry, even if the metric itself has dynamical degrees of freedom. Although it does not have a direct quantum interpretation, the higher-dimensional version of the $\TTb$ flow provides important geometric insights into the space of classical field theories. All of these characteristics are ultimately related to the existence of the auxiliary flow (equation \eqref{flowh4d}) in the space of metrics, which becomes even more important when a dynamical theory of gravity is involved.
In particular, note that, for generic $d>2$, the Einstein equations \eqref{einstein_equations} can be equivalently written as
\begin{equation}\label{einst_eoms_in_d_rhs}
 R_{\mu\nu}[h^{(\coupling)}] = T_{\mu\nu}^{(\coupling)} - \frac{1}{d-2}(h^{-1})^{(\coupling)\alpha\beta} T_{\alpha\beta}^{(\coupling)}h^{(\coupling)}_{\mu\nu}\,.    
\end{equation}
In addition, one has
\begin{equation}
T_{\mu\nu}^{(\coupling)} = - \widehat{T}_{\mu\nu}^{(\coupling)} + \frac{2}{d}   (h^{-1})^{(\coupling)\alpha\beta} \widehat{T}_{\alpha\beta}^{(\coupling)}h^{(\coupling)}_{\mu\nu}\,, 
\end{equation}
which, combined with \eqref{einst_eoms_in_d_rhs}, yields
\begin{equation}
    R_{\mu\nu}[h^{(\coupling)}] = -\widehat{T}^{(\coupling)}_{\mu\nu} + \frac{d-4}{d(d-2)} (h^{-1})^{(\coupling)\alpha\beta}\widehat{T}^{(\coupling)}_{\alpha\beta} h^{(\coupling)}_{\mu\nu}\,.
\end{equation}
Taking the trace of the above equation, we see that
\begin{equation}
(h^{-1})^{(\coupling)\alpha\beta}\widehat{T}^{(\coupling)}_{\alpha\beta}  = \frac{2-d}{2}R[h^{(\coupling)}] \,,
\end{equation}
which implies
\begin{equation}
 \widehat{T}^{(\coupling)}_{\mu\nu} =  - R_{\mu\nu}[h^{(\coupling)}]  - \frac{d-4}{2d} R[h^{(\coupling)}]h^{(\coupling)}_{\mu\nu}\,.
\end{equation}
Assuming $d>2$, from the flow equation \eqref{flowh4d}, we finally obtain \footnote{In the previous section, we gave extensive importance to the eigenvalue structure of the stress-energy tensor. However, it should be noted that the metric flow \eqref{flowh4d} is satisfied independently of the particular form of $T^{(0)}_{\mu\nu}$.}
\begin{equation}\label{ricci4d}
\frac{\mathrm{d} h_{\mu\nu}^{(\coupling)}}{\mathrm{d}\coupling} = \frac{4}{d}R_{\mu\nu}[h^{(\coupling)}]+ \frac{2(d-4)}{d^2} R[h^{(\coupling)}]h^{(\coupling)}_{\mu\nu}\,,\quad h_{\mu\nu}^{(0)}= g_{\mu\nu}\,.  
\end{equation}
The equation above describes the purely geometric evolution of the space-time manifold, which is known as the Ricci-Bourguignon flow \cite{rbf::15}. In this context, the explicit dependence of the deformed metric $h_{\mu \nu}^{(\coupling)}$ on the matter sector is absent from the flow equation, suggesting that the evolution of the space-time geometry is self-regulated to some extent. The residual influence on the matter content is simply manifested through the boundary condition $R_{\mu \nu}[h^{(0)}]-1/2 R[h^{(0)}]h^{(0)}_{\mu\nu}={T}_{\mu \nu}^{(0)}$ at $\coupling=0$. Notice also that, in $d=4$, equation \eqref{ricci4d} simplifies into
\begin{equation}
\frac{\mathrm{d} h^{(\coupling)}_{\mu\nu}}{\mathrm{d}\coupling} = R_{\mu\nu}[h^{(\coupling)}]\,,\quad h_{\mu\nu}^{(0)} = g_{\mu\nu}\,,    
\end{equation}
thus inducing a Ricci flow \cite{Hamilton1982} of the space-time metric. In Section \ref{sec:black_holes}, we will see how Ricci flows emerge once again when studying the thermodynamic properties of the deformed solutions to the field equations.

To restore the standard conventions used in the context of geometric flows, we define the auxiliary parameter $\sigma := -2\coupling/d$, and introduce the variable
\begin{equation}\label{kappa_and_d}
    \kappa  := \frac{4-d}{2d}\,.
\end{equation}
In this way, equation \eqref{ricci4d} assumes the familiar form
\begin{equation}
\frac{\mathrm{d} h^{(\sigma)}_{\mu\nu}}{\mathrm{d}\sigma} = -2R_{\mu\nu}[h^{(\sigma)}]+ 2\kappa  R[h^{(\sigma)}]h^{(\sigma)}_{\mu\nu}\,,\quad h_{\mu\nu}^{(0)} = g_{\mu\nu}\,.     
\end{equation}
As shown in \cite{rbf::15}, as long as $\kappa$ is below its critical value,
\begin{equation}\label{condition_kappa}
  \kappa < \frac{1}{2(d-1)} := \kappa_{\text{crit}}\,,  
\end{equation}
every initial compact Riemannian manifold admits a unique smooth solution for finite $\sigma$. From the definition \eqref{kappa_and_d}, we see that the condition \eqref{condition_kappa} is indeed satisfied for every integer value $d>2$.

Within this framework, it is moreover possible to prove that, if the undeformed space-time manifold has vanishing scalar curvature, yet $R_{\mu\nu}[h^{(0)}]$ is non-vanishing, \footnote{For simplicity, let $d=4$. If $R_{\mu\nu}[h^{(0)}] = 0$, the flow is trivially static: this merely happens because the associated matter sector has a vanishing stress-energy tensor. In contrast, consider, for example, the space-time structure induced by the Einstein-Maxwell equations: the matter sector is conformal in $d=4$, and $T^{(0)}_{\mu\nu}$ is traceless. Since $R_{\mu\nu}[h^{(0)}] = - \widehat{T}^{(0)}_{\mu\nu}$ in $\coupling =0$, one has $ R[h^{(0)}] = 0$, even though the space-time manifold is not Ricci-flat.} the Ricci scalar becomes positive for any given positive value of the auxiliary parameter $\sigma$, provided that $\kappa \leq \kappa_{\text{crit}}$ (see Remark 4.3 of \cite{rbf::15}). Again, the latter condition is always verified for $d>2$. This is consistent with the results presented in \cite{Conti_2022}, where it was shown that higher-dimensional $\TTb$-like flows do not preserve, in general, the original flatness of the space-time manifold.
\subsection{Einstein-Ricci solitons and the singular conformal limit}
As briefly noted in the previous section, the stress tensor degeneracy \eqref{eig_degeneracy} characterises a wide range of physical theories in four dimensions. Among such theories, some show an even greater degeneracy in the eigenvalues of the energy-momentum tensor: a notable example is that of a vacuum energy term, whose action functional reads \footnote{The minus sign in \eqref{cosm_const} is introduced to adhere to the usual conventions.}
\begin{equation}\label{cosm_const}
  S^{(0)}_m[g] = -\int \mathrm{d}^4x \sqrt{-g}\cosm\,,
\end{equation}
where $\cosm$ is assumed to be constant throughout space-time. In this short section, we analyse the evolution of the above matter action under the $\TTb$-like flow \eqref{ttbd}, exploring its singular limit though the result presented in \cite{Aldrovandi:2001vx, Aldrovandi:2004km}. The Hilbert stress tensor associated with \eqref{cosm_const} is simply computed as
\begin{equation}\label{t_for_cosmo}
    T^{(0)}_{\mu\nu} = -\cosm g_{\mu\nu}\,,
\end{equation}
and the metric deformation \eqref{integrate4} reduces to a global rescaling of the original space-time metric:
\begin{equation}\label{def_h_cosm}
    h_{\mu\nu}^{(\coupling)} = (1+\coupling \cosm)g_{\mu\nu}\,.
\end{equation}
Noting that 
\begin{equation}
    -\sqrt{-g} \left(\Lambda+\coupling T^{(0)\mu\nu}\widehat{T}_{(0)\mu\nu}\right)= -\frac{\sqrt{-h^{(\coupling)}}}{(1+\coupling \Lambda)^2}\left(\Lambda+\coupling\Lambda^2\right)\,,
\end{equation}
the $\TTb$-like deformed action is easily obtained by means of equation \eqref{dressing_4}: \footnote{It is easy to explicitly verify that \eqref{def_cosm} satisfies the $\TTb$ flow. However, remember that in this context the deformed metric $h_{\mu\nu}^{(\coupling)}$ is considered to be the actual space-time metric, and the matter sector depends on $\coupling$ only explicitly.}
\begin{equation}\label{def_cosm}
    S^{(\coupling)}_m [h^{(\coupling)}] =- \int \mathrm{d}^4x \sqrt{-h^{(\coupling)}}\left(\frac{\cosm}{1+\coupling\cosm}\right)\,.
\end{equation}
When the above action is coupled to general relativity, the Einstein field equations read
\begin{equation}\label{einst_eoms_for_ads}
    R_{\mu\nu}[h^{(\coupling)}] =\frac{\cosm}{1+\coupling \cosm} h_{\mu\nu}^{(\coupling)} := \Delta h_{\mu\nu}^{(\coupling)}\,,
\end{equation}
or, in terms of the scalar curvature, $R[h^{(\coupling)}] = 4\Delta$. Note that, in terms of the Ricci flows described in the previous section, the above equation defines an Einstein-Ricci soliton. A Ricci soliton is simply a space-time manifold whose metric $h^{(\coupling)}_{\mu\nu}$ evolves under the Ricci flow in an inessential manner, in the sense
that all the stages $h_{\mu\nu}^{(\coupling)}$ coincides with $h_{\mu\nu}^{(0)} = g_{\mu\nu}$ up to diffeomorphisms and multiplications by some positive constant \cite{sym12020289}. In static coordinates, a general solution to equation \eqref{einst_eoms_for_ads} can be written as
\begin{equation}\label{h_for_ads}
h_{\mu\nu}^{(\coupling)}\mathrm{d}x^\mu\mathrm{d}x^\nu =  -\left(1-\frac{2\Delta r^2}{3}\right)\mathrm{d}t^2 + \left(1-\frac{2\Delta r^2}{3}\right)^{-1}\!\!\mathrm{d}\rad^2 +  r^2\left(\mathrm{d}\theta^2+\sin^2\theta \mathrm{d}\phi^2\right)\,. 
\end{equation}
The metric \eqref{h_for_ads} presents an event horizon in $r = \sqrt{3/(2\Delta)}$ for positive values of $\Delta$, whose value is, in turn, determined both by the seed geometry and by the value of $\coupling$. Assuming $\cosm>0$, there exists a critical value of $\coupling$ for which the sign of $\Delta$ is flipped, and the singularity disappears.

\begin{figure}[h]
    \centering
    \includegraphics[scale=0.31]{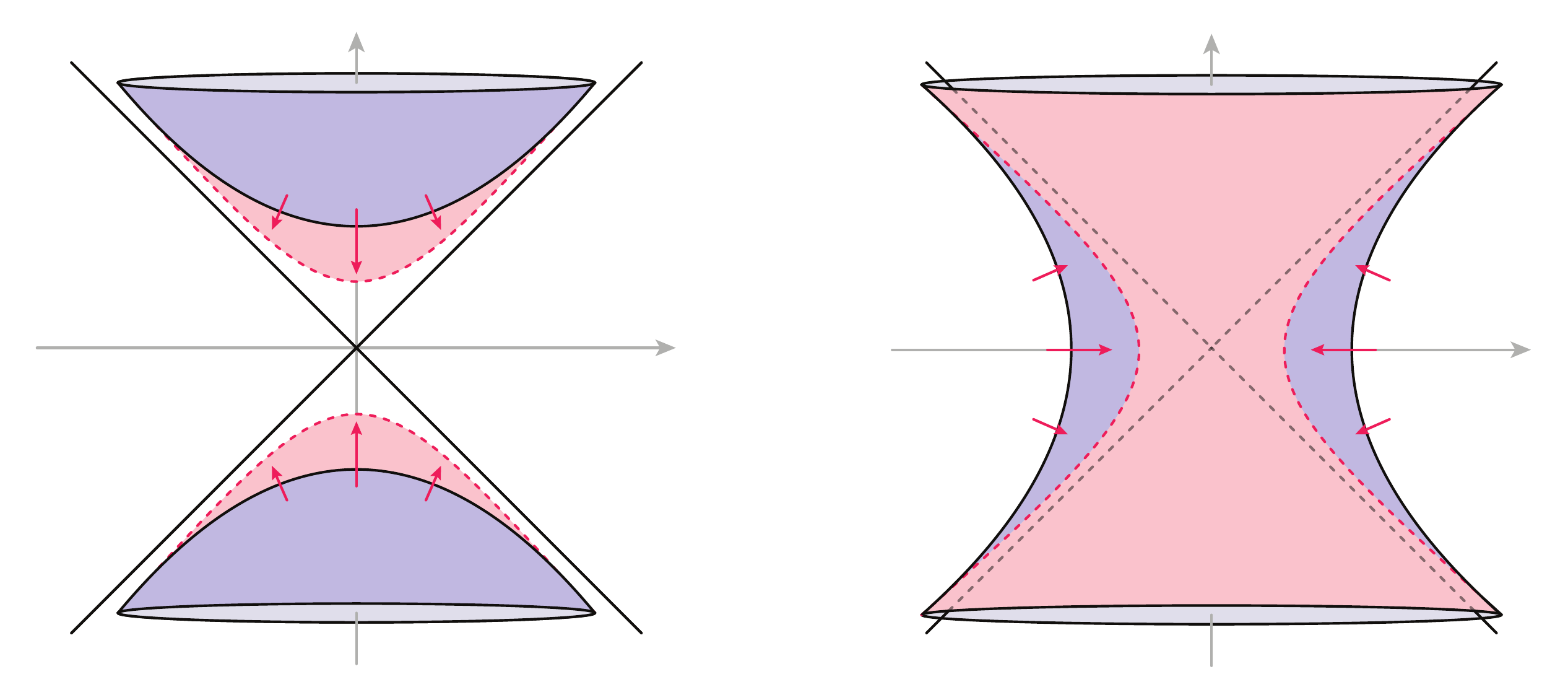}
    \caption{On the left, the evolution of the AdS$_4$ space-time along the flow. As $\coupling \to -1/\cosm$, the Anti-de Sitter space approaches a conical geometry. On the right, a similar depiction of the evolution of the dS$_4$ space-time along the flow.}
    \label{fig:mesh6}
\end{figure}
Space-times with constant scalar curvature are called maximally symmetric, since they can accommodate the highest possible number of Killing vectors. Moreover, once a specific metric signature is fixed, such space-times are unique for each value of $R[h^{(\coupling)}]$. Note that the deformation trivialises the geometry in the strong coupling limits $\coupling \to \pm \infty$, as the symmetry group of space-time trivialises to the Poincaré group $\operatorname{ISO}(1,3) = \mathbb{R}^{1,3} \rtimes \operatorname{O}(1,3)$, and the space-time manifold itself reduces to the Minkowski space-time. When the curvature of the manifold is finite and positive (respectively, negative), the solution to the equations of motion is known as the de Sitter (respectively, Anti-de Sitter) space-time, and its associated symmetry groups is known as the de Sitter (respectively, Anti-de Sitter) group $\mathrm{dS}_4$ (respectively, $\mathrm{AdS}_4$). Both $\mathrm{dS}_4$ and $\mathrm{AdS}_4$ can be defined as hyper-surfaces in $\mathbb{R}^{1,4}$ and $\mathbb{R}^{2,3}$, whose points in Cartesian coordinates $\xi^A$, $A=0,\dots,4$, satisfy, respectively,
\begin{equation}
\begin{aligned}
 \mathrm{dS_4}:&\quad  \left(\xi^0\right)^2-\left(\xi^1\right)^2-\left(\xi^2\right)^2-\left(\xi^3\right)^2-\left(\xi^4\right)^2=-l^2\,, \\
 \mathrm{AdS_4}:&\quad  \left(\xi^0\right)^2-\left(\xi^1\right)^2-\left(\xi^2\right)^2-\left(\xi^3\right)^2+\left(\xi^4\right)^2=l^2 \,,
\end{aligned}
 \end{equation}
where $l := -3 \eta_{44}/\Delta$ is known as the de Sitter length parameter. 

As the flow parameter $\coupling$ approaches $-1/\cosm$, the effective cosmological constant $\Delta$ grows towards infinity, the de Sitter length $l$ vanishes, and the group deformation produces simultaneous changes in the embedding geometries \cite{Aldrovandi:2001vx, Aldrovandi:2004km}. Specifically, in the limit of an infinite cosmological term, the de Sitter and the Anti-de
Sitter spaces are both led to a four-dimensional cone-space, singular at the cone vertex (see Figure \ref{fig:mesh6} for a pictorial visualisation). Note that, as $l \to 0$, the metric $h_{\mu\nu}^{(\coupling)}$ ends up being singular everywhere on the cone, and the Ricci scalar curvature becomes infinite.

As shown in \cite{Aldrovandi:2004km}, in the limit of an infinite cosmological constant, the group which defines the kinematics of the cone space-time reduces to the so-called conformal Poincaré group \cite{Aldrovandi:1998ux}, the semi-direct product of Lorentz and the proper conformal group. Moreover, despite the singular nature of the metric tensor, preventing the definition of space distances and time intervals, a conformally invariant metric can be introduced, which allows the definition of conformal concepts of space distances and time intervals.

Note that the above discussion can be easily extended to arbitrary dimensions. For generic $d$, the deformed metric reads
\begin{equation}\label{def_h_cosm_d}
    h^{(\coupling)}_{\mu\nu} = (1+\coupling \cosm)^{\frac{4}{d}}g_{\mu\nu}\,,
\end{equation}
and the corresponding deformed matter sector is simply given by
\begin{equation}\label{def_cosm_d}
    S^{(\coupling)}_m [h^{(\coupling)}] = -\int \mathrm{d}^dx \sqrt{-h^{(\coupling)}}\left(\frac{\cosm}{1+\coupling\cosm}\right)\,,
\end{equation}
reproducing the structure obtained in \eqref{def_cosm} for the $d=4$ case. When the deformed matter sector is coupled to General Relativity, the Einstein field equations read
\begin{equation}
    R_{\mu\nu}[h^{(\coupling)}] = \frac{2}{d-2}\Delta h_{\mu\nu}^{(\coupling)}\,,
\end{equation}
where, as before, we introduced a deformed cosmological constant $\Delta = \cosm/(1+\coupling \cosm)$. Note that we did not discuss the effects of additional $\sqrt{\TTb}$-like deformations of the matter sector \eqref{cosm_const}, as they have no effect on the seed theory. This can be easily seen from the dressing perspective, since $\Lambda$ does not depend on the space-time metric, and, due to the unimodularity of the $\sqrt{\TTb}$ deformation matrix, one has $\sqrt{-g} = \sqrt{-h^{(0,\gamma)}}$.

\subsection{Classical $(d-1)$-form field theories as vacuum energy theories}
A large class of physical models can be expressed in the language of Abelian $p$-form field theories, with the scalar ($0$-form) field and the Maxwell ($1$-form) field being some of the most notable examples. The other possibilities in four space-time dimensions are $p=2$ and $p=3$-form field theories, and the latter will be the main focus of this brief section. As pointed out in \cite{Mitskievich:1998uk}, both the $p = 2$ and the $p = 3$ case inherently produce stress-energy tensors typical of perfect fluids. Specifically, they can describe non-rotating fluids ($p = 2$) and rotating fluids (with a combination of $p = 2$ and $p = 3$ fields), while the pure $p = 3$ case reproduces the effects of a cosmological constant. This phenomenon can be extended to arbitrary space-time dimensions \cite{Mitskievich:2002nj}, with $(d-2)$-form field theories being responsible for non-rotating perfect fluids, and $(d-1)$-form field theories accounting for vacuum energy contributions.

In $d$ space-time dimensions, we consider the rank $d-1$ potential
\begin{equation}
 A = \frac{1}{(d-1)!} A_{\mu_1\dots \mu_{d-1}} dx^{\mu^1}\wedge\dots \wedge dx^{\mu^{d-1}}\,, \quad \mu_i = 1,\dots,d\,, 
\end{equation}
and introduce the field strength $F = \mathrm{d}A$. In what follows, we will consider matter actions $S_m^{(0)}[A,g]$ constructed from the Lorentz invariant
\begin{equation}
    K := d! \star( F \wedge \star F ) = F_{\mu_1\dots\mu_d}F^{\mu_1\dots\mu_d}\,.
\end{equation}
Denoting by $\mathcal{L}_m^{(0)}$ the Lagrangian density of the seed theory, the stress-energy tensor associated with the $(d-1)$-form field theory takes the form
\begin{equation}\label{stress_d-1}
    T^{(0)}_{\mu\nu} = \left(\mathcal{L}_m^{(0)} - 2 K \frac{\partial \mathcal{L}_m^{(0)}}{\partial K}\right)g_{\mu\nu}\,.
\end{equation}
From \eqref{integrate_d}, we immediately see that the $\TTb$ deformed metric is given by
\begin{equation}
    h_{\mu\nu}^{(\coupling)} = \left(1-\coupling\mathcal{L}_m^{(0)} + 2\coupling K \frac{\partial \mathcal{L}_m^{(0)}}{\partial K}\right)^{\frac{4}{d}}g_{\mu\nu}\,.
\end{equation}
In addition, one can show that the equations of motion for the potential $A_{\mu_1\dots \mu_{d-1}}$ reduce to \cite{Mitskievich:1998uk, Mitskievich:2002nj}
\begin{equation}\label{eoms_d-1}
 \sqrt{K}\,\frac{\partial \mathcal{L}_m^{(0)}}{\partial K} = c^{(0)}\,,
\end{equation}
where $c^{(0)}$ is a constant. If $\mathcal{L}_m^{(0)} = a\sqrt{K}$ for some real coefficient $a$, equation \eqref{eoms_d-1} is trivially satisfied under any configuration of the $(d-1)$-form field, and the stress-energy tensor \eqref{stress_d-1} identically vanishes. Furthermore, if we assume that $\mathcal{L}_m^{(0)} \neq a\sqrt{K}$, we observe that equation \eqref{eoms_d-1} forces $K$ itself to be a constant. Defining
\begin{equation}
\cosm_K:=2 K \frac{\partial \mathcal{L}_m^{(0)}}{\partial K}-\mathcal{L}_m^{(0)} \,,    
\end{equation}
we see that the structure of the energy-momentum tensor \eqref{t_for_cosmo} is replicated by the equations of motion of the theory. 

Along the $\TTb$-like flow, we expect the deformed Lagrangian to depend on the $(d-1)$-form field only through $K$. In particular, denoting by $\mathcal{L}_m^{(\coupling)}$ the deformed Lagrangian density, we will have
\begin{equation}\label{t_d-1_deformed}
   T^{(\coupling)}_{\mu\nu} = \left(\mathcal{L}_m^{(\coupling)} - 2 K \frac{\partial \mathcal{L}_m^{(\coupling)}}{\partial K}\right)h^{(\coupling)}_{\mu\nu}\,,   
\end{equation}
together with
\begin{equation}\label{eoms_d-1_deformed}
  \sqrt{K}\,\frac{\partial \mathcal{L}_m^{(\coupling)}}{\partial K} = c^{(\coupling)}\,.   
\end{equation}
The implications of \eqref{t_d-1_deformed} and \eqref{eoms_d-1_deformed} are the same as in the undeformed case, with the deformed theory in $\coupling$ reproducing the effects of a (deformed) cosmological constant. Moreover, one has
\begin{equation}
\begin{split}
 \frac{\mathrm{d}c^{(\coupling)}}{\mathrm{d}\coupling} &=  \sqrt{K} \frac{\partial}{\partial K}\frac{\partial \mathcal{L}_m^{(\coupling)}}{\partial \coupling} 
 = \frac{1}{d}\sqrt{K}\frac{\partial}{\partial K}\left(\widehat{T}^{(\coupling)\mu\nu}T_{\mu\nu}^{(\coupling)}\right)
= \frac{2}{d}\sqrt{K}\widehat{T}^{(\coupling)\mu\nu} \frac{\partial T_{\mu\nu}^{(\coupling)}}{\partial K}\,.
\end{split}
\end{equation}
Using \eqref{t_d-1_deformed}, after some algebraic manipulations, we arrive at
\begin{equation}
 \frac{\mathrm{d}c^{(\coupling)}}{\mathrm{d}\coupling} = -2\sqrt{K}\left(\mathcal{L}_m^{(\coupling)} - 2 K \frac{\partial \mathcal{L}_m^{(\coupling)}}{\partial K}\right)\left(\frac{\partial \mathcal{L}_m^{(\coupling)}}{\partial K} + 2 K \frac{\partial^2 \mathcal{L}_m^{(\coupling)}}{\partial K^2}\right)\,.    
\end{equation}
As far as the explicit structure of the deformed theory is concerned,
it is important to note that the initial boundary conditions for the $(d-1)$-form field theory are generally different from those encountered when dealing with a pure cosmological constant. Additionally, the extra involvement of the metric tensor $g_{\mu\nu}$ in constructing the Lorentz invariant $K$ makes the dressing procedure significantly more challenging than before.
Therefore, even though classical $(d-1)$-form field theories behave like vacuum energy theories, their deformation does not immediately replicate the typical structure of the Einstein-Ricci solitons that are characteristic of pure (A)dS space-times. However, on-shell, the evolution of the deformed space-time geometry still exhibits self-similarity. Setting $\mathcal{L}^{(0)}_m = K$ and solving the flow perturbatively in $\coupling$, one can verify that the associated deformed action is
\begin{equation}\label{deformed_full}
S^{(\coupling)}[A,h^{(\coupling)}] =2\sum_{n=0}^{\infty}\int \mathrm{d}^d x \sqrt{-h^{(\coupling)}}\left[\frac{(4 n+1)!}{(n+1)! (3 n+2)!}\right] \coupling^{n} K^{n+1}\,.   
\end{equation}
The above series can be equivalently expressed in terms of generalised hypergeometric functions, yielding
\begin{equation}\label{deformed_d-1_form_field}
S^{(\coupling)}[A,h^{(\coupling)}]  =\frac{3}{4\coupling} \int\mathrm{d}^d x \sqrt{-h^{(\coupling)}}   \left[ {}_3F_2\left(-\frac{1}{2},-\frac{1}{4},\frac{1}{4};\frac{1}{3},\frac{2}{3};\frac{256 }{27}\coupling K\right)-1\right]\,.    
\end{equation}
Among several other models, Maxwell's theory in $d=2$ belongs to the class of $(d-1)$-form field theories, and the action \eqref{deformed_d-1_form_field} correctly reproduces the results presented in  \cite{Conti:2018jho} for the two-dimensional case. There is an intuitive reason behind the somehow non-trivial observation that the deformation yields the same result independently of the space-time dimension. Note that, at each perturbative order in $\coupling$, equation \eqref{t_d-1_deformed} ensures that $T_{\mu\nu}^{(\coupling)}= \widehat{T}_{\mu\nu}^{(\coupling)}$. Moreover, being $T_{\mu\nu}^{(\coupling)}$ proportional to the metric tensor, at each order the matter action is deformed by the operator
\begin{equation}
    \frac{1}{d}T^{(\coupling)\mu\nu}\widehat{T}_{\mu\nu}^{(\coupling)} = \left(\mathcal{L}_m^{(\coupling)} - 2 K \frac{\partial \mathcal{L}_m^{(\coupling)}}{\partial K}\right)^2\,,
\end{equation}
which does not depend on the specific value of $d$.

It is also interesting to observe that the coefficients in \eqref{deformed_full} correspond, for each $n$, to the number of rooted 3-connected triangulations with $2n+2$ faces. \footnote{A graph is said to be 3-connected if there are no two vertices whose removal would disconnect the graph, and it is called rooted when one of its vertices, edges, or faces is distinguished in some way.} As a final remark since, at any order, the $\sqrt{\TTb}$-like flow is driven by the difference of the two independent eigenvalues of the energy-momentum tensor, and given the degenerate stucture of \eqref{t_d-1_deformed}, the marginal deformation does not affect the theory.
\section{$\TTb$ and $\sqrt{\TTb}$-deformed black holes in $d=4$}\label{sec:black_holes}
The discussion in Section \ref{sec:ricciflow} focuses on the case of Einstein-Ricci solitons, which is just one example of how the effects of $\TTb$-like deformations on solutions to the Einstein field equations can be analyzed analytically.

In this section, using the formalism developed in the previous sections, we study how black hole solutions are modified when a $\TTb$-like deformation is induced on the matter sector of the theory. \footnote{
In the case of $d=4$, the dual Palatini description \eqref{eibi} of the gravity sector connected to the deformation can be used alternatively to study the deformed solutions. This is possible due to the extensive body of work in cosmology literature concerning Eddington-inspired Born-Infeld black holes (see, for example, \cite{Olmo:2013gqa}).}
Focusing on the extremal solutions associated with $\TTb$-like deformations of Reissner-Nordström black holes, we show that the evolution of the Bekenstein-Hawking entropy along the stress tensor flow is governed by a Ricci flow induced on the horizon surface. 
Additionally, using the commutativity of the $\TTb$ and $\sqrt{\TTb}$-like perturbation, we illustrate how electrovacuum solutions are deformed by their combined action.  
\subsection{$\TTb$-deformed electrovacuum solutions}
We start by considering Maxwell's theory of electrodynamics as the seed theory for the deformation. The initial action is
\begin{equation}\label{maxwell}
    S_m^{(0)}[A,g] = \frac{1}{4}\int \mathrm{d}^4 x \sqrt{-g}\, F^{\mu\nu}F_{\mu\nu}\,,
\end{equation}
where $F_{\mu\nu} = \partial_\mu A_\nu - \partial_\nu A_\mu$ is the field strength associated to the Abelian gauge field $A_\mu$. The stress-energy tensor of the theory is computed as
\begin{equation}
    T_{\mu\nu}^{(0)} = \frac{1}{4} F^{\alpha\beta}F_{\alpha\beta} g_{\mu\nu} - F_{\mu}{}^{\alpha} F_{\nu\alpha}\,,
\end{equation}
and the matrix $\BT^{(0)}$ admits two distinct eigenvalues $\eigen_+$ and $\eigen_-$, each of multiplicity $2$, and given by 
\begin{equation}\label{max_eigenv}
    \eigen_\pm = \pm \sqrt[4]{\det\BT^{(0)}}\,.
\end{equation}
As briefly stated in section \ref{sec::4d}, this type of degeneracy extends well beyond Maxwell's theory and characterizes any Abelian gauge theory in four space-time dimensions. Therefore, the findings of this section can be readily extended to derive the deformed version of a much broader class of theories. By implementing the dressing procedure introduced in \eqref{dressing_4}, it is possible to show that the $\TTb$-deformed variant of \eqref{maxwell} is the Born-Infeld theory\cite{Conti_Iannella, Conti_2022, Morone:2024ffm}\footnote{The full identification is achieved by setting $\coupling = -1/(2\beta^2)$, where $\beta$ is known as the Born-Infeld parameter, corresponding to the maximal value of the electromagnetic field strength.} 
\begin{equation}\label{bi}
     S_{m}^{(\coupling)}[A,h^{(\coupling)}] = \frac{1}{2\coupling}\int \mathrm{d}^4x\sqrt{-h^{(\coupling)}}\left(\sqrt{1+\coupling\mathcal{F} -\frac{\coupling^2}{4} \mathcal{G}^2}-1\right) \,,
\end{equation}
where we introduced the invariants $\mathcal{F}:= F^{\mu \nu}{F}_{\mu \nu}$ and $\mathcal{G}:= F^{\mu \nu}  \left(\star {F}   \right)_{\mu \nu}$. Coupling the above functional to General Relativity, one obtains the Einstein-Born-Infeld action
\begin{equation}\label{deformedEH}
    S[A,h^{(\coupling)}] = S_{EH}[h^{(\coupling)}] + S_{m}^{(\coupling)}[A,h^{(\coupling)}]\,.
\end{equation}
As discussed in the previous section, the equations of motion for \eqref{deformedEH} are equivalent to those obtained by coupling the undeformed theory \eqref{maxwell} to the Eddington-Inspired Born-Infeld functional \eqref{eibi}:
\begin{equation}
      S[A,h^{(\coupling)}]\simeq  \frac{1}{\coupling}\int \mathrm{d}^4 x \left[\sqrt{-\det\left(g_{\mu\nu}+\coupling R_{\mu\nu}[\Gamma]\right)}-\sqrt{-g}\right]+\frac{1}{4}\int \mathrm{d}^4 x \sqrt{-g}\, F^{\mu\nu}F_{\mu\nu} \,, 
\end{equation}
where the two metrics $h^{(\coupling)}_{\mu\nu}$ and $g_{\mu\nu}$ are related by \eqref{integrate4}. We now aim at studying spherically symmetric, stationary solutions to the Einstein equations for the deformed theory, namely,
\begin{equation}\label{hard_eq}
    R_{\mu\nu}[h^{(\coupling)}] = -\widehat{T}_{\mu\nu}^{(\coupling)}\,.
\end{equation}
Making the most out of \eqref{tthat}, solving the non-trivial system of equations \eqref{hard_eq} can be reduced to solving
\begin{equation}\label{eoms_with_t0}
    R_{\mu\nu}[h^{(\coupling)}\circ g] = -\widehat{T}^{(0)}_{\mu\nu}\,.
\end{equation}
Additionally, the diagonalisation procedure of $\widehat{\BT}^{(0)}$ is automatically implemented when considering reference frames in which the solutions enjoy spherical symmetry. In particular, for a given radial function $r$, we write the gauge field as $A = A_t \mathrm{d}t$, where \footnote{For the rest of this paper, we will assume $Q>0$.}
\begin{equation}\label{gauge_field}
 A_t  = -\frac{Q}{r} + \text{constant terms}\,,
\end{equation}
ensuring that the only non-vanishing component of the field strength $F_{\mu\nu}$ is $F_{rt} = Q/r^2$. The stress tensor of the seed theory can be compactly written as 
\begin{equation}\label{t_free}
    \BT^{(0)} = -\frac{Q^2}{2 r^4} \bm{\sigma}_3\otimes \mathbf{1}_{2\times 2}\,, 
\end{equation}
and, looking back at \eqref{max_eigenv}, we can identify $\eigen_\pm = \pm {Q^2}/{2 r^4}$. We then make an ansatz for the metric $g_{\mu\nu}$, following the proposal of \cite{Olmo:2013gqa}:
\begin{equation}\label{gmetric}
g_{\mu\nu}\mathrm{d}x^\mu\mathrm{d}x^\nu =  - A(x)\mathrm{d}t^2 + \frac{\mathrm{d}x^2}{A(x)} +  r^2 (x)\left(\mathrm{d}\theta^2+\sin^2\theta \mathrm{d}\phi^2\right)\,. 
\end{equation}
The function $r(x)$ can be alternatively chosen as a radial coordinate, which turns the above expression into
\begin{equation}
g_{\mu\nu}\mathrm{d}x^\mu\mathrm{d}x^\nu =  - A(x)\mathrm{d}t^2 + \frac{\mathrm{d}r^2}{B(x)} +  r^2 \left(\mathrm{d}\theta^2+\sin^2\theta \mathrm{d}\phi^2\right)\,.    
\end{equation}
In general, this replacement is possible as long
as the relation between the coordinate $x$ and the radial function $r$ is monotonic \cite{Olmo:2013gqa}. The deformed metric $h_{\mu\nu}^{(\coupling)}$ is then obtained by acting on $g_{\mu\nu}$ with the deformation matrix $\BO = \mathbf{1}-\coupling\widehat{\BT}^{(0)}$: denoting its two distinct eigenvalues by $\omega_\pm$, where
\begin{equation}\label{omegapm}
    \omega_{\pm} = 1\pm \frac{\coupling Q^2}{2r^4}\,,
\end{equation}
such that $\BO = \operatorname{diag}(\omega_-,\omega_+)\otimes \mathbf{1}_{2\times 2}$, one has
\begin{equation}\label{h_line}
\begin{split}
h_{\mu\nu}^{(\coupling)}\mathrm{d}x^\mu\mathrm{d}x^\nu  = - \omega_-A(x)\mathrm{d}t^2 + \omega_-\frac{\mathrm{d}x^2}{A(x)} +  \omega_+r^2 (x)\left(\mathrm{d}\theta^2+\sin^2\theta\mathrm{d}\phi^2\right)\,.
\end{split}
\end{equation}
Note that the action of $\BO$ deforms the effective radius of the spherical element of the metric: defining the auxiliary coordinate $\rad := \sqrt{\omega_+} r$, such that
\begin{equation}\label{r in terms of z}
    r^2 = \frac{1}{2}\left(z^2+\sqrt{z^4-2\coupling Q^2}\right)\,,
\end{equation}
it is possible to rewrite equation \eqref{h_line} as
\begin{equation}\label{h_in_z}
h_{\mu\nu}^{(\coupling)}\mathrm{d}x^\mu\mathrm{d}x^\nu =  - C(\rad)\mathrm{d}t^2 + \frac{\mathrm{d}\rad^2}{C(\rad)} +  \rad^2\left(\mathrm{d}\theta^2+\sin^2\theta \mathrm{d}\phi^2\right)\,, 
\end{equation}
where we introduced the metric function $C(\rad) := \omega_-A(x)$. Note that the square root term in \eqref{r in terms of z} is well-defined only for negative values of the flow parameter $\coupling$. With the identification $\coupling=-1/(2\beta^2)$, requiring $\coupling <0$ corresponds to a positivity condition for the Born-Infeld bound. For this reason, it is convenient to study the solution in terms of the auxiliary flow parameter 
\begin{equation}
    \ecoupling:=-\coupling\,.
\end{equation}
In the deformed frame, the geometry of space-time is governed by $T_{\mu\nu}^{(\ecoupling)}$: in particular, with the usual ansatz $C(\rad) = 1-{2 m(\rad)}/{\rad}$, the mass term $m(z)$ is required to satisfy \cite{Bronnikov:2000vy} \footnote{The minus sign in equation \eqref{define mass func} originates from the definition of stress-energy tensor given in \eqref{define_T}, which carries an extra minus sign if compared with the usual definition of energy density.}
\begin{equation}\label{define mass func}
    m(z) = -\int_0^z \mathrm{d}\xi\,\xi^2\, (h^{-1})^{(\ecoupling)0\alpha}T_{\alpha 0}^{(\ecoupling)}(\xi):= -\int_0^z \mathrm{d}\xi\,\xi^2\, \mathcal{H}^{(\ecoupling)}(\xi)\,.
\end{equation}
We then define the asymptotic mass $m_0 := m(\infty)$, which allows us to rewrite \eqref{define mass func} as
\begin{equation}\label{dm2}
  m(z) = m_0 +\int_z^\infty \mathrm{d}\xi\,\xi^2\,\mathcal{H}^{(\ecoupling)}(\xi)\,.
\end{equation}
At this point, using the equivalence \eqref{tthat}, we see that
\begin{equation}
(h^{-1})^{(\ecoupling)\mu\alpha}\widehat{T}_{\alpha\nu}^{(\ecoupling)} =  (\Omega^{-1})^\mu{}_\beta\, g^{\beta\gamma}\widehat{T}_{\gamma\nu}^{(\ecoupling)} = (\Omega^{-1})^\mu{}_\beta\, g^{\beta\gamma}\widehat{T}_{\gamma\nu}^{(0)} \,,  
\end{equation}
which explicitly yields
\begin{equation}\label{energydens}
  \mathcal{H}^{(\ecoupling)}(\xi)  = -\frac{Q^2 }{2 r^4(\xi)-\ecoupling Q^2} = -\frac{Q^2}{\xi^4+\xi^2\sqrt{\xi^4 + 2 \ecoupling Q^2}}\,.
\end{equation}
The mass function $m(z)$ is then obtained as
\begin{equation}
    m(z) =m_0 -\int_z^\infty \mathrm{d}\xi\, \frac{Q^2}{\xi^2+\sqrt{\xi^4 + 2 \ecoupling Q^2}}\,,
\end{equation}
or, equivalently, by solving the differential equation
\begin{equation}\label{dm}
    \frac{\mathrm{d}m}{\mathrm{d}z} = \frac{Q^2}{z^2+\sqrt{z^4 + 2 \ecoupling Q^2}}\,.
\end{equation}
In $\ecoupling=0$, where $\rad = r$, equation \eqref{dm} reduces to
\begin{equation}
  \lim_{\ecoupling \to 0}\frac{\mathrm{d}m}{\mathrm{d}\rad}  = \frac{Q^2}{2\rad^2} \,,
\end{equation}
which, upon integration, reproduces the Reissner-Nordström structure of the associated undeformed solution, namely,
\begin{equation}\label{crn}
    \lim_{\ecoupling \to 0}{C}(\rad) = 1-\frac{2m_0}{\rad} + \frac{Q^2}{\rad^2}\,.
\end{equation}
The structure of the event horizon associated with the asymptotic solution \eqref{crn} is related to the roots of $C(z)$, given by
\begin{equation}
    z_{\pm} = m_0 \pm \sqrt{m_0^2 - Q^2}\,.
\end{equation}
The two solutions coincide in the extremal case, for which $z_\pm := \bar{z} = Q$. Conversely, in the strong coupling regime, for $\ecoupling\to \infty$, the energy density \eqref{energydens} vanishes, the mass function goes to $0$, and space-time becomes Minkowskian. For generic values of $\ecoupling$, solving the differential equation \eqref{dm}, one obtains
\begin{equation}\label{general_c}
    C(z) = 1-\frac{2m_0}{z} + \frac{z^2}{3\ecoupling} \left(1-\sqrt{1+\frac{2\ecoupling Q^2}{z^4}}\right) +\frac{4Q^2}{3z}\int_z^\infty \frac{\mathrm{d}\xi }{\sqrt{\xi^4+2\ecoupling Q^2}}\,.
\end{equation}
The last term is an elliptic integral of the first kind, which can be written in terms of hypergeometric functions:
\begin{equation}
    \begin{aligned}
\int_z^{\infty} \frac{\mathrm{d} \xi}{\sqrt{\xi^4+2\ecoupling Q^2}}  =\frac{1}{z}\left[{ }_2 F_1\left(\frac{1}{4}, \frac{1}{2} ; \frac{5}{4} ;-\frac{2\ecoupling Q^2}{ z^4}\right)\right] .
\end{aligned}
\end{equation}
It is worth noting that, with the identification $\ecoupling=-\coupling = 1/(2\beta^2)$, this result aligns with the Einstein-Born-Infeld black hole solutions found in the literature \cite{Gunasekaran:2012dq}. From the discussion of Section \ref{sec:ricciflow}, we also conclude that the metric \eqref{h_in_z}, with $C(z)$ given by \eqref{general_c}, is a solution of the Ricci flow equation
\begin{equation}
\frac{\mathrm{d}h_{\mu\nu}^{(\ecoupling)}}{\mathrm{d}\ecoupling} = -R_{\mu\nu}[h^{(\ecoupling)}]\,.
\end{equation}
The quite complicated structure of the function $C(z)$ associated with the $\TTb$-like deformed metric makes it difficult to study the solution purely relying on analytical tools. However, the extremal solution corresponding to \eqref{general_c} admits particularly simple expressions as far as the horizon data are concerned.
\subsection{Ricci flows of the extremal Bekenstein-Hawking entropy}
For generic values of the flow parameter $\coupling$, the black hole solution described by \eqref{general_c} can have either zero, one or two event horizons, depending on the values of $m_0$ and $Q$. In the extremal case, when the two solutions coincide, the horizon radius $\bar{z}$ of the $\TTb$-deformed black hole is obtained as \cite{Gunasekaran:2012dq}
\begin{equation}\label{barzQ}
    \bar{z} = \sqrt{Q^2-\frac{\ecoupling}{2}}\,.
\end{equation}
Identifying $Q$ as the horizon radius $\bar{r}$ of the (undeformed) extremal Reissner-Nordström black hole, equation \eqref{barzQ} can be written in a more suggestive form as
\begin{equation}\label{ricci_1}
  \bar{z} = \sqrt{\bar{r}^2 -\frac{\ecoupling}{2}}\,.   
\end{equation}
Introducing the two-dimensional metric tensor $H_{ij}^{(\ecoupling)}$ associated to the spherical element of the four-dimensional metric \eqref{h_in_z}, namely,
\begin{equation}    H_{ij}^{(\ecoupling)}(z,\psi) \mathrm{d}\psi^{i}\mathrm{d}\psi^{j}:=\rad^2\left(\mathrm{d}\theta^2+\sin^2\theta \mathrm{d}\phi^2\right)\,, 
\end{equation}
where $(\psi^1,\psi^2) = (\theta,\phi)$, one can observe that equation \eqref{ricci_1} is a solution to the flow equation
\begin{equation}\label{ricciflow}
    \frac{\mathrm{d}}{\mathrm{d}\ecoupling}  H_{ij}^{(\ecoupling)}(\bar{z},\psi) = -\frac{1}{2}R_{ij}\left[H^{(\ecoupling)}(\bar{z},\psi)\right]\,,
\end{equation}
where $R_{ij}[H^{(\ecoupling)}]$ denotes the Ricci curvature tensor associated to the metric $H_{ij}^{(\ecoupling)}$. Together with the boundary condition 
\begin{equation}
   H_{ij}^{(0)} (\bar{z},\psi) =  H_{ij} (\bar{r},\psi)\,,
\end{equation}
which ensures that, in $\ecoupling=0$, the horizon surface reduces to the Reissner-Nordström-like one, equation \eqref{ricciflow} defines a Ricci flow of the event horizon associated with the $\TTb$-deformed extremal black hole. The evolution of the Bekenstein-Hawking entropy $\mathcal{S}$ of the extremal solution, given by the area law 
\begin{equation}\label{bek_haw}
    \mathcal{S} = \frac{A_{\text{hor}}}{4}=\pi \bar{z}^2\,,
\end{equation}
where $A_{\text{hor}}$ denotes the area of the black hole horizon, is in turn fully governed by the Ricci flow equation \eqref{ricciflow}. This shows that, for spherically symmetric, static, electrovacuum solutions, a $\TTb$-like deformation of the associated four-dimensional theory is reflected on the extremal geometry of the black hole's event horizon in terms of a Ricci flow, which determines the evolution of the entropy area law \eqref{bek_haw} as the coupling parameter $\ecoupling$ is varied. In particular, combining \eqref{bek_haw} with \eqref{ricci_1}, one has
\begin{equation}
    \frac{\partial \mathcal{S}}{\partial \ecoupling} = -\frac{\pi}{2}\,,\quad \ecoupling\leq2 Q^2\,.
\end{equation}
Flowing towards positive values of $ \ecoupling$ — or, equivalently, towards negative values of $\coupling$ — we observe that there is a critical value of the auxiliary flow parameter, $\ecoupling_{\text{crit}} =2 Q^2$, at which the singularity disappears, and the entropy of the black hole vanishes.

 \begin{figure}[h]
    \centering
    \includegraphics[scale=0.37]{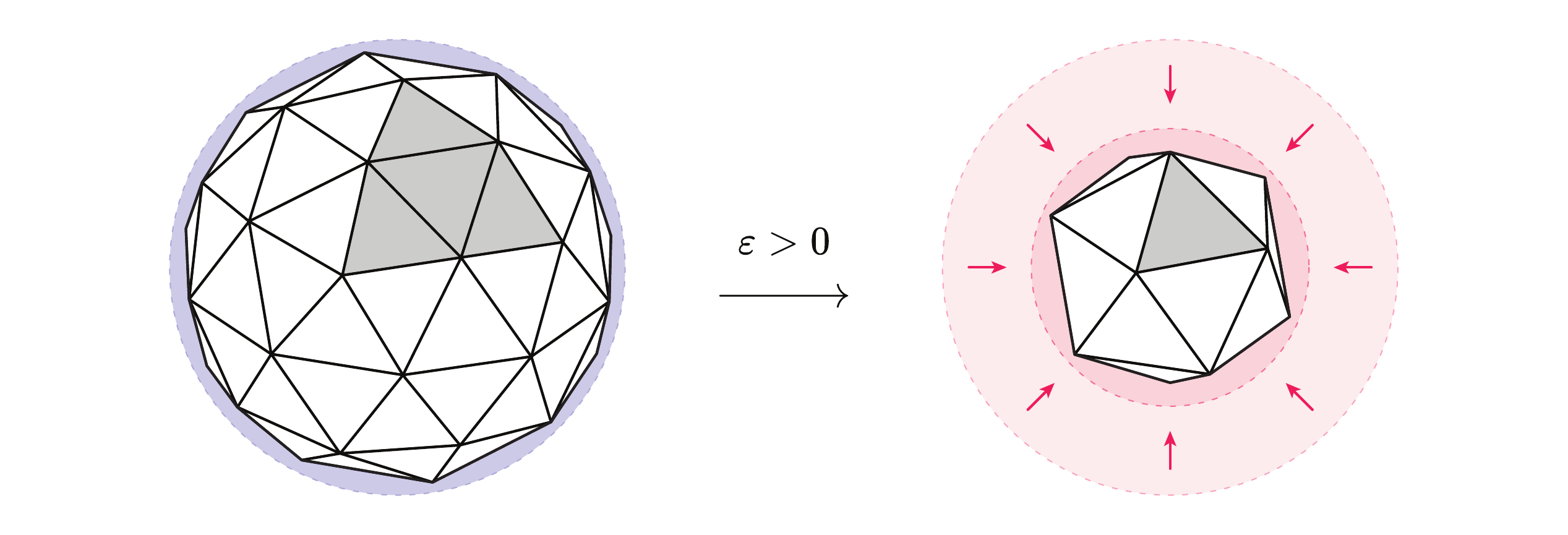}
    \caption{Evolution of the extremal black hole event horizon along the flow, where each triangle represents one fundamental domain (induced by the Planck scale, here taken to be adimensional). Note that, because of \eqref{bek_haw}, four triangles constitute one bit of Bekenstein-Hawking entropy. Conversely, flowing towards positive values of $\ecoupling$, the horizon shrinks, and the black hole entropy is reduced.}
    \label{fig:mesh63}
\end{figure}

\subsection{$\sqrt{\TTb}$-deformed electrovacuum solutions}
We now examine how static, spherically symmetric electrovacuum solutions to the Einstein-Maxwell theory are deformed along the marginal $\sqrt{\TTb}$ flow. Notably, it has been shown \cite{Conti_2022, Babaei-Aghbolagh:2022uij,Ferko:2022iru} that the Maxwell theory \eqref{maxwell} is deformed along the $\sqrt{\TTb}$ flow into ModMax electrodynamics \cite{Bandos:2020jsw}, whose action is
\begin{equation}\label{modmax_action}
    S^{(0,\gamma)}_m[A,h^{(0,\gamma)}] = \frac{1}{4}\int \mathrm{d}^4 x \sqrt{-h^{(0,\gamma)}} \left(\cosh \gamma\, \mathcal{F} - \sinh \gamma \,\sqrt{\mathcal{F}^2+\mathcal{G}^2}\right)\,,
\end{equation}
where $\mathcal{F}:= F^{\mu \nu}{F}_{\mu \nu}$ and $\mathcal{G}:= F^{\mu \nu}  \left(\star {F}   \right)_{\mu \nu}$. As in the pure $\TTb$ case, we start by making an ansatz for the undeformed metric $g_{\mu\nu}$, which we write as
\begin{equation}\label{g_root_bh}
g_{\mu\nu}\mathrm{d}x^\mu\mathrm{d}x^\nu =  - A(x)\mathrm{d}t^2 + \frac{\mathrm{d}r^2}{B(x)} +  r^2 \left(\mathrm{d}\theta^2+\sin^2\theta \mathrm{d}\phi^2\right)\,.    
\end{equation}
The $\sqrt{\TTb}$ deformed line element is obtained by acting on \eqref{g_root_bh} with the deformation matrix $\BP$ introduced in \eqref{pimatr_4d}: denoting by $\pi_\pm$ its eigenvalues, we have
\begin{equation}\label{k_bh}
\begin{split}
h_{\mu\nu}^{(0,\gamma)}\mathrm{d}x^\mu\mathrm{d}x^\nu  &= - \pi_+A(x)\mathrm{d}t^2 + \pi_+\frac{\mathrm{d}x^2}{A(x)} +  \pi_-r^2 (x)\left(\mathrm{d}\theta^2+\sin^2\theta\mathrm{d}\phi^2\right)\\
&:= C(w)\mathrm{d}t^2 + \frac{\mathrm{d}z^2}{C(w)} +  w^2 \left(\mathrm{d}\theta^2+\sin^2\theta\mathrm{d}\phi^2\right) \,,    
\end{split}  
\end{equation}
where $  \pi_\pm = e^{\pm \frac{\gamma}{2}}$. Note that, similarly as before, in the second line of \eqref{k_bh} we introduced the auxiliary coordinate $w := \sqrt{\pi_-} r$, as well as the metric function $C(z):= \pi_+ A(x)$\,. In particular, we see that $r^2 =e^{\frac{\gamma}{2}}w^2$.
We now aim to obtain an expression for the matrix $(h^{-1})^{(0,\gamma)\mu\alpha}T_{\alpha\nu}^{(0,\gamma)}$ in terms of the seed theory data. This is not easily done in general, but when the undeformed theory is conformal — as in the case of \eqref{maxwell} — one has \cite{Ebert:2023tih}
\begin{equation}\label{identity_root_tens}
    T^{(0,\gamma)}_{\mu\nu} = \cosh \frac{\gamma}{2}\, T_{\mu\nu}^{(0,0)} + \frac{1}{2} {\sinh \frac{\gamma}{2}}\,\sqrt{T^{(0)\alpha\beta}T^{(0,0)}_{\alpha\beta}}\, g_{\mu\nu}\,,
\end{equation}
which, in turn, implies that the diagonal entries of the matrices $g^{\mu\alpha}T_{\alpha\nu}^{(0,0)}$ and $(h^{-1})^{(0,\gamma)\mu\alpha}T_{\alpha\nu}$  are the same. In particular, we have 
\begin{equation}\label{plug_in_mass}
 (h^{-1})^{(0,\gamma)\mu\alpha}T_{\alpha\nu}^{(0,\gamma)} = -\frac{Q^2}{2r^4}\bm{\sigma}_3\otimes \mathbf{1}_{2\times 2} =  -\frac{e^{-\gamma}Q^2}{2w^4}\bm{\sigma}_3\otimes \mathbf{1}_{2\times 2}\,.  
\end{equation}
Assuming $C(w) = 1- 2m(w)/w$, one has
\begin{equation}\label{mass_root}
    m(z) = m_0 + \int_z^\infty \mathrm{d}\xi \, \xi^2\, \mathcal{H}^{(0,\gamma)}(\xi)\,,
\end{equation}
where
\begin{equation}
\mathcal{H}^{(0,\gamma)}(\xi) =- \frac{Q^2}{2r^4(\xi)} = -\frac{e^{-\gamma}Q^2}{2\xi^4}\,.      
\end{equation}
The mass function $m(w)$ is then easily obtained once \eqref{plug_in_mass} is substituted back into \eqref{mass_root}, yielding
\begin{equation}
m(w) = m_0 - \frac{e^{-\gamma}Q^2}{w}\,. 
\end{equation}
We thus conclude that the metric function $C(w)$ associated to the ModMax black hole solution is
\begin{equation}
    {C}(w) = 1-\frac{2m_0}{w} + \frac{e^{-\gamma}Q^2}{w^2}\,,
\end{equation}
which reproduces the results presented in \cite{Flores-Alfonso:2020euz}. Observe that the global effect of the $\sqrt{\TTb}$ deformation simply amounts to a rescaling of the Reissner-Nordström charge, with \begin{equation}
\label{eq:above}
Q^2 \mapsto e^{-\gamma}Q^2 \,.  
\end{equation}
Relation (\ref{eq:above}) can be also obtained by the following argument, proposed in \cite{Flores-Alfonso:2020euz}, which directly leads to the deformed solution for the electrically charged black hole: given the spherical symmetry of the solution, the invariant $\mathcal{G}$ vanishes, and, on-shell, the ModMax Lagrangian reduces to
\begin{equation}
    \mathcal{L}^{(0,\gamma)}_m = \frac{1}{4}(\cosh \gamma - \sinh \gamma)\mathcal{F} =e^{-\gamma}\mathcal{L}^{(0,0)}_m\,,
\end{equation}
where $\mathcal{L}^{(0,0)}_m = \mathcal{F}/4$ is the Maxwell Lagrangian. From this viewpoint, recalling that higher-dimensional $\sqrt{\TTb}$ and $\TTb$-like flows commute, it is easy to combine the two deformations. The $\TTb$+$\sqrt{\TTb}$-deformed matter sector reproduces the ModMax-Born-Infeld theory, described by the action
\begin{equation}\label{mmbi}
S^{(\coupling,\gamma)}_m[A,h^{(\coupling,\gamma)}] =     \frac{1}{2\coupling}\int \mathrm{d}^4x\sqrt{-h^{(\coupling,\gamma)}}\left(\sqrt{1-4\coupling\mathcal{L}_m^{(0,\gamma)} -\frac{\coupling^2}{4} \mathcal{G}^2}-1\right) \,,
\end{equation}
where
\begin{equation}
 \mathcal{L}_m^{(0,\gamma)} = \frac{1}{4} \left(\cosh \gamma\, \mathcal{F} - \sinh \gamma \,\sqrt{\mathcal{F}^2+\mathcal{G}^2}\right)\,.  
\end{equation}
Explicitly, after rescaling the black hole charge in \eqref{general_c} by a factor of $e^{-\frac{\gamma}{2}}$, and reintroducing the auxiliary parameter $\ecoupling=-\coupling$, we obtain that the solution to the Einstein field equations for the matter sector \eqref{mmbi} is given by
\begin{equation}\label{hmmbi}
       h_{\mu\nu}^{(\ecoupling,\gamma)}\mathrm{d}x^\mu\mathrm{d}x^\nu =  - C(\rad)\mathrm{d}t^2 + \frac{\mathrm{d}\rad^2}{C(\rad)} +  \rad^2\left(\mathrm{d}\theta^2+\sin^2\theta \mathrm{d}\phi^2\right)\,,
\end{equation}
where the metric function $C(z)$ reads:
\begin{equation}
   C(z) = 1-\frac{2m_0}{z} + \frac{z^2}{3\ecoupling} \left(1-\sqrt{1+\frac{2\ecoupling e^{-\gamma}Q^2}{z^4}}\right) +\frac{4e^{-\gamma}Q^2}{3z^2}\left[{ }_2 F_1\left(\frac{1}{4}, \frac{1}{2} ; \frac{5}{4} ;-\frac{2\ecoupling e^{-\gamma}Q^2}{ z^4}\right)\right] .  
\end{equation}
For any fixed value of $\gamma$, the metric \eqref{hmmbi} satisfies the Ricci flow
\begin{equation}
    \frac{\mathrm{d}h^{(\ecoupling,\gamma)}_{\mu\nu}}{\mathrm{d}\ecoupling} = -R_{\mu\nu}[h^{(\ecoupling,\gamma)}]\,.
\end{equation}
From \eqref{barzQ}, we see that the horizon radius of the extremal solution is obtained as
\begin{equation}\label{barzQ2}
    \bar{z} = \sqrt{e^{-\gamma}Q^2 -\frac{\ecoupling}{2}}\,,
\end{equation}
with the auxiliary $\TTb$-like coupling $\ecoupling$ inducing a Ricci flow of the horizon surface. The corresponding Bekenstein-Hawking entropy of the extremal ModMax-Born-Infeld black hole reads
\begin{equation}
    \mathcal{S} = \pi \left(e^{-\gamma}Q^2 -\frac{\ecoupling}{2}\right)\,.
\end{equation}
Note that, for any given value of $\gamma$, there exists a critical value  $\ecoupling_{\mathrm{crit}}= 2e^{-\gamma}Q^2$ such that $\mathcal{S}=0$.
\subsection{Charged black holes in asymptotically curved space-time}
So far, we have studied $\TTb$ and $\sqrt{\TTb}$-deformed electrovacuum solutions in asymptotically flat geometries. However, this formalism can easily be extended to include asymptotically (A)dS$_4$ space-times as well. Importantly, adding a cosmological term to the seed theory does not spoil the desired stress tensor degeneracy. Consider the action:
\begin{equation}\label{S_cosmo}
 {\underline{S}}^{(0)}_m[A,g] = S^{(0)}_m [A,g] - \int \mathrm{d}^4x \sqrt{-g}\cosm\,,
\end{equation}
where $S^{(0)}_m [A,g]$ is defined as in \eqref{maxwell}, and $\cosm$ accounts for vacuum energy contributions. The stress tensor associated with \eqref{S_cosmo} is simply computed as
\begin{equation}\label{underT}
    \underline{T}^{(0)}_{\mu\nu} = {T}^{(0)}_{\mu\nu} - \cosm g_{\mu\nu}\,,\quad T^{(0)}_{\mu\nu} := -\frac{2}{\sqrt{-g}}\frac{\delta S^{(0)}_m}{\delta g^{\mu\nu}}\,,
\end{equation}
and the eigenvalues $\underline{\eigen}_{\pm}$ of the matrix $\underline{\BT}^{(0)} = \{g^{\mu\alpha}\underline{T}^{(0)}_{\alpha\nu}\}_{\mu,\nu = 0,3}$ are obtained from the ones of ${\BT}^{(0)}$ (see \eqref{max_eigenv}) as
\begin{equation}
    \underline{\eigen}_{\pm} = \eigen_\pm - \cosm\,.
\end{equation}
A generic solution to the $\TTb$-like flow \eqref{flowh4d} associated to the seed theory \eqref{S_cosmo} can be written in terms of the deformed metric $h^{(\coupling)}_{\mu\nu}$ as \cite{Conti_Iannella}
\begin{equation}  {\underline{S}}^{(\coupling)}_m[A,h^{(\coupling)}]  = {S}^{(\auxcoupling )}_m[A,h^{(\coupling)}] - \int \mathrm{d}^4 x\sqrt{-h^{(\coupling)}}\left(\frac{\cosm}{1+\coupling \cosm}\right)\,,\quad \auxcoupling  := \coupling(1+\coupling\cosm)\,,
\end{equation}
where ${S}^{(\auxcoupling )}_m[A,h^{(\coupling)}]$ coincides with the Born-Infeld functional \eqref{bi} with an effective flow parameter $\auxcoupling= \coupling(1+\coupling\cosm)$.
Note that the cosmological term gets modified along the flow, and, in the deformed theory, the full contribution to the vacuum energy of the system is given by
\begin{equation}\label{eff_l}
  \Delta = \frac{\cosm}{1+\coupling \cosm}\,.
\end{equation}
From \eqref{underT}, we see that the deformation matrix
\begin{equation}
\underline{\BO} = \mathbf{1}-\coupling \underline{\widehat{\BT}}^{(0)}    
\end{equation} admits two independent eigenvalues $\underline{\omega}_\pm$, such that $\underline{\BO} =\operatorname{diag}(\underline{\omega}_-,\underline{\omega}_+)\otimes \mathbf{1}_{2\times 2}$, which are obtained as
\begin{equation}
 \underline{\omega}_{\pm} = 1\pm \frac{\coupling Q^2}{2r^4}+ \coupling \cosm \,.
\end{equation}
We can now couple the deformed theory to the Einstein-Hilbert term, and study spherically symmetric, static solutions of the following action:
\begin{equation}
    S[A, h^{(\coupling)}] = S_{EH}[h^{(\coupling)}] + {S}^{(\auxcoupling )}_m[A,h^{(\coupling)}] - \int \mathrm{d}^4 x\sqrt{-h^{(\coupling)}}\Delta\,.
\end{equation}
Echoing the ansatz \eqref{gmetric} for the metric $g_{\mu\nu}$, we can write the associated deformed metric $h^{(\coupling)}_{\mu\nu}$ through the action of $\underline{\BO}$ as
\begin{equation}
     h^{(\coupling)}_{\mu\nu}\mathrm{d}x^\mu\mathrm{d}x^\nu =  - C(\rad)\mathrm{d}t^2 + \frac{\mathrm{d}\rad^2}{C(\rad)} +  \rad^2\left(\mathrm{d}\theta^2+\sin^2\theta \mathrm{d}\phi^2\right)\,,   
\end{equation}
where, this time, we introduced $C(z):= \underline{\omega}_- A(x)$, together with the auxiliary coordinate $z := \sqrt{\underline{\omega}_+} r$. Explicitly, setting $\ecoupling=-\coupling$, $z$ and $r$ are related by the following equation:
\begin{equation}
    r^2 = \frac{1}{2(1-\ecoupling\Lambda)}\left(z^2 + \sqrt{z^4 + 2\ecoupling(1-\ecoupling \Lambda)Q^2}\right)\,.
\end{equation}
Assuming $C(z) = 1-2m(z)/z$, and computing the deformed energy density as
\begin{equation}
  \mathcal{H}^{(\ecoupling)}(\xi) =- \frac{Q^2}{\xi^4 +\xi^2\sqrt{\xi^4 + 2\eauxcoupling Q^2}}-\Delta\,,\quad \eauxcoupling := \ecoupling(1-\ecoupling\cosm)\,,
\end{equation}
we see that the mass function $m(z)$ is determined by the solution to the differential equation
\begin{equation}\label{dm3}
   \frac{\mathrm{d}m}{\mathrm{d}\rad} = \frac{Q^2}{z^2 +\sqrt{z^4 + 2\eauxcoupling Q^2}}+{z^2}{\Delta}\,,
\end{equation}
which translates into
\begin{equation}\label{ads_bh_1}
      C(z) = 1-\frac{2m_0}{z} - \frac{2\Delta z^2}{3} + \frac{z^2}{3\eauxcoupling } \left(1-\sqrt{1+\frac{2\eauxcoupling  Q^2}{z^4}}\right) +\frac{4Q^2}{3z^2}\left[{ }_2 F_1\left(\frac{1}{4}, \frac{1}{2} ; \frac{5}{4} ;-\frac{2\eauxcoupling Q^2}{ z^4}\right)\right]. 
\end{equation}
Note that the above result is consistent with the Einstein-Born-Infeld black hole solutions present in the literature (see, for example, \cite{Gunasekaran:2012dq}).

We conclude by observing that, despite the seed theory being not conformal due to the presence of $\Lambda$, it is still possible to take into account $\sqrt{\TTb}$-like deformations of the matter action \eqref{S_cosmo}. To this end, it is sufficient to note that since the cosmological constant produces a homogeneous shift in the stress-energy tensor eigenvalues: since, at any order, the $\sqrt{\TTb}$-like deformation is driven by the difference of the two independent eigenvalues of the energy-momentum tensor, the $\Lambda$ term is in no way affected by the deformation, and we find
\begin{equation}  {\underline{S}}^{(0,\gamma)}_m[A,h^{(0,\gamma)}]  = {S}^{(0,\gamma)}_m[A,h^{(0,\gamma)}] - \int \mathrm{d}^4 x\sqrt{-h^{(0,\gamma)}}\Lambda\,.
\end{equation}
The energy density associated to the ModMax theory in asymptotically curved space is then obtained as
\begin{equation}
 \mathcal{H}^{(0,\gamma)}(\xi) =- \frac{Q^2}{2r^4(\xi)}-\Lambda = -\frac{e^{-\gamma}Q^2}{2\xi^4}-\Lambda\,,      
\end{equation}
implying that, once again, the global effect of the $\sqrt{\TTb}$-like deformation of the solution to the Einstein equations comes down to a rescaling of the black hole charge. Combining this result with \eqref{ads_bh_1}, we see that the ModMax-Born-Infeld solution in asymptotically curved space-time is characterised by the metric function
\begin{equation}
 C(z) = 1-\frac{2m_0}{z}- \frac{2\Delta z^2}{3} + \frac{z^2}{3\eauxcoupling } \left(1-\sqrt{1+\frac{2\eauxcoupling e^{-\gamma} Q^2}{z^4}}\right) +\frac{4e^{-\gamma}Q^2}{3z^2}\left[{ }_2 F_1\left(\frac{1}{4}, \frac{1}{2} ; \frac{5}{4} ;-\frac{2\eauxcoupling e^{-\gamma}Q^2}{ z^4}\right)\right].    
\end{equation}

\section{Conclusions}
In this work, primarily focusing on the eigenvalue structure of the stress-energy tensor, together with the dressing mechanism which characterised irrelevant $\TTb$ and $\TTb$-like deformations, we illustrated a method for associating gravity functionals to a broad class of stress tensor deformations, both in two and in higher space-time dimensions. After coupling the deformed theory to General Relativity, we showed how the metric flow associated with ${\TTb}$-like deformations in higher-dimensional space-times can be equivalently interpreted in terms of Ricci and Ricci-Bourguignon flows of the space-time geometry. Similar strategies were applied in the context of marginal $\sqrt{\TTb}$ and $\sqrt{\TTb}$-like flows. We then showed how the metric deformations arising from the dressing formalism can be employed to study specific solutions to the field equations, which led us to examine $\TTb$ and $\sqrt{\TTb}$-like deformed black hole solutions in four dimensions, reproducing spherically symmetric, static solution to the ModMax-Einstein-Born-Infeld theory. In addition, we discussed how the entropy associated with the extremal $\TTb$-deformed solution evolves along the flow, revealing a connection with the Ricci flow of two-dimensional surfaces.

The newly introduced methods for studying integrable deformations of two-dimensional non-linear sigma models using auxiliary fields \cite{Ferko:2024ali, Bielli:2024khq, Bielli:2024ach} provides fresh insights within the context of $\TTb$ and $\sqrt{\TTb}$-deformed two-dimensional theories. Understanding how this approach might be related to their gravity interpretation presents an important challenge that has yet to be fully addressed.

While the dressing-type mechanism associated with the stress-energy tensor structure \eqref{eig_degeneracy_generic} are well-understood for the case $n = m = d/2$, similar frameworks for generic values of $n$, $m$ are still under investigation. Among the wide range of scenarios, we believe the $n=d-1$, $m=1$ case deserves the most attention, as it encompasses a large variety of physical theories.

From the gravity side, while the algebraic simplifications which arise in $d=4$ space-times allow recasting the bimetric gravity theories associated to $\TTb$ and $\sqrt{\TTb}$-like deformations within a Palatini framework, significant complications show up in $d\geq 4$, and, at the moment, no higher-dimensional counterpart to \eqref{eibi} is known. The $d=6$ case is of particular interest, as chiral two-form theories display the desired eigenvalue degeneracy of the stress-energy tensor \cite{Ferko:2024zth, Deger:2024jfh}.

\bigskip 

\textbf{Acknowledgements} -- We thank Hossein Babaei-Aghbolagh, Song He, Stefano Negro and Hao Ouyang for their insightful discussions and valuable contributions to our previous collaborations on related research projects. We thank Christian Ferko, Jue Hou and Gabriele Tartaglino-Mazzucchelli for their valuable insights and comments. We also thank Hossein Babaei-Aghbolagh for sharing notes which independently discussed $\sqrt{\TTb}$-deformed Reissner-Nordström solutions. 

\medskip

This project received partial support from the INFN project SFT and the PRIN Project No. 2022ABPBEY, with the title ``Understanding quantum field theory through its deformations''. Roberto Tateo expresses gratitude to the participants of the MATRIX Research Program ``New Deformations of Quantum Field and Gravity Theories'' for the stimulating atmosphere and discussions, and to the Mathematical Research Institute MATRIX and the Sydney Mathematical Research Institute for the invitation and financial support during his Australian visit.

\bibliography{Biblio3}

\providecommand{\href}[2]{#2}\begingroup\raggedright\begin{thebibliography}{100}

\bibitem{Zamolodchikov:1991vx}
A.~B. Zamolodchikov, {\it {From tricritical Ising to critical Ising by thermodynamic Bethe ansatz}},  {\em Nucl. Phys. B} {\bf 358} (1991) 524--546.

\bibitem{Mussardo:1999aj}
G.~Mussardo and P.~Simon, {\it {Bosonic type S matrix, vacuum instability and CDD ambiguities}},  {\em Nucl. Phys. B} {\bf 578} (2000) 527--551 [\href{http://arXiv.org/abs/hep-th/9903072}{{\tt hep-th/9903072}}].

\bibitem{Zamolodchikov:2004ce}
A.~B. Zamolodchikov, {\it {Expectation value of composite field {$T$} anti-{$T$} in two-dimensional quantum field theory}},  \href{http://arXiv.org/abs/hep-th/0401146}{{\tt hep-th/0401146}}.

\bibitem{Dubovsky:2012sh}
S.~Dubovsky, R.~Flauger and V.~Gorbenko, {\it {Effective String Theory Revisited}},  {\em JHEP} {\bf 09} (2012) 044 [\href{http://arXiv.org/abs/1203.1054}{{\tt 1203.1054}}].

\bibitem{Dubovsky:2012wk}
S.~Dubovsky, R.~Flauger and V.~Gorbenko, {\it {Solving the Simplest Theory of Quantum Gravity}},  {\em JHEP} {\bf 09} (2012) 133 [\href{http://arXiv.org/abs/1205.6805}{{\tt 1205.6805}}].

\bibitem{Caselle:2013dra}
M.~Caselle, D.~Fioravanti, F.~Gliozzi and R.~Tateo, {\it {Quantisation of the effective string with TBA}},  {\em JHEP} {\bf 07} (2013) 071 [\href{http://arXiv.org/abs/1305.1278}{{\tt 1305.1278}}].

\bibitem{Cavaglia:2016oda}
A.~Cavagli\`a, S.~Negro, I.~M. Sz\'ecs\'enyi and R.~Tateo, {\it {{$\text{T}{\overline{\text{T}}}$}-deformed 2D Quantum Field Theories}},  {\em JHEP} {\bf 10} (2016) 112 [\href{http://arXiv.org/abs/1608.05534}{{\tt 1608.05534}}].

\bibitem{Smirnov:2016lqw}
F.~A. Smirnov and A.~B. Zamolodchikov, {\it {On space of integrable quantum field theories}},  {\em Nucl. Phys. B} {\bf 915} (2017) 363--383 [\href{http://arXiv.org/abs/1608.05499}{{\tt 1608.05499}}].

\bibitem{Bonelli:2018kik}
G.~Bonelli, N.~Doroud and M.~Zhu, {\it {{$\text{T}{\overline{\text{T}}}$}-deformations in closed form}},  {\em JHEP} {\bf 06} (2018) 149 [\href{http://arXiv.org/abs/1804.10967}{{\tt 1804.10967}}].

\bibitem{Conti_Iannella}
R.~Conti, L.~Iannella, S.~Negro and R.~Tateo, {\it Generalised {B}orn-{I}nfeld models, {L}ax operators and the {$\text{T}{\overline{\text{T}}}$} perturbation},  {\em JHEP} {\bf 2018} (11, 2018) [\href{http://arXiv.org/abs/1806.11515v2}{{\tt 1806.11515v2}}].

\bibitem{Cardy:2018sdv}
J.~Cardy, {\it {The {$\text{T}{\overline{\text{T}}}$} deformation of quantum field theory as random geometry}},  {\em JHEP} {\bf 10} (2018) 186 [\href{http://arXiv.org/abs/1801.06895}{{\tt 1801.06895}}].

\bibitem{Datta:2018thy}
S.~Datta and Y.~Jiang, {\it {$T\bar{T}$ deformed partition functions}},  {\em JHEP} {\bf 08} (2018) 106 [\href{http://arXiv.org/abs/1806.07426}{{\tt 1806.07426}}].

\bibitem{Baggio:2018gct}
M.~Baggio and A.~Sfondrini, {\it {Strings on NS-NS Backgrounds as Integrable Deformations}},  {\em Phys. Rev. D} {\bf 98} (2018), no.~2 021902 [\href{http://arXiv.org/abs/1804.01998}{{\tt 1804.01998}}].

\bibitem{Dei:2018jyj}
A.~Dei and A.~Sfondrini, {\it {Integrable S matrix, mirror TBA and spectrum for the stringy AdS$_{3}$ $\times$ S$^{3}$ $\times$ S$^{3}$ $\times$ S$^{1}$ WZW model}},  {\em JHEP} {\bf 02} (2019) 072 [\href{http://arXiv.org/abs/1812.08195}{{\tt 1812.08195}}].

\bibitem{Chakraborty:2019mdf}
S.~Chakraborty, A.~Giveon and D.~Kutasov, {\it {$T\overline{T}$, $J\overline{T}$, $T\bar{J}$ and String Theory}},  {\em J. Phys. A} {\bf 52} (2019), no.~38 384003 [\href{http://arXiv.org/abs/1905.00051}{{\tt 1905.00051}}].

\bibitem{Callebaut:2019omt}
N.~Callebaut, J.~Kruthoff and H.~Verlinde, {\it {$ T\overline{T} $ deformed CFT as a non-critical string}},  {\em JHEP} {\bf 04} (2020) 084 [\href{http://arXiv.org/abs/1910.13578}{{\tt 1910.13578}}].

\bibitem{McGough:2016lol}
L.~McGough, M.~Mezei and H.~Verlinde, {\it {Moving the {CFT} into the bulk with {$\text{T}{\overline{\text{T}}}$}}},  {\em JHEP} {\bf 04} (2018) 010 [\href{http://arXiv.org/abs/1611.03470}{{\tt 1611.03470}}].

\bibitem{Giveon:2017nie}
A.~Giveon, N.~Itzhaki and D.~Kutasov, {\it {$ \mathrm{T}\overline{\mathrm{T}} $ and LST}},  {\em JHEP} {\bf 07} (2017) 122 [\href{http://arXiv.org/abs/1701.05576}{{\tt 1701.05576}}].

\bibitem{Gorbenko:2018oov}
V.~Gorbenko, E.~Silverstein and G.~Torroba, {\it {dS/dS and $ T\overline{T} $}},  {\em JHEP} {\bf 03} (2019) 085 [\href{http://arXiv.org/abs/1811.07965}{{\tt 1811.07965}}].

\bibitem{Kraus:2018xrn}
P.~Kraus, J.~Liu and D.~Marolf, {\it {Cutoff AdS$_{3}$ versus the $ T\overline{T} $ deformation}},  {\em JHEP} {\bf 07} (2018) 027 [\href{http://arXiv.org/abs/1801.02714}{{\tt 1801.02714}}].

\bibitem{Hartman:2018tkw}
T.~Hartman, J.~Kruthoff, E.~Shaghoulian and A.~Tajdini, {\it {Holography at finite cutoff with a $T^2$ deformation}},  {\em JHEP} {\bf 03} (2019) 004 [\href{http://arXiv.org/abs/1807.11401}{{\tt 1807.11401}}].

\bibitem{Guica:2019nzm}
M.~Guica and R.~Monten, {\it {$T\overline T$ and the mirage of a bulk cutoff}},  {\em SciPost Phys.} {\bf 10} (2021), no.~2 024 [\href{http://arXiv.org/abs/1906.11251}{{\tt 1906.11251}}].

\bibitem{Jiang:2019tcq}
Y.~Jiang, {\it {Expectation value of $T\overline{T}$ operator in curved spacetimes}},  {\em JHEP} {\bf 02} (2020) 094 [\href{http://arXiv.org/abs/1903.07561}{{\tt 1903.07561}}].

\bibitem{Jafari:2019qns}
G.~Jafari, A.~Naseh and H.~Zolfi, {\it {Path Integral Optimization for $T\overline{T}$ Deformation}},  {\em Phys. Rev. D} {\bf 101} (2020), no.~2 026007 [\href{http://arXiv.org/abs/1909.02357}{{\tt 1909.02357}}].

\bibitem{Griguolo:2021wgy}
L.~Griguolo, R.~Panerai, J.~Papalini and D.~Seminara, {\it {Nonperturbative effects and resurgence in Jackiw-Teitelboim gravity at finite cutoff}},  {\em Phys. Rev. D} {\bf 105} (2022), no.~4 046015 [\href{http://arXiv.org/abs/2106.01375}{{\tt 2106.01375}}].

\bibitem{Fichet:2023xbu}
S.~Fichet, E.~Megias and M.~Quiros, {\it {Holography of Linear Dilaton Spacetimes from the Bottom Up}},  \href{http://arXiv.org/abs/2309.02489}{{\tt 2309.02489}}.

\bibitem{He:2024xbi}
S.~He, Y.-z. Li and Y.~Xie, {\it {Holographic stress tensor correlators on higher genus Riemann surfaces}},  \href{http://arXiv.org/abs/2406.04042}{{\tt 2406.04042}}.

\bibitem{Doyon:2021tzy}
B.~Doyon, J.~Durnin and T.~Yoshimura, {\it {The Space of Integrable Systems from Generalised $T\overline{T}$-Deformations}},  {\em SciPost Phys.} {\bf 13} (2022), no.~3 072 [\href{http://arXiv.org/abs/2105.03326}{{\tt 2105.03326}}].

\bibitem{Cardy:2020olv}
J.~Cardy and B.~Doyon, {\it {$ T\overline{T} $ deformations and the width of fundamental particles}},  {\em JHEP} {\bf 04} (2022) 136 [\href{http://arXiv.org/abs/2010.15733}{{\tt 2010.15733}}].

\bibitem{Doyon:2023bvo}
B.~Doyon, F.~H\"ubner and T.~Yoshimura, {\it {New classical integrable systems from generalized $T\overline{T}$-deformations}},  \href{http://arXiv.org/abs/2311.06369}{{\tt 2311.06369}}.

\bibitem{Jiang:2019epa}
Y.~Jiang, {\it {A pedagogical review on solvable irrelevant deformations of 2D quantum field theory}},  {\em Commun. Theor. Phys.} {\bf 73} (2021), no.~5 057201 [\href{http://arXiv.org/abs/1904.13376}{{\tt 1904.13376}}].

\bibitem{Dubovsky:2023lza}
S.~Dubovsky, S.~Negro and M.~Porrati, {\it {Topological gauging and double current deformations}},  {\em JHEP} {\bf 05} (2023) 240 [\href{http://arXiv.org/abs/2302.01654}{{\tt 2302.01654}}].

\bibitem{Conti:2018tca}
R.~Conti, S.~Negro and R.~Tateo, {\it {The $T\overline{T} $ perturbation and its geometric interpretation}},  {\em JHEP} {\bf 02} (2019) 085 [\href{http://arXiv.org/abs/1809.09593}{{\tt 1809.09593}}].

\bibitem{Dubovsky_2017}
S.~Dubovsky, V.~Gorbenko and M.~Mirbabayi, {\it Asymptotic fragility, near {AdS}2 holography and {$\text{T}{\overline{\text{T}}}$}},  {\em JHEP} {\bf 2017} (sep, 2017).

\bibitem{Tolley:2019nmm}
A.~J. Tolley, {\it {$ T\overline{T} $ deformations, massive gravity and non-critical strings}},  {\em JHEP} {\bf 06} (2020) 050 [\href{http://arXiv.org/abs/1911.06142}{{\tt 1911.06142}}].

\bibitem{Caputa:2020lpa}
P.~Caputa, P.~Caputa, S.~Datta, S.~Datta, Y.~Jiang, Y.~Jiang, P.~Kraus and P.~Kraus, {\it {Geometrizing $ T\overline{T} $}},  {\em JHEP} {\bf 03} (2021) 140 [\href{http://arXiv.org/abs/2011.04664}{{\tt 2011.04664}}]. [Erratum: JHEP 09, 110 (2022)].

\bibitem{Frolov:2019nrr}
S.~Frolov, {\it {$T\overline{T}$ Deformation and the Light-Cone Gauge}},  {\em Proc. Steklov Inst. Math.} {\bf 309} (2020) 107--126 [\href{http://arXiv.org/abs/1905.07946}{{\tt 1905.07946}}].

\bibitem{Frolov:2019xzi}
S.~Frolov, {\it {$T{\overline T}$, $\widetilde JJ$, $JT$ and $\widetilde JT$ deformations}},  {\em J. Phys. A} {\bf 53} (2020), no.~2 025401 [\href{http://arXiv.org/abs/1907.12117}{{\tt 1907.12117}}].

\bibitem{Conti_2022}
R.~Conti, J.~Romano and R.~Tateo, {\it Metric approach to a {$\text{T}{\overline{\text{T}}}$}-like deformation in arbitrary dimensions},  {\em JHEP} {\bf 2022} (9, 2022) [\href{http://arXiv.org/abs/2206.03415}{{\tt 2206.03415}}].

\bibitem{Ferko:2022cix}
C.~Ferko, A.~Sfondrini, L.~Smith and G.~Tartaglino-Mazzucchelli, {\it {Root-${T}\overline{{T}}$ Deformations in Two-Dimensional Quantum Field Theories}},  {\em Phys. Rev. Lett.} {\bf 129} (2022), no.~20 201604 [\href{http://arXiv.org/abs/2206.10515}{{\tt 2206.10515}}].

\bibitem{Ferko:2023ruw}
C.~Ferko, L.~Smith and G.~Tartaglino-Mazzucchelli, {\it {Stress Tensor Flows, Birefringence in Non-Linear Electrodynamics, and Supersymmetry}},  {\em SciPost Phys.} {\bf 15} (2023) 198 [\href{http://arXiv.org/abs/2301.10411}{{\tt 2301.10411}}].

\bibitem{Babaei-Aghbolagh:2022uij}
H.~Babaei-Aghbolagh, K.~B. Velni, D.~M. Yekta and H.~Mohammadzadeh, {\it {Emergence of non-linear electrodynamic theories from $T\overline{T}$-like deformations}},  {\em Phys. Lett. B} {\bf 829} (2022) 137079 [\href{http://arXiv.org/abs/2202.11156}{{\tt 2202.11156}}].

\bibitem{Babaei-Aghbolagh:2022leo}
H.~Babaei-Aghbolagh, K.~Babaei~Velni, D.~Mahdavian~Yekta and H.~Mohammadzadeh, {\it {Marginal $T\overline{T}$-like deformation and modified Maxwell theories in two dimensions}},  {\em Phys. Rev. D} {\bf 106} (2022), no.~8 086022 [\href{http://arXiv.org/abs/2206.12677}{{\tt 2206.12677}}].

\bibitem{Hadasz:2024pew}
L.~Hadasz and R.~von Unge, {\it {Defining Root-$T\overline{T}$}},  \href{http://arXiv.org/abs/2405.17945}{{\tt 2405.17945}}.

\bibitem{Borsato:2022tmu}
R.~Borsato, C.~Ferko and A.~Sfondrini, {\it {Classical integrability of root-$T\overline{T}$ flows}},  {\em Phys. Rev. D} {\bf 107} (2023), no.~8 086011 [\href{http://arXiv.org/abs/2209.14274}{{\tt 2209.14274}}].

\bibitem{Taylor:2018xcy}
M.~Taylor, {\it {TT deformations in general dimensions}},  \href{http://arXiv.org/abs/1805.10287}{{\tt 1805.10287}}.

\bibitem{Babaei-Aghbolagh:2020kjg}
H.~Babaei-Aghbolagh, K.~Babaei~Velni, D.~M. Yekta and H.~Mohammadzadeh, {\it {$ T\overline{T} $-like flows in non-linear electrodynamic theories and S-duality}},  {\em JHEP} {\bf 04} (2021) 187 [\href{http://arXiv.org/abs/2012.13636}{{\tt 2012.13636}}].

\bibitem{Hou:2022csf}
J.~Hou, {\it {$ T\overline{T} $ flow as characteristic flows}},  {\em JHEP} {\bf 03} (2023) 243 [\href{http://arXiv.org/abs/2208.05391}{{\tt 2208.05391}}].

\bibitem{Ferko:2023sps}
C.~Ferko, Y.~Hu, Z.~Huang, K.~Koutrolikos and G.~Tartaglino-Mazzucchelli, {\it {{$\text{T}{\overline{\text{T}}}$}-Like {F}lows and $3d$ {N}onlinear {S}upersymmetry}},  \href{http://arXiv.org/abs/2302.10410}{{\tt 2302.10410}}.

\bibitem{Bandos:2020jsw}
I.~Bandos, K.~Lechner, D.~Sorokin and P.~K. Townsend, {\it {A non-linear duality-invariant conformal extension of Maxwell's equations}},  {\em Phys. Rev. D} {\bf 102} (2020) 121703 [\href{http://arXiv.org/abs/2007.09092}{{\tt 2007.09092}}].

\bibitem{Ferko:2022iru}
C.~Ferko, L.~Smith and G.~Tartaglino-Mazzucchelli, {\it {On Current-Squared Flows and ModMax Theories}},  {\em SciPost Phys.} {\bf 13} (2022), no.~2 012 [\href{http://arXiv.org/abs/2203.01085}{{\tt 2203.01085}}].

\bibitem{Sorokin:2021tge}
D.~P. Sorokin, {\it {Introductory Notes on Non-linear Electrodynamics and its Applications}},  {\em Fortsch. Phys.} {\bf 70} (2022), no.~7-8 2200092 [\href{http://arXiv.org/abs/2112.12118}{{\tt 2112.12118}}].

\bibitem{Lechner:2022qhb}
K.~Lechner, P.~Marchetti, A.~Sainaghi and D.~P. Sorokin, {\it {Maximally symmetric nonlinear extension of electrodynamics and charged particles}},  {\em Phys. Rev. D} {\bf 106} (2022), no.~1 016009 [\href{http://arXiv.org/abs/2206.04657}{{\tt 2206.04657}}].

\bibitem{Ferko:2019oyv}
C.~Ferko, H.~Jiang, S.~Sethi and G.~Tartaglino-Mazzucchelli, {\it {Non-linear supersymmetry and $ T\overline{T} $-like flows}},  {\em JHEP} {\bf 02} (2020) 016 [\href{http://arXiv.org/abs/1910.01599}{{\tt 1910.01599}}].

\bibitem{Morone:2024ffm}
T.~Morone, S.~Negro and R.~Tateo, {\it {Gravity and $\TTb$ flows in higher dimensions}},  {\em Nucl. Phys. B} {\bf 1005} (2024) 116605 [\href{http://arXiv.org/abs/2401.16400}{{\tt 2401.16400}}].

\bibitem{Babaei-Aghbolagh:2024hti}
H.~Babaei-Aghbolagh, S.~He, T.~Morone, H.~Ouyang and R.~Tateo, {\it {Geometric formulation of generalized root-$\TTb$ deformations}},  \href{http://arXiv.org/abs/2405.03465}{{\tt 2405.03465}}.

\bibitem{Tsolakidis:2024wut}
E.~Tsolakidis, {\it {Massive gravity generalization of $T\overline{T}$ deformations}},  \href{http://arXiv.org/abs/2405.07967}{{\tt 2405.07967}}.

\bibitem{Floss:2023nod}
T.~Fl\"oss, D.~Roest and T.~Westerdijk, {\it {Non-linear Electrodynamics from Massive Gravity}},  \href{http://arXiv.org/abs/2308.04349}{{\tt 2308.04349}}.

\bibitem{Vollick:2003qp}
D.~N. Vollick, {\it {Palatini approach to Born-Infeld-Einstein theory and a geometric description of electrodynamics}},  {\em Phys. Rev. D} {\bf 69} (2004) 064030 [\href{http://arXiv.org/abs/gr-qc/0309101}{{\tt gr-qc/0309101}}].

\bibitem{Banados:2008rm}
M.~Banados, {\it {Eddington-Born-Infeld action for dark matter and dark energy}},  {\em Phys. Rev. D} {\bf 77} (2008) 123534 [\href{http://arXiv.org/abs/0801.4103}{{\tt 0801.4103}}].

\bibitem{Banados:2008fj}
M.~Banados, P.~G. Ferreira and C.~Skordis, {\it {Eddington-Born-Infeld gravity and the large scale structure of the Universe}},  {\em Phys. Rev. D} {\bf 79} (2009) 063511 [\href{http://arXiv.org/abs/0811.1272}{{\tt 0811.1272}}].

\bibitem{Banados_2010}
M.~Bañados and P.~G. Ferreira, {\it Eddington’s theory of gravity and its progeny},  {\em Physical Review Letters} {\bf 105} (2010), no.~1.

\bibitem{jimenez_2018}
J.~Beltrán~Jiménez, L.~Heisenberg, G.~J. Olmo and D.~Rubiera-Garcia, {\it Born–infeld inspired modifications of gravity},  {\em Physics Reports} {\bf 727} (1, 2018) 1–129 [\href{http://arXiv.org/abs/1704.03351v2}{{\tt 1704.03351v2}}].

\bibitem{Olmo:2020fnk}
G.~J. Olmo, E.~Orazi and D.~Rubiera-Garcia, {\it {Multicenter solutions in Eddington-inspired Born\textendash{}Infeld gravity}},  {\em Eur. Phys. J. C} {\bf 80} (2020), no.~11 1018 [\href{http://arXiv.org/abs/2006.08180}{{\tt 2006.08180}}].

\bibitem{Guerrero:2020azx}
M.~Guerrero, G.~Mora-P\'erez, G.~J. Olmo, E.~Orazi and D.~Rubiera-Garcia, {\it {Rotating black holes in Eddington-inspired Born-Infeld gravity: an exact solution}},  {\em JCAP} {\bf 07} (2020) 058 [\href{http://arXiv.org/abs/2006.00761}{{\tt 2006.00761}}].

\bibitem{Nascimento:2019qor}
J.~R. Nascimento, G.~J. Olmo, P.~J. Porf\'\i{}rio, A.~Y. Petrov and A.~R. Soares, {\it {Nonlinear $\sigma$-models in the {E}ddington-inspired {B}orn-{I}nfeld {G}ravity}},  {\em Phys. Rev. D} {\bf 101} (2020), no.~6 064043 [\href{http://arXiv.org/abs/1912.10779}{{\tt 1912.10779}}].

\bibitem{Afonso:2019fzv}
V.~I. Afonso, G.~J. Olmo, E.~Orazi and D.~Rubiera-Garcia, {\it {New scalar compact objects in {R}icci-based gravity theories}},  {\em JCAP} {\bf 12} (2019) 044 [\href{http://arXiv.org/abs/1906.04623}{{\tt 1906.04623}}].

\bibitem{Pani:2012qb}
P.~Pani, T.~Delsate and V.~Cardoso, {\it {Eddington-inspired Born-Infeld gravity. Phenomenology of non-linear gravity-matter coupling}},  {\em Phys. Rev. D} {\bf 85} (2012) 084020 [\href{http://arXiv.org/abs/1201.2814}{{\tt 1201.2814}}].

\bibitem{Banerjee:2021auy}
P.~Banerjee, D.~Garain, S.~Paul, R.~Shaikh and T.~Sarkar, {\it {A Stellar Constraint on Eddington-inspired Born\textendash{}Infeld Gravity from Cataclysmic Variable Binaries}},  {\em Astrophys. J.} {\bf 924} (2022), no.~1 20 [\href{http://arXiv.org/abs/2105.09172}{{\tt 2105.09172}}].

\bibitem{Olmo:2013gqa}
G.~J. Olmo, D.~Rubiera-Garcia and H.~Sanchis-Alepuz, {\it {Geonic black holes and remnants in Eddington-inspired Born-Infeld gravity}},  {\em Eur. Phys. J. C} {\bf 74} (2014) 2804 [\href{http://arXiv.org/abs/1311.0815}{{\tt 1311.0815}}].

\bibitem{Pereira:2023bxt}
C.~F.~S. Pereira, A.~R. Soares, R.~L.~L. Vit\'oria and H.~Belich, {\it {Bosonic quantum dynamics in Eddington-inspired Born\textendash{}Infeld gravity global monopole spacetime}},  {\em Eur. Phys. J. C} {\bf 83} (2023), no.~4 270.

\bibitem{Hamilton1982}
R.~S. Hamilton, {\it Three-manifolds with positive {R}icci curvature},  {\em Journal of Differential Geometry} {\bf 17} (Jan., 1982).

\bibitem{rbf::15}
G.~Catino, L.~Cremaschi, Z.~Djadli, C.~Mantegazza and L.~Mazzieri, {\it The {R}icci-{B}ourguignon flow},  {\em Pacific Journal of Mathematics} {\bf 287} (2017), no.~2 337--370 [\href{http://arXiv.org/abs/1507.00324}{{\tt 1507.00324}}].

\bibitem{Aramini:2022wbn}
F.~Aramini, N.~Brizio, S.~Negro and R.~Tateo, {\it {Deforming the ODE/IM correspondence with $ \textrm{T}\overline{\textrm{T}} $}},  {\em JHEP} {\bf 03} (2023) 084 [\href{http://arXiv.org/abs/2212.13957}{{\tt 2212.13957}}].

\bibitem{Ebert:2022ehb}
S.~Ebert, C.~Ferko, H.-Y. Sun and Z.~Sun, {\it {$T\overline{T}$ in JT Gravity and BF Gauge Theory}},  {\em SciPost Phys.} {\bf 13} (2022), no.~4 096 [\href{http://arXiv.org/abs/arXiv:2205.07817}{{\tt arXiv:2205.07817}}].

\bibitem{Conti:2018jho}
R.~Conti, L.~Iannella, S.~Negro and R.~Tateo, {\it {Generalised Born-Infeld models, Lax operators and the $ \mathrm{T}\overline{\mathrm{T}} $ perturbation}},  {\em JHEP} {\bf 11} (2018) 007 [\href{http://arXiv.org/abs/1806.11515}{{\tt 1806.11515}}].

\bibitem{DiFrancesco:1997nk}
P.~Di~Francesco, P.~Mathieu and D.~Senechal, {\em {Conformal Field Theory}}.
\newblock Graduate Texts in Contemporary Physics. Springer-Verlag, New York, 1997.

\bibitem{Griffiths}
G.~Griffiths and W.~Schiesser, {\it Linear and nonlinear waves},  {\em Scholarpedia} {\bf 4} (01, 2009) 4308.

\bibitem{Hirota:1971zz}
R.~Hirota, {\it {Exact Solution of the Korteweg-de Vries Equation for Multiple Collisions of Solitons}},  {\em Phys. Rev. Lett.} {\bf 27} (1971) 1192--1194.

\bibitem{Faddeev:1974em}
L.~D. Faddeev, L.~A. Takhtajan and V.~E. Zakharov, {\it {Complete description of solutions of the Sine-Gordon equation}},  {\em Dokl. Akad. Nauk Ser. Fiz.} {\bf 219} (1974) 1334--1337.

\bibitem{Conti:2019dxg}
R.~Conti, S.~Negro and R.~Tateo, {\it {Conserved currents and $\text{T}\overline{\text{T}}_s$ irrelevant deformations of 2D integrable field theories}},  {\em JHEP} {\bf 11} (2019) 120 [\href{http://arXiv.org/abs/1904.09141}{{\tt 1904.09141}}].

\bibitem{Dorey:1998pt}
P.~Dorey and R.~Tateo, {\it {Anharmonic oscillators, the thermodynamic Bethe ansatz, and nonlinear integral equations}},  {\em J. Phys. A} {\bf 32} (1999) L419--L425 [\href{http://arXiv.org/abs/hep-th/9812211}{{\tt hep-th/9812211}}].

\bibitem{Bazhanov:1998wj}
V.~V. Bazhanov, S.~L. Lukyanov and A.~B. Zamolodchikov, {\it {Spectral determinants for Schrodinger equation and Q operators of conformal field theory}},  {\em J. Statist. Phys.} {\bf 102} (2001) 567--576 [\href{http://arXiv.org/abs/hep-th/9812247}{{\tt hep-th/9812247}}].

\bibitem{Dorey:2007zx}
P.~Dorey, C.~Dunning and R.~Tateo, {\it {The ODE/IM Correspondence}},  {\em J. Phys. A} {\bf 40} (2007) R205 [\href{http://arXiv.org/abs/hep-th/0703066}{{\tt hep-th/0703066}}].

\bibitem{Dorey:2019ngq}
P.~Dorey, C.~Dunning, S.~Negro and R.~Tateo, {\it {Geometric aspects of the ODE/IM correspondence}},  {\em J. Phys. A} {\bf 53} (2020), no.~22 223001 [\href{http://arXiv.org/abs/1911.13290}{{\tt 1911.13290}}].

\bibitem{Lukyanov:2010rn}
S.~L. Lukyanov and A.~B. Zamolodchikov, {\it {Quantum Sine(h)-Gordon Model and Classical Integrable Equations}},  {\em JHEP} {\bf 07} (2010) 008 [\href{http://arXiv.org/abs/1003.5333}{{\tt 1003.5333}}].

\bibitem{Ebert:2023tih}
S.~Ebert, C.~Ferko and Z.~Sun, {\it {Root-$\TTb$ deformed boundary conditions in holography}},  {\em Phys. Rev. D} {\bf 107} (2023), no.~12 126022 [\href{http://arXiv.org/abs/2304.08723}{{\tt 2304.08723}}].

\bibitem{Mitskievich:1998uk}
N.~V. Mitskievich, {\it {Modeling general relativistic perfect fluids in field theoretic language}},  {\em Int. J. Theor. Phys.} {\bf 38} (1999) 997--1016 [\href{http://arXiv.org/abs/gr-qc/9811077}{{\tt gr-qc/9811077}}].

\bibitem{BeltranJimenez:2017doy}
J.~Beltran~Jimenez, L.~Heisenberg, G.~J. Olmo and D.~Rubiera-Garcia, {\it {Born-{I}nfeld inspired modifications of gravity}},  {\em Phys. Rept.} {\bf 727} (2018) 1--129 [\href{http://arXiv.org/abs/1704.03351}{{\tt 1704.03351}}].

\bibitem{BeltranJimenez:2019acz}
J.~Beltr\'an~Jim\'enez and A.~Delhom, {\it {Ghosts in metric-affine higher order curvature gravity}},  {\em Eur. Phys. J. C} {\bf 79} (2019), no.~8 656 [\href{http://arXiv.org/abs/1901.08988}{{\tt 1901.08988}}].

\bibitem{Ferko:2024yhc}
C.~Ferko and C.~Luke~Martin, {\it {Field-Dependent Metrics and Higher-Form Symmetries in Duality-Invariant Theories of Non-Linear Electrodynamics}},  6, 2024.
\newblock \href{http://arXiv.org/abs/2406.17194}{{\tt 2406.17194}}.

\bibitem{2002math11159P}
G.~{Perelman}, {\it {The entropy formula for the Ricci flow and its geometric applications}},  {\em arXiv Mathematics e-prints} (Nov., 2002) math/0211159 [\href{http://arXiv.org/abs/math/0211159}{{\tt math/0211159}}].

\bibitem{Aldrovandi:2001vx}
R.~Aldrovandi, A.~L. Barbosa, M.~Calcada and J.~G. Pereira, {\it {Kinematics of a space-time with an infinite cosmological constant}},  {\em Found. Phys.} {\bf 33} (2003) 613--624 [\href{http://arXiv.org/abs/gr-qc/0105068}{{\tt gr-qc/0105068}}].

\bibitem{Aldrovandi:2004km}
R.~Aldrovandi, J.~P. Almeida and J.~G. Pereira, {\it {A Singular conformal universe}},  {\em J. Geom. Phys.} {\bf 56} (2006) 1042--1056 [\href{http://arXiv.org/abs/gr-qc/0403099}{{\tt gr-qc/0403099}}].

\bibitem{sym12020289}
S.~Deshmukh and H.~Alsodais, {\it A note on {R}icci solitons},  {\em Symmetry} {\bf 12} (2020), no.~2.

\bibitem{Aldrovandi:1998ux}
R.~Aldrovandi and J.~G. Pereira, {\it {A Second Poincare group}},  in {\em {Conference on Topics in Theoretical Physics II: Festschrift for A.H. Zimerman}}, 9, 1998.
\newblock \href{http://arXiv.org/abs/gr-qc/9809061}{{\tt gr-qc/9809061}}.

\bibitem{Mitskievich:2002nj}
N.~V. Mitskievich, {\it {Space-times, electromagnetism and fluids: A Revision of traditional concepts}},  {\em Rev. Mex. Fis.} {\bf 49S2} (2003) 39--51 [\href{http://arXiv.org/abs/gr-qc/0202032}{{\tt gr-qc/0202032}}].

\bibitem{Bronnikov:2000vy}
K.~A. Bronnikov, {\it {Regular magnetic black holes and monopoles from nonlinear electrodynamics}},  {\em Phys. Rev. D} {\bf 63} (2001) 044005 [\href{http://arXiv.org/abs/gr-qc/0006014}{{\tt gr-qc/0006014}}].

\bibitem{Gunasekaran:2012dq}
S.~Gunasekaran, R.~B. Mann and D.~Kubiznak, {\it {Extended phase space thermodynamics for charged and rotating black holes and Born-Infeld vacuum polarization}},  {\em JHEP} {\bf 11} (2012) 110 [\href{http://arXiv.org/abs/1208.6251}{{\tt 1208.6251}}].

\bibitem{Flores-Alfonso:2020euz}
D.~Flores-Alfonso, B.~A. Gonz\'alez-Morales, R.~Linares and M.~Maceda, {\it {Black holes and gravitational waves sourced by non-linear duality rotation-invariant conformal electromagnetic matter}},  {\em Phys. Lett. B} {\bf 812} (2021) 136011 [\href{http://arXiv.org/abs/2011.10836}{{\tt 2011.10836}}].

\bibitem{Ferko:2024ali}
C.~Ferko and L.~Smith, {\it {An Infinite Family of Integrable Sigma Models Using Auxiliary Fields}},  \href{http://arXiv.org/abs/2405.05899}{{\tt 2405.05899}}.

\bibitem{Bielli:2024khq}
D.~Bielli, C.~Ferko, L.~Smith and G.~Tartaglino-Mazzucchelli, {\it {T-Duality and $T \overline{T}$-like Deformations of Sigma Models}},  \href{http://arXiv.org/abs/2407.11636}{{\tt 2407.11636}}.

\bibitem{Bielli:2024ach}
D.~Bielli, C.~Ferko, L.~Smith and G.~Tartaglino-Mazzucchelli, {\it {Integrable Higher-Spin Deformations of Sigma Models from Auxiliary Fields}},  \href{http://arXiv.org/abs/2407.16338}{{\tt 2407.16338}}.

\bibitem{Ferko:2024zth}
C.~Ferko, S.~M. Kuzenko, K.~Lechner, D.~P. Sorokin and G.~Tartaglino-Mazzucchelli, {\it {Interacting Chiral Form Field Theories and $T\overline T$-like Flows in Six and Higher Dimensions}},  \href{http://arXiv.org/abs/2402.06947}{{\tt 2402.06947}}.

\bibitem{Deger:2024jfh}
N.~S. Deger, A.~J. Murcia and D.~P. Sorokin, {\it {Waves and strings in an interacting conformal chiral 2-form theory in six dimensions}},  \href{http://arXiv.org/abs/2405.20375}{{\tt 2405.20375}}.

\end{thebibliography}\endgroup

\end{document}